\documentclass[useAMS,usedcolumn,usenatbib,usegraphicx]{mn2e}

\usepackage{epsfig,graphicx,natbib,subfig}
\usepackage{./reference/mycite}
\usepackage{amssymb}
\usepackage{gensymb}
\usepackage{amsfonts}
\usepackage{amsmath}
\usepackage{color}
\usepackage{lscape,longtable}
\usepackage{times}
 
\def \kms{\ensuremath{{\rm \,km\,s}^{-1}}}

\begin{document}

\title[{Heavily Reddened Type 1 Quasars at $z>2$}]{Heavily Reddened Type 1 Quasars at \boldmath{z $>$ 2} I: Evidence for Significant Obscured Black-Hole Growth at the Highest Quasar Luminosities}
\author[M. Banerji et al.]{ \parbox{\textwidth}
{Manda Banerji$^{1,2,3}$\thanks{E-mail: m.banerji@ucl.ac.uk}, S. Alaghband-Zadeh$^{2,3}$, Paul C. Hewett$^{2}$, Richard G. McMahon$^{2,3}$}
  \vspace*{6pt} \\
$^{1}$Department of Physics \& Astronomy, University College London, Gower Street, London WC1E 6BT, UK. \\
$^{2}$Institute of Astronomy, University of Cambridge, Madingley Road, Cambridge, CB3 0HA, UK.\\
$^{3}$Kavli Institute for Cosmology, University of Cambridge, Madingley Road, Cambridge, CB3 0HA, UK.\\
} 

\maketitle

\begin{abstract} 

We present a new population of $z>2$ dust-reddened, Type 1
quasars with $0.5 \lesssim E(B-V) \lesssim 1.5$, selected using near infra-red (NIR) imaging data
from the UKIDSS-LAS, ESO-VHS and \textit{WISE} surveys.  NIR
spectra obtained using the Very Large Telescope (VLT) for 24 new objects bring our total sample of
spectroscopically confirmed hyperluminous ($>$10$^{13}$L$_\odot$), high-redshift dusty quasars to 38.  There is no evidence for reddened quasars
having significantly different H$\alpha$ equivalent widths relative to unobscured quasars. The average black-hole masses ($\sim$10$^{9}$-10$^{10}$M$_\odot$)
and bolometric luminosities ($\sim$10$^{47}$ erg/s) are comparable to the most luminous
unobscured quasars at the same redshift, but with a tail extending to very high luminosities of
$\sim$10$^{48}$ erg/s. Sixty-six per cent of the reddened quasars are
detected at $>3\sigma$ at 22$\mu$m by \textit{WISE}. The average
6$\mu$m rest-frame luminosity is log$_{10}$(L$_{6\mu\rm{m}}$/erg
s$^{-1}$)=47.1$\pm$0.4, making the objects among the mid-infrared
brightest AGN currently known.  The extinction-corrected space-density estimate now extends over three magnitudes ($-30 < M_i < -27$) and demonstrates
that the reddened quasar luminosity function is significantly flatter than that of the unobscured quasar population at
$z=2-3$. At the brightest magnitudes, $M_i \lesssim -29$, the space
density of our dust-reddened population exceeds that of unobscured
quasars. A model where the probability that a quasar becomes
dust-reddened increases at high luminosity is consistent with the
observations and such a dependence could be explained by an increase
in luminosity and extinction during AGN-fuelling phases.  The
properties of our obscured Type 1 quasars are distinct from the
heavily obscured, Compton-thick AGN that have been identified at much
fainter luminosities and we conclude that they likely correspond to a
brief evolutionary phase in massive galaxy formation.
\end{abstract}


\section{INTRODUCTION}

Luminous dust-obscured quasars have been postulated to represent the
missing evolutionary link between merger-induced starburst galaxies and ultraviolet
(UV) bright quasars \citep{Sanders:88} in theoretical models of galaxy formation
(e.g. \citealt{Hopkins:08, Narayanan:10}). Samples of obscured active galactic
nuclei (AGN) have been selected both in the hard X-ray
(e.g. \citealt{Brusa:10, Civano:12, LaMassa:13}) and the mid-infrared
(e.g. \citealt{Stern:12, Assef:13, Donley:12}). However, these surveys
do not distinguish between a centrally concentrated obscuring medium,
on parsec scales, close to the black-hole accretion disk
\citep{Urry:95, Antonucci:93}, and larger-scale obscuration due to gas
and dust clouds in the quasar host galaxy \citep{Martinez:05}. Most
hard X-ray and mid-infrared surveys (e.g. with the \textit{Spitzer} Space Telescope) have covered relatively small
areas of sky, probing obscured black-hole growth in relatively
low-luminosity AGN, although the situation is now changing at mid infrared wavelengths with data from the \textit{WISE} All-Sky Survey \citep{Wright:10}. 
In these faint samples, the number of obscured AGN
is found to decrease with increasing luminosity. Constraints on
obscured black-hole growth at the highest quasar luminosities however
remain highly uncertain and the relationship between hyperluminous ($>10^{13}$L$_\odot$)
obscured quasars and the now well-studied obscured AGN at low
luminosities is also uncertain.

X-ray surveys also show a curious lack of high Eddington ratio
($L/L_{\rm Edd}>0.1$) objects with N$_{\rm{H}}$ between 3$\times$10$^{21}$
and 3$\times$10$^{22}$ cm$^{-2}$ \citep{Fabian:12,
Raimundo:11}. Objects with such properties, i.e. high luminosity
combined with a significant gas/dust column, could possess massive
AGN-driven outflows. The apparent upper limit to the distribution of
Eddington ratios could be a selection bias, as most X-ray surveys do
not cover a large enough area to detect the rarest and brightest
quasars with high Eddington ratios. Such objects have traditionally
been selected through their rest-frame UV emission which is very
sensitive to dust extinction and therefore biased against detecting
luminous quasars with even moderate gas column densities.  The Sloan
Digital Sky Survey (SDSS) Data Release (DR) 7 quasar sample
\citep{Schneider:10} now mitigates the situation to a degree.  The
flux limit in the redder $i$-band allows some intrinsically very
luminous objects with $E(B-V)\simeq$0.5 to be included
\citep{Richards:03}. The new SDSS-III BOSS DR10 quasar survey
\citep{Paris:13} reaches fainter optical magnitudes than the SDSS and
has significantly increased the number of quasars known at the main
epoch of galaxy formation at $z\sim 2-3$. Populations of red quasars
are detected in the BOSS data \citep{Ross:14} including the first
large sample of narrow-line Type 2 obscured quasars at high redshifts
\citep{Alexandroff:13}, some of which also show evidence for broad-lines in the near infrared \citep{Greene:14}. 
However, even with the fainter flux-limit,
BOSS cannot find quasars with $E(B-V) \gtrsim 0.6$.

Wide-field infrared surveys now cover much of the observable sky,
providing the sensitivity to detect dusty, intrinsically luminous,
broad-line quasars. The 2MASS survey has identified
optically obscured quasars that are also radio-loud (e.g. \citealt{Glikman:07,
Urrutia:09, Glikman:12}); the radio selection using FIRST data is
necessary to eliminate contaminant populations but means that the 
majority of AGN are excluded by the selection. In radio-loud red quasars (e.g. \citealt{Webster:95}), the red colours can arise
as a result of the presence of a red synchrotron component \citep{Francis:01, Whiting:01} rather than
due to dust. The flux-limit of the
2MASS survey has also meant that these red quasars are predominantly at low-redshift although
with some of the most luminous quasars seen out to $z>2$ \citep{Glikman:12}. A concern at low-redshifts
is that host galaxy contamination can redden the observed colours of these quasars. The
galaxy spectral energy distribution (SED) peaks at $\sim$1$\mu$m, corresponding to the inflection
point in the AGN SED. This feature is in the NIR $JHK$ filters at $z\sim 0.2-1$. However, at $z\sim$0.7, the FIRST-2MASS
red quasars are seen to be associated with merging galaxy hosts
\citep{Urrutia:08}, consistent with an interpretation where the
objects are in a transition phase from starburst to unobscured quasar. Extending these investigations to focus on the main epoch of 
galaxy and black-hole formation at $z\sim2-3$ will be in instructive in understanding the importance of this obscured
phase in galaxy formation.

The \textit{WISE} All-Sky Survey operating at mid-infrared
wavelengths, has recently led to the identification of a new
population of hyperluminous dust-obscured galaxies (HyLIRGs), whose
spectral energy distributions (SEDs) are consistent with a dominant
contribution from AGN \citep{Eisenhardt:12, Wu:12}. These HyLIRGs are
also postulated to represent the evolutionary link between massive
starbursts and luminous quasars. The \textit{WISE} samples are complementary to near infrared selected samples
as they enable the selection of much more highly obscured AGN. The AGN that have so far been 
identified by \textit{WISE} are seen to be Compton thick
\citep{Stern:14}. Broad emission lines originating from gas
close to the black-hole are therefore completely concealed and estimating virial
black-hole masses is no longer possible.  

The wide-field NIR UKIDSS Large Area Survey (LAS) \citep{Lawrence:07}
and VISTA Hemisphere Survey (VHS; \citealt{McMahon:13}) allow us to
probe much fainter than was possible using 2MASS and enable selection of complementary 
samples of obscured quasars to the \textit{WISE} survey. In a survey over
$>$1000 deg$^2$ of the UKIDSS and VHS data we were able to identify a
population of $z\sim2-3$, $E(B-V)\sim 0.8$ quasars
(\citealt{Banerji:12, Banerji:13} - hereafter B12 and B13
respectively), demonstrating that contaminant populations can be
eliminated without having to rely on radio pre-selection.  Our new
quasars are among the most bolometrically luminous quasars known
(L$_{\rm{bol}}>$10$^{47}$ erg/s) but, as a result of their significant
dust-extinctions, are often invisible in wide-field optical surveys
like SDSS. The sample represents a population of hyperluminous
quasars at $z\sim2-3$ whose extinction properties are intermediate
between UV-selected, essentially unobscured, quasars and the very
heavily obscured AGN emerging from \textit{WISE}. 
Similar obscured luminous quasars showing strong evidence for corresponding
to the AGN feedback phase, are also now emerging from multi-wavelength degree-scale
galaxy surveys such as COSMOS \citep{Brusa:10, Bongiorno:14, Brusa:14}. Indeed the brightest obscured quasar
selected from the COSMOS survey is the faintest obscured quasar selected in our wide-field UKIDSS search for
similar objects (B12). The ability to isolate the luminous AGN population with intermediate
levels of extinction means that it is now possible to investigate
objects with extinctions corresponding to reddenings of $0 \lesssim
E(B-V) \lesssim 3$. New insight into the importance of non-spherical
geometries (i.e. `unified schemes') and evolution, involving specific
phases during AGN lifetimes, for explaining the distribution of
extinction properties can be expected.  

Our survey also led to the identification of the prototype quasar:
ULASJ1234+0907, the reddest broad-line quasar currently known. With an
Eddington ratio of 0.65 (B12) and a neutral hydrogen column-density of
9$\times10^{21}$ cm$^{-2}$ \citep{Banerji:14}, ULASJ1234+0907 lies in
the sparsley-populated high $L/L_{\rm Edd}$ and intermediate gas column
regime found from the statistics of hard X-ray sources.  Among the
most luminous quasars known at both X-ray and far-infrared
wavelengths, ULASJ1234+0907 may represent the emerging phase of a
supermassive black-hole shortly after a starburst \citep{Banerji:14}.
Larger surveys for NIR-selected, dust-obscured, luminous
quasars may unearth more objects like ULASJ1234+0907. 

The current paper presents a much larger survey for heavily
dust-reddened, hyperluminous broad-line quasars at $z=2-3$, where the
abundance of gas-rich, star-forming galaxies provides a natural
explanation for the presence of extended obscuring media in these
luminous quasars.  Combined with objects presented in B12 and B13, the
new sample allows us to characterise in detail the demographics of the
luminous reddened quasar population at $z\sim 2.5$, including the
direct calculation of the space density over a three-magnitude
interval.  We emphasise that the parameter space sampled by our survey
is very different from that of X-ray surveys for obscured AGN in deep
fields covering only a few square degrees.  A companion paper
(Alaghband-Zadeh et al., in preparation) investigates the spatially
resolved properties of the new reddened quasars from the IFU data
presented here. 

The structure of the paper is as follows. First, the observational
data, including the photometric selection (Section 2) and
spectroscopic follow-up (Section 3), is described. The generation of
the key physical properties of the quasars, including extinction
estimates, luminosities and black-hole masses, is covered in Section
4, prior to consideration of the object properties derived from WISE
photometric data in Section 5. The primary scientific result, the
calculation of the population space density and luminosity function is
presented in Section 6, followed by a summary of the main results in
Section 7.  Throughout the paper we assume a flat $\Lambda$CDM
cosmology with h$_0$=0.70, $\Omega_m=0.30$ and
$\Omega_\Lambda$=0.70. The VHS magnitudes are on the AB system where
the conversions from Vega to AB used are: $J=+0.937$, $H=+1.384$,
$K=1.839$. All \textit{WISE} magnitudes and colours are on the Vega
system.

\section{PHOTOMETRIC SELECTION}

\label{sec:phot}

\subsection{Colour Selection of Reddened Quasars}

We make use of new wide-field infrared photometry over $\sim$1500
deg$^2$ from VHS and \textit{WISE}, in order to identify luminous
reddened quasar candidates in the southern hemisphere. In addition, we
also use $>$3000 deg$^2$ of imaging data from the UKIDSS Large Area
Survey (LAS) \citep{Lawrence:07} in the northern hemisphere, to select a
sample of the reddest quasars, such as ULASJ1234+0907
\citep{Banerji:14}. The details of the colour selection has been
presented in B13 but we provide a summary here. Candidates must be
morphologically classified as point sources in the $K$-band (i.e. with $kclass=-1$; \citealt{Gonzalez-Solares:11})
and possess extremely red NIR colours of $(J-K)_{\rm{AB}}>1.6$. The
morphology cut is necessary to ensure that we isolate a population of
quasars at high redshifts corresponding to the peak epoch of
black-hole accretion activity, where the host galaxy light in the NIR
is sub-dominant. The red colours of these quasars in the NIR, can then
be attributed to dust-reddening rather than galaxy starlight. A
\textit{WISE} colour-cut of $(W1-W2)>0.85$ is then applied to separate
quasars from stars. Candidates are required to be detected at a
signal-to-noise-ratio (S/N) of $>$5 in both the \textit{WISE} $W1$ and
$W2$ bands. As shown in B13, this selection is extremely effective in
isolating rare populations of heavily reddened quasars at high
redshifts.

Our quasar sample encompasses several different survey regions. Below we
describe the candidate selection criteria used in each of these regions
as well as ancillary multi-wavelength data that were used to
prioritise targets for spectroscopy. After applying the colour and
morphology selections in each of our survey areas, all candidates were
visually inspected in order to remove spurious sources. The most
common such spurious sources are instances of close neighbours in the
VHS data with separations of $<$3\,arcsec and where the \textit{WISE}
photometry appears to be blended, making the \textit{WISE}
colours unreliable. The survey regions, selection criteria and number
of candidates are summarised in Table \ref{tab:area}.

\begin{itemize}

\item{\textit{The VHS-DES Stripe82 Region}: This region
(20h$<$RA$<$0h, $-2\degree<DEC<2\degree$) overlaps the SDSS
Stripe82 multi-epoch survey where the coadded data reaches
$\simeq$2\,mag fainter than the main SDSS survey \citep{Annis:11}. 
As discussed in B12, we used the deep coadd SDSS Stripe82
photometry to select only the subset of red $(J-K)$ candidates that are
also faint and red at optical wavelengths. Candidates were therefore
required to satisfy the additional selection criteria:
$i_{\rm{AB}}>20.5$ and $(i-K)_{\rm{AB}}>2.5$.

Many of our infrared-selected candidates that failed this cut and are
blue and bright in the optical, possess SDSS or BOSS spectra. Later,
in Section \ref{sec:optical}, we show the SDSS spectra to demonstrate
which object populations are eliminated by the optical cuts and to
determine what the contaminants may be in regions where we do not yet
have optical coverage. Our selection produced 10 candidates down to a
flux limit of $K_{\rm{AB}}<18.8$ over 129\,deg$^2$, after visual
inspection to remove spurious and blended sources. Two of the 10,
ULASJ2150$-$0055 and ULASJ2224$-$0015, were observed as part of the
sample described in B12. One more candidate (SDSSJ211805.27+010344.7) has an optical spectrum from
SDSS where it is classified as a $z\sim5$ quasar, although with low
confidence. The optical spectrum does not show any obvious emission
lines. The remaining seven sources were added to our spectroscopic
target list.}

\item{\textit{The VHS-DES SPT Region}: In this region (20h$<$RA$<$0h,
$-60\degree<DEC<-45\degree$) we identify a total of 29 candidates down to a flux
limit of $K_{AB}<18.8$ over 498\,deg$^2$. In addition, over the
83\,deg$^2$ area of the `SPT Deep Field' (centred at 23h$<$RA$<$00h and
$-60\degree<DEC<-50\degree$), the selection was extended to
$K_{AB}<19.3$, giving four additional candidates. The SPT Deep Field
benefits from multi-wavelength coverage from \textit{Spitzer}
\citep{Ashby:13}, \textit{Herschel} and \textit{XMM-Newton}, which can
be used to investigate the multi-wavelength properties of our
quasars.}

\item{\textit{The VHS-ATLAS SGC Region:} In this region covered by VHS (20h$<$RA$<$0h,
$-20\degree<DEC<-2\degree$)  we only selected candidates over the $\sim$488\,deg$^2$ area with optical photometry from
the SDSS. The VHS-ATLAS regions have shallower $J$-band coverage
compared to VHS-DES \citep{McMahon:13} and the selection of candidates
with extremely red $(J-K)$ colours is prone to contamination from
larger numbers of spurious sources. However, unlike the VHS-DES
region, VHS-ATLAS also benefits from $Y$-band coverage, which we used
to impose an additional selection criterion: $Y>J$. Three candidates were
selected and verified to be faint and red at optical wavelengths based
on the SDSS data.}

\item{\textit{The VHS-ATLAS NGC Region:} This region
(13h40$<$RA$<$16h20 and $-20\degree<DEC<-2\degree$) encompasses the area used for
target selection in our pilot survey (B13). Nine
bright targets with $K_{AB}<18.4$ were selected from 421\,deg$^2$. Two
of which, VHSJ1350-0503 and VHSJ1409-0830, were observed in B13. The
remaining seven candidates were added to our spectroscopic
target list.}

\item{\textit{The UKIDSS-LAS Data Release 10 $(H-K)>1.4$ Sample:}
Finally, using the UKIDSS-LAS DR10, covering a total area of
3131\,deg$^2$ at DEC$<$20$\degree$ (observable from the Very Large
Telescope), we also constructed a candidate list of the reddest
quasars which satisfy the additional colour selection criterion of
$(H-K)>1.4$ and have $K<18.9$. These would correspond to the dustiest
hyperluminous quasars at these high redshifts with $E(B-V) \gtrsim
1.2$ and the objective was to find more quasars like ULASJ1234+0907
\citep{Banerji:14}. Three of these extremely red candidates were
visible at the time of observations and were also added to our
spectroscopic target list.}

\end{itemize}

Our final candidate list comprised a total of 53 spectroscopic
targets, excluding candidates already observed by B12 and B13. As
described below, 48 of these targets were successfully observed, with
24 classified as broad-line quasars. The candidate list includes a
complete sample of 41 objects, with $K_{\rm{AB}}<18.9$, over a region
of 1115\,deg$^2$, selected to provide quantitative constraints on the
space density of the reddened quasar population.

\begin{table*}
\begin{center}
\centering
\caption{Summary of survey regions and selection criteria for reddened
quasars presented in this work plus B12 and B13. Several survey
regions are spectroscopically incomplete. Column 4 gives the number of
candidates observed, with the total number of candidates and the
percentage completness in brackets. Column 5 gives the corresponding
information for the number of candidates yielding spectroscopic
redshifts, with the total number of spectroscopic targets and the
percentage redshift yield in brackets.}
\label{tab:area}
\begin{tabular}{llccc}
Survey Region & Selection Criterion & Area/deg$^2$ & Nobs (Ncand) & Nw/z (Nobs) \\
\hline \hline
\multicolumn{5}{c}{{$K<18.4$ Bright Sample}} \\
\hline 
VHS-DES SPT$^{a}$ & $(J-K)>1.6$, $(W1-W2)>0.85$ & 498 & 12(12,100\%) & 7(12,58\%) \\
VHS-ATLAS NGC$^{a,c}$ & $(J-K)>1.6$, $(W1-W2)>0.85$, $Y>J$ & 421 & 5(9,56\%) & 3(5,60\%) \\
VHS-ATLAS SGC$^{a}$ & $(J-K)>1.6$, $(W1-W2)>0.85$, $Y>J$ & 488 & 1(1,100\%) & 0(1,0\%) \\
VHS-DES S82-W$^{a,b}$ & $(J-K)>1.6$, $(W1-W2)>0.85$ & 129 & 5(5,100\%) & 1(5,20\%) \\
& $i > 20.5$, $(i-K)>2.5$ & & & \\
UKIDSS-LAS S82-E$^{b}$ & $(J-K)>1.6$, $(W1-W2)>0.85$ & 116 & 5(7,71\%) & 5(5,100\%) \\
& $i > 20.5$, $(i-K)>2.5$ & & & \\
UKIDSS-LAS DR10$^{a,b}$ & $(J-K)>1.6$, $(W1-W2)>0.85$, $(H-K)>1.4$ & 3131 & 6(5,83\%) & 3(5,60\%) \\
\hline
\multicolumn{5}{c}{\textbf{$18.4<K<18.9$ Intermediate Sample}} \\
\hline 
VHS-DES SPT$^{a}$ & $(J-K)>1.6$, $(W1-W2)>0.85$ & 498 & 17(17,100\%) & 9(17,53\%) \\
VHS-DES S82-W$^{a,b}$ & $(J-K)>1.6$, $(W1-W2)>0.85$ & 129 & 4(4,100\%) & 2(4,50\%) \\
& $i > 20.5$, $(i-K)>2.5$ & & & \\
VHS-ATLAS SGC$^{a}$ & $(J-K)>1.6$, $(W1-W2)>0.85$, $Y>J$ & 488 & 2(2,100\%) & 2(2,100\%) \\
 \hline
\multicolumn{5}{c}{\textbf{$18.9<K<19.3$ Faint Sample}} \\
\hline
VHS-DES SPT Deep$^{a}$ & $(J-K)>1.6$, $(W1-W2)>0.85$ & 83 & 4(4,100\%) & 1(4,25\%) \\
\hline
\small{$^{a}$This Work $^{b}$Banerji et al. 2012 $^{c}$Banerji et al. 2013} 
\end{tabular}
\end{center}
\end{table*}

\subsection{Optically Bright Reddened Quasar Candidates with SDSS Spectroscopy}

\label{sec:optical}

\begin{figure*}
\begin{center}
\begin{tabular}{cc}
\large{SDSS J220325.00-004002.8} & \large{SDSS J210050.13-005752.5} \\
\includegraphics[scale=0.4,angle=0]{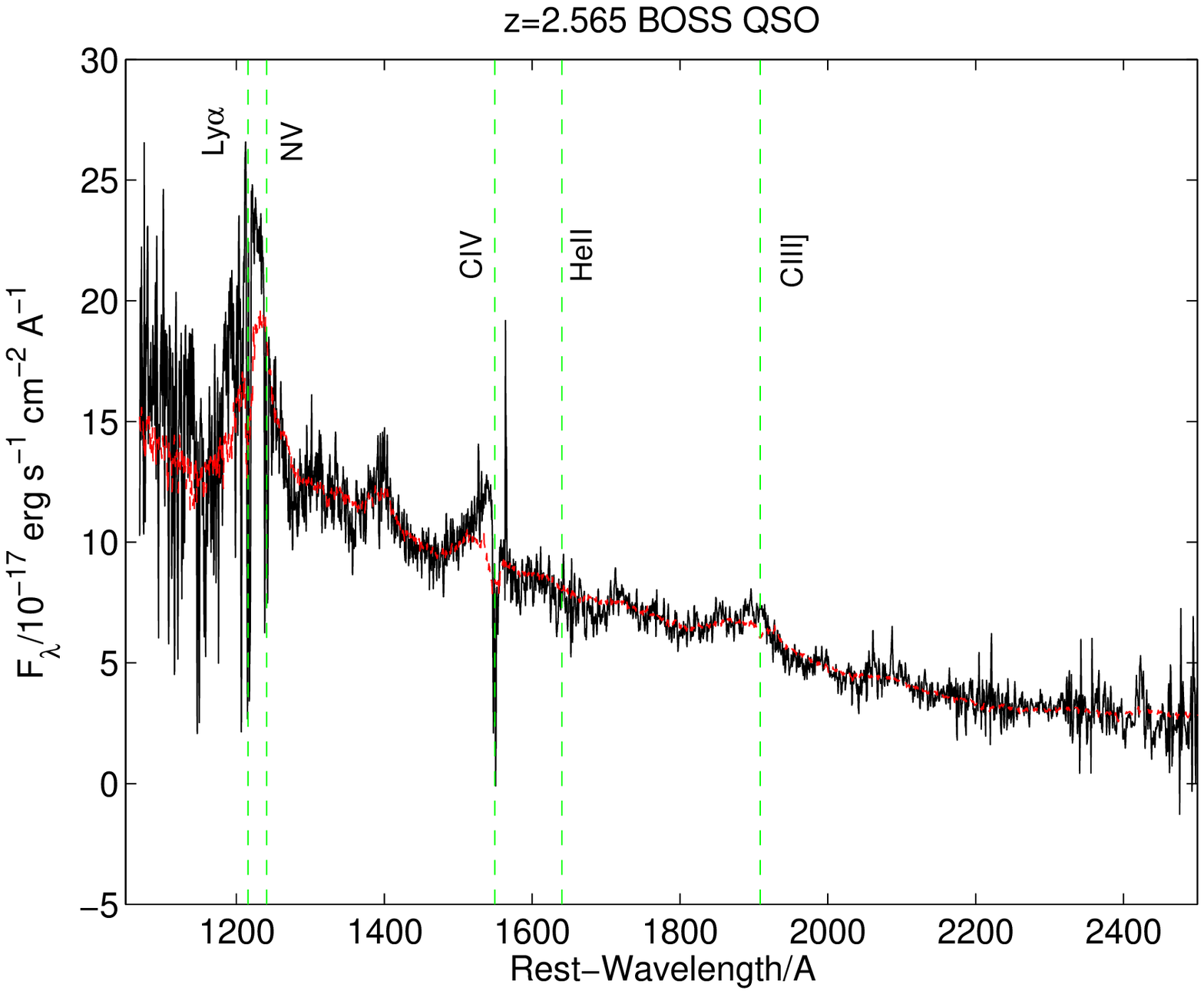} & \includegraphics[scale=0.4,angle=0]{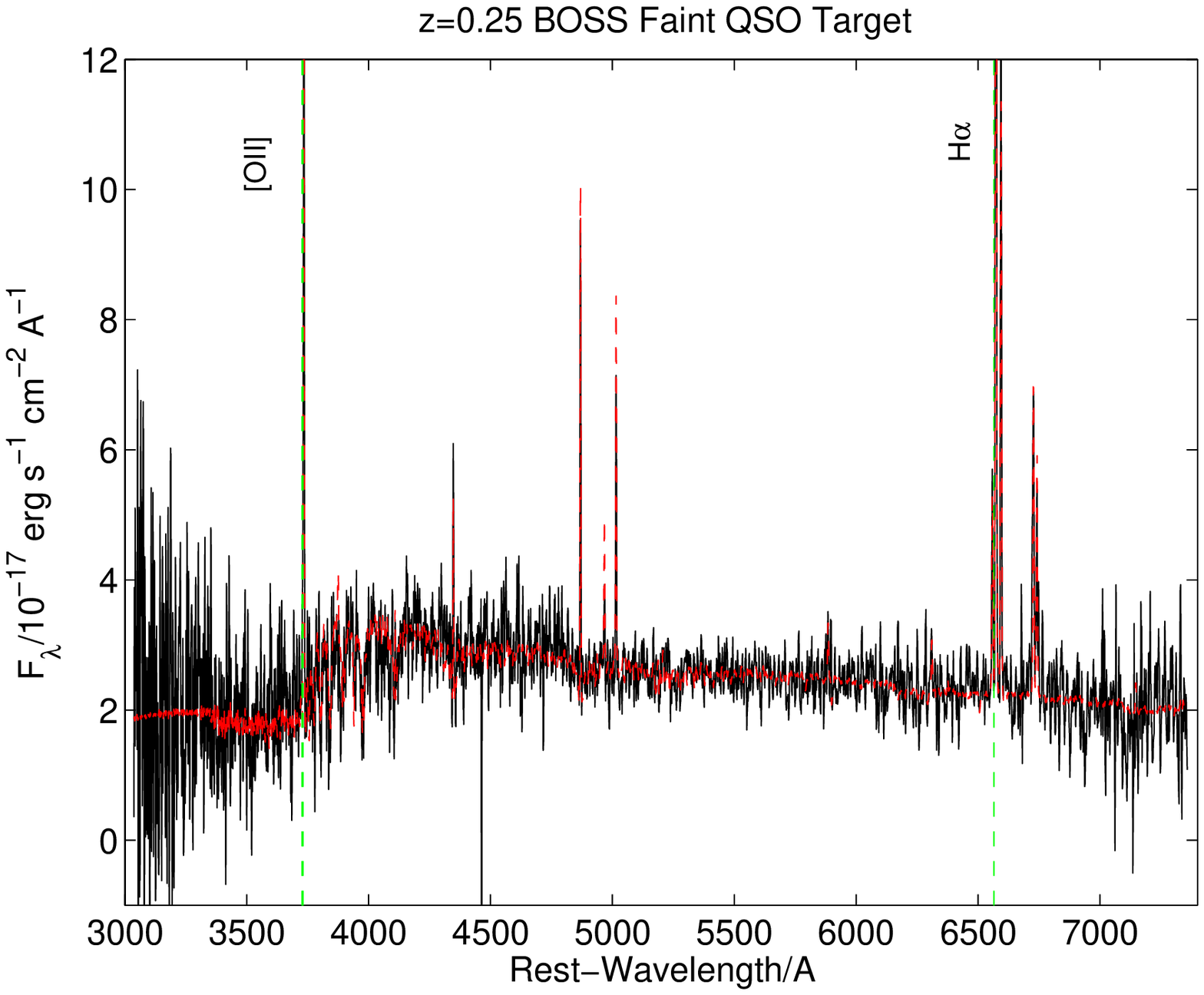} \\
\large{SDSS J204837.25-002437.2} & \large{SDSS J205601.68-001613.3} \\
\includegraphics[scale=0.4,angle=0]{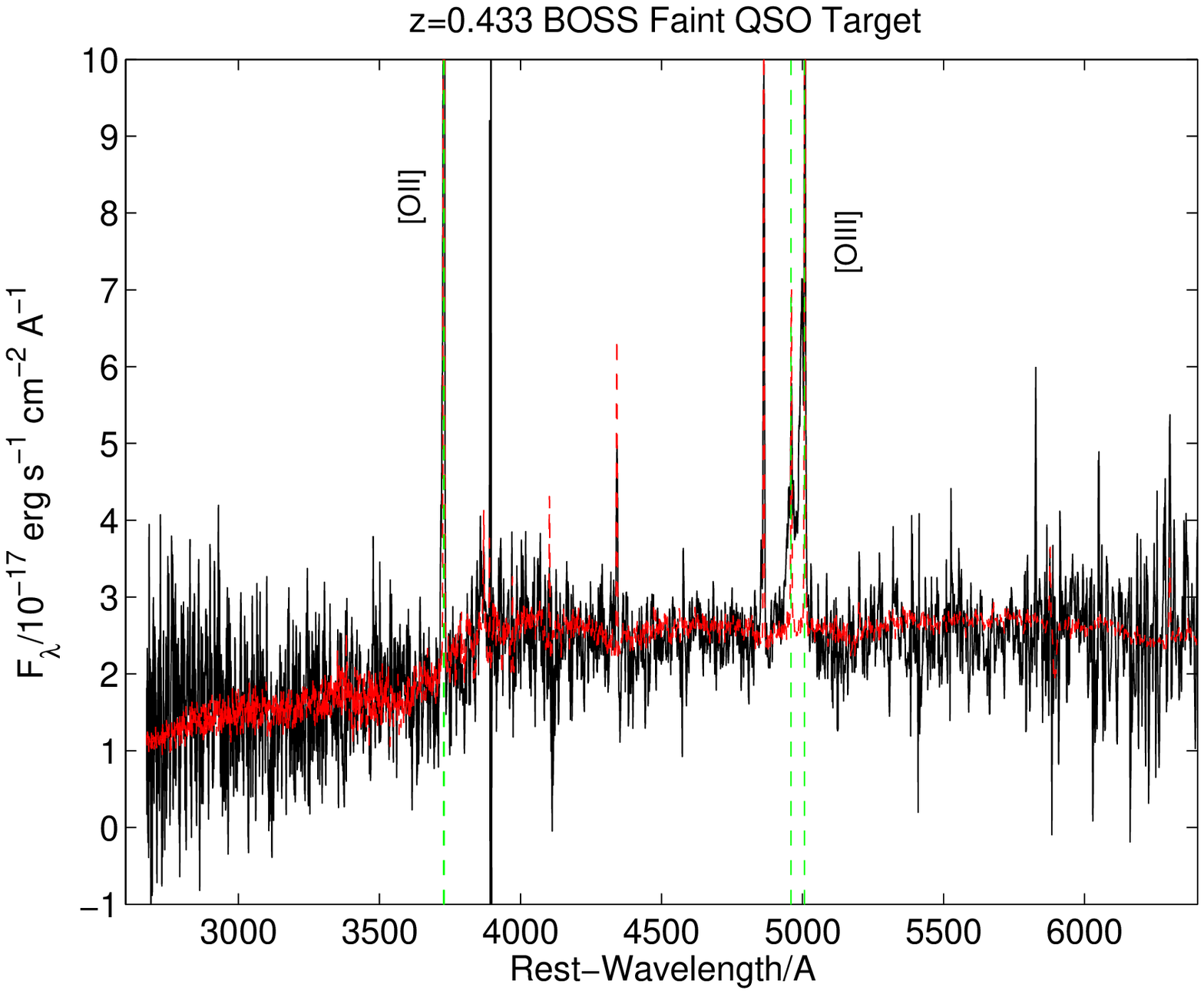} & \includegraphics[scale=0.4,angle=0]{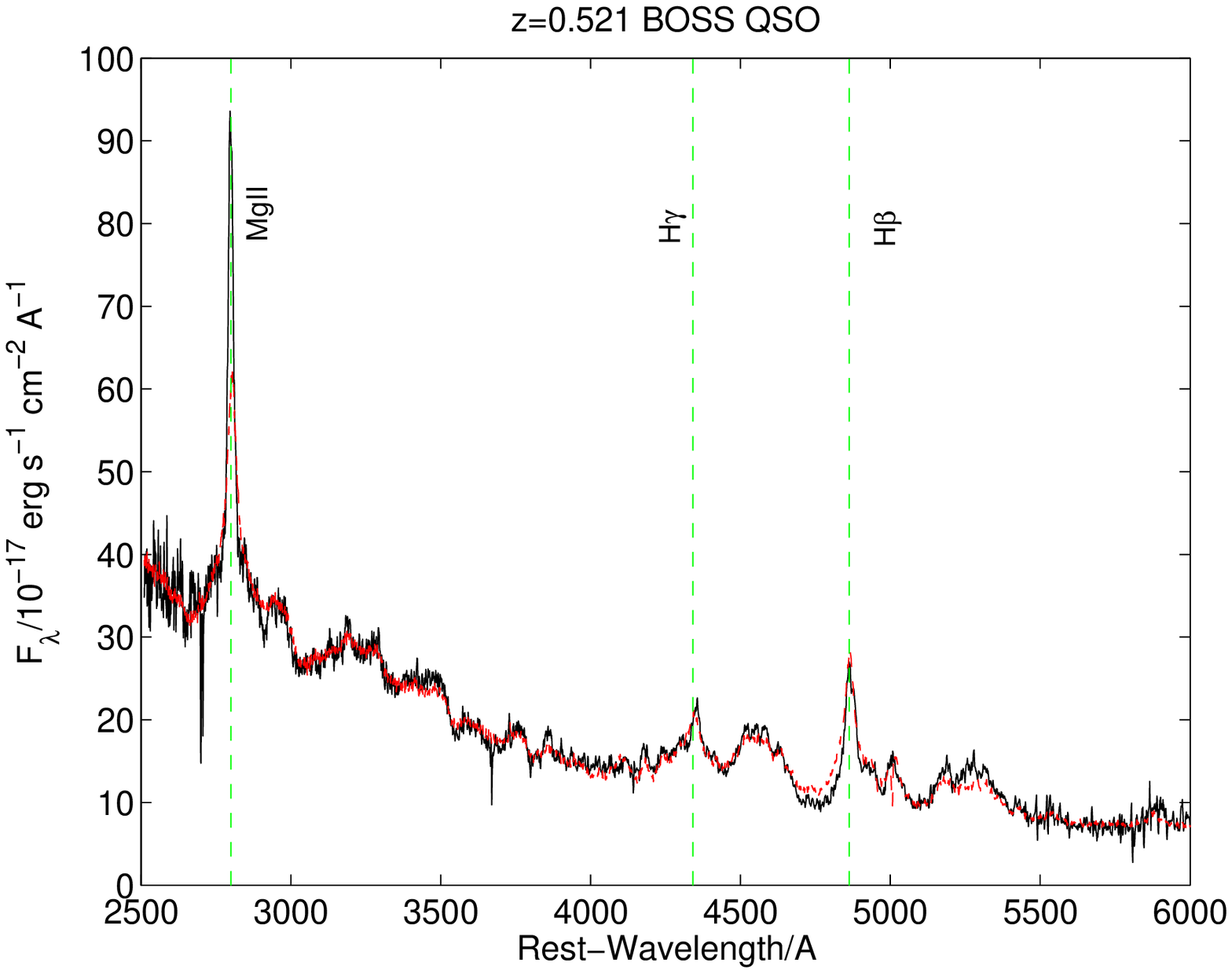} \\
\end{tabular}
\begin{tabular}{c}
\large{SDSS J221930.42+003626.3} \\
\includegraphics[scale=0.4,angle=0]{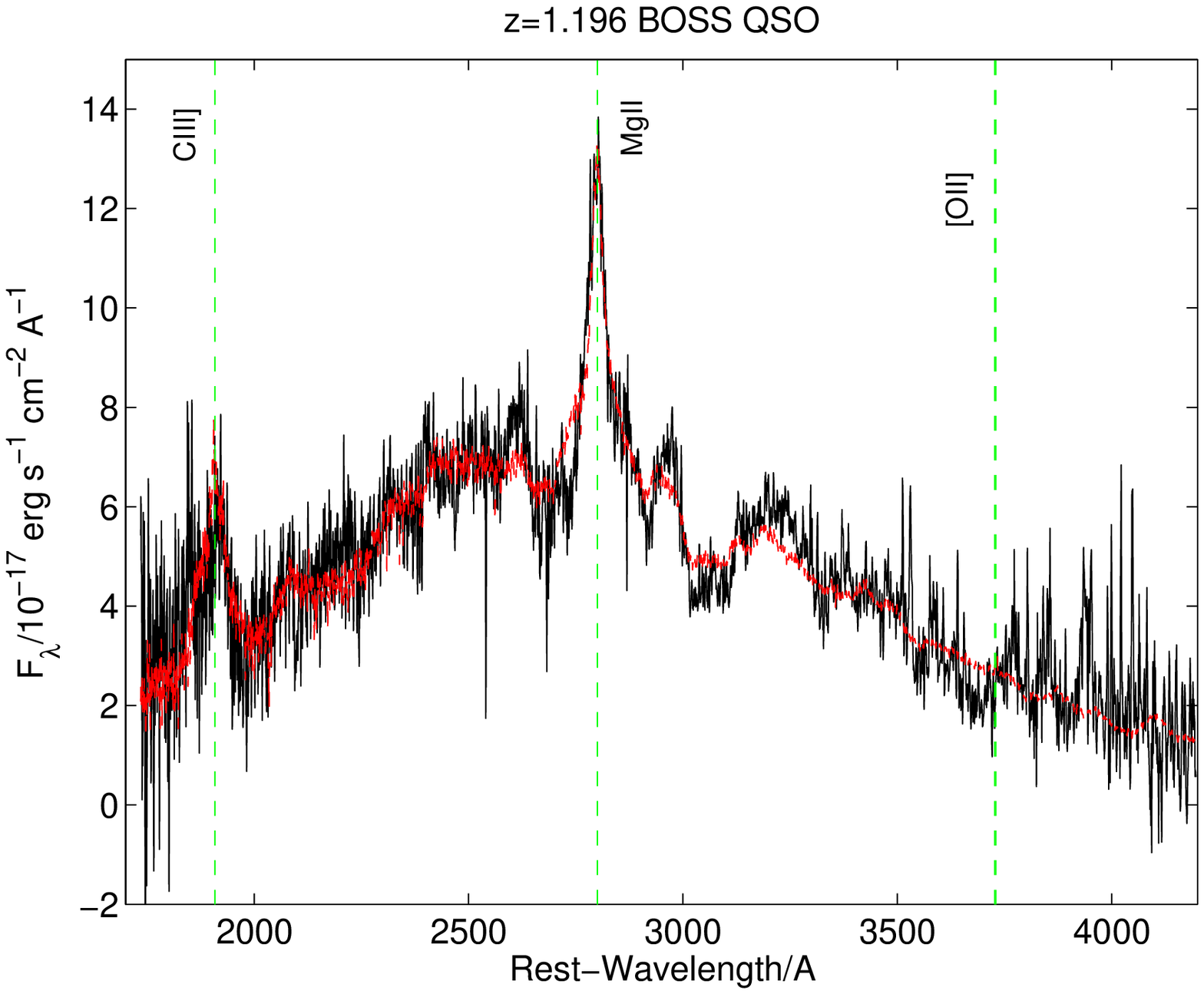} \\
\end{tabular}
\caption{SDSS spectra of optically bright reddened quasar candidates
(black solid) along with the best model-fit used for the redshift
determinations (dashed red). Note, all of these were quasar targets
within BOSS.}
\label{fig:boss}
\end{center}
\end{figure*}

Before describing our NIR spectroscopic observations, we examine the
SDSS/BOSS optical spectra for the optically-bright objects in the VHS-DES
Stripe82. There are five reddened quasar candidates, based on their
$(J-K)$ and $(W1-W2)$ colours, which have $i<20.5$ and/or $(i-K)<2.5$
and were therefore removed from our final candidate list. Spectra for
these sources are shown in Fig. \ref{fig:boss}. Three objects are at
$z<0.6$; two appear to be morphologically compact star-forming
galaxies while one is a low-redshift quasar.

The two remaining optically bright candidates with SDSS spectra are
both quasars at higher redshifts.  ULASJ2219+0036 (SDSS J221930.42+003626.3), a quasar at
$z=1.196$, was also identified in B12, where NIR spectroscopy was
presented. No emission lines were detected, confirming that our NIR
spectroscopy can fail to identify reddened quasars at low
redshifts. SDSS J220325.00-004002.8, a quasar at $z=2.565$, possesses a spectrum with
a blue UV continuum and is therefore not a heavily dust-reddened
quasar. We note that both these high redshift quasars have spectroscopy from the BOSS survey but not from the SDSS Data Release 7. 

As demonstrated in \citet{Ross:14}, red quasars selected from the SDSS/BOSS
surveys constitute a heterogenous population including reddened Type 1
quasars, Type 2 quasars, Broad Absorption Line quasars, as well as
sources with extreme emission line equivalent widths. Each of these
populations is interesting in its own right but in this work, we are
specifically interested in investigating the properties of the
reddened Type 1 quasars and our target selection is designed to
isolate this subset of red objects.

\section{SPECTROSCOPIC FOLLOW-UP}\label{sec:specobs}

Follow-up of our reddened quasar spectroscopic targets was conducted in
Visitor Mode over three full nights in 2013 July using the SINFONI
spectrograph on the Very Large Telescope (VLT). The seeing ranged from
0.5 to 1.0\,arcsec with a median of 0.78\,arcsec. The quasar
candidates were expected to cover a broad range in redshift, similar
to the samples in B12 \& B13, and use of the SINFONI $H+K$ filters was
therefore requested. However, due to technical difficulties, only the
SINFONI R=4000 $K$-band grating was available. The $K$-filter has a
wavelength range of 1.9-2.5$\mu$m, corresponding to a redshift range
of 2.1$<$$z$$<$2.7 for H$\alpha$ and redshifts 3.1$<$$z$$<$4.1 for
H$\beta$ and [OIII]$\lambda\lambda$4959,5007. At redshifts $0.5<z<2.1$
however, no strong emission lines are visible in the $K$-band so this
represents a \textit{redshift desert} for our survey. ULASJ2219+0036
(see Section~\ref{sec:optical}) is an example of a candidate in the
redshift desert.

The dispersion is 2.5\AA\ per pixel. Observations were carried out in noAO mode, utilising the
largest field of view of the instrument, which is
8$\times$8\,arcsec. The targets were offset by $\pm$1.5\,arcsec in RA
and DEC from the centre of the IFU for sky-subtraction purposes. Apart
from the choice of filter, the observational setup is identical to
that in B12. The total integration time was adjusted depending on the
$K$-band magnitude of each source and the prevailing conditions at the
time of observation. Fainter targets therefore have much longer
exposure times, chosen such that any broad-lines present should
be detected at high S/N in the binned spectra.

All data were reduced using standard ESO pipeline recipes using the
\textit{gasgano} data reduction package. Data reduction steps included
dark subtraction, non-linearity corrections, flat-fielding, sky
subtraction, extraction and wavelength calibration. Relative flux
calibration was performed using telluric standards observed at similar
airmass to the target. Typically, four to five telluric standards were
observed through the night. Due to the absence of spectrophotometric
standards in the NIR, absolute flux calibration was achieved by
normalising the spectra using the $K$-band magnitudes of the
targets.

From our final list of 53 spectroscopic targets, 48 objects were
successfully observed with SINFONI over the three nights. Twenty-four
show broad-lines in the NIR $K$-band spectrum and are therefore
spectroscopically confirmed to be quasars, with redshifts assigned
assuming that the emission line is H$\alpha$.  We note that the observed equivalent widths of these lines
are inconsistent with corresponding to other Balmer lines (e.g. Pa$\alpha$, Pa$\beta$) at lower redshifts, which are expected to be at least an order of magnitude weaker \citep{Glikman:06}. The spectroscopically
confirmed quasars together with their coordinates, $K$-band
magnitudes, redshifts and exposure times are listed in Table
\ref{tab:sample}. One dimensional spectra, extracted over regions maximising the S/N of the 
H$\alpha$ emission line, are presented for all objects in 
Fig.~\ref{fig:spectra}.

\begin{table*}
\begin{center}
\caption{Summary of Spectroscopically Confirmed Reddened Quasars.}
\label{tab:sample}
\begin{tabular}{lcccccc}
\hline \hline
Name & RA & DEC & K$_{\rm{AB}}$ & Redshift & Exposure time (s) \\
\hline
\multicolumn{6}{c}{\textbf{VHS-DES SPT Region}} \\
\hline
VHSJ2024-5623 & 306.1074 & $-56.3898$ & 18.76 & 2.282 & 1800 \\
VHSJ2028-4631 & 307.0083 & $-46.5325$ & 18.87 & 2.464 & 1800 \\
VHSJ2028-5740 & 307.2092 & $-$57.6681 & 17.25 & 2.121 & 800 \\
VHSJ2048-4644 & 312.0435 & $-46.7387$ & 18.78 & 2.182 & 1600 \\
VHSJ2100-5820 & 315.1403 & $-58.3354$ & 18.47 & 2.360 & 1600 \\
VHSJ2101-5943 & 315.3311 & $-$59.7291 & 16.68 & 2.313 & 320 \\
VHSJ2115-5913 & 318.8818 & $-59.2188$ & 17.63 & 2.115 & 800 \\
VHSJ2130-4930 & 322.7490 & $-49.5032$ & 18.47 & 2.448 & 1600 \\
VHSJ2141-4816 & 325.3530 & $-48.2830$ & 18.54 & 2.655 & 1600 \\
VHSJ2212-4624 & 333.0796 & $-46.4101$ & 18.72 & 2.141 & 1800 \\
VHSJ2220-5618 & 335.1398 & $-56.3107$ & 16.72 & 2.220 & 320 \\
VHSJ2227-5203 & 336.9491 & $-52.0582$ & 18.80 & 2.656 & 2200 \\
VHSJ2235-5750 & 338.9331 & $-57.8372$ & 17.95 & 2.246 & 1000 \\
VHSJ2256-4800 & 344.1444 & $-48.0088$ & 17.80 & 2.250 & 1000 \\
VHSJ2257-4700 & 344.2589 & $-47.0157$ & 18.51 & 2.156 & 1600 \\
VHSJ2306-5447 & 346.5011 & $-54.7881$ & 18.17 & 2.372 & 1000 \\
VHSJ2332-5240 & 353.0387 & $-52.6780$ & 19.16 & 2.450 & 2400 \\
\hline
\multicolumn{6}{c}{\textbf{VHS-ATLAS SGC and NGC Regions}} \\
\hline
VHSJ1556-0835 & 239.1571 & $-8.5952$ & 18.20 & 2.188 & 1200 \\
VHSJ2143-0643 & 325.8926 & $-6.7206$ & 18.48 & 2.383 & 1600 \\
VHSJ2144-0523 & 326.2394 & $-5.3881$ & 18.71 & 2.152 & 1600 \\
\hline
\multicolumn{6}{c}{\textbf{VHS-DES Stripe82 Region}} \\
\hline
VHSJ2109-0026 & 317.3630 & $-0.4497$ & 18.59 & 2.344 & 1600 \\
VHSJ2355-0011 & 358.9394 & $-0.1893$ & 18.42 & 2.531 & 1600 \\
\hline
\multicolumn{6}{c}{\textbf{UKIDSS-LAS DR10 $(H-K)>1.4$}} \\
\hline
ULASJ0123+1525 & 20.8022 & $+15.4230$ & 18.59 & 2.629 & 1200 \\
ULASJ2315+0143 & 348.9843 & $+1.7307$ & 18.38 & 2.560 & 400 \\
\hline
\end{tabular}
\end{center}
\end{table*}

\section{DEMOGRAPHICS OF THE REDDENED QUASAR POPULATION}

\subsection{Spectral Energy Distributions, Dust Extinctions \& Bolometric Luminosities}

\label{sec:sed}

To provide estimates of dust extinctions and luminosities we fit the
multi-wavelength photometric data for our confirmed reddened quasars
using the quasar SED models of \citet{Maddox:08, Maddox:12}. In the
VHS-DES SPT region the data consists of photometry in $JHK$ and
\textit{WISE} $W1$ and $W2$.  For the Stripe82, VHS-ATLAS and UKIDSS
regions, optical photometry (from the Stripe82 coadd survey and SDSS)
is also available and has been utilised in the SED fitting. The fitting procedure is described in B12 and
accounts for the effect of the observed H$\alpha$ equivalent width on
the $(J-K)$ colours as follows. The equivalent width of the H$\alpha$ emission line in our base SED
model has been derived by fitting to the observed colours
($ugrizYJHK$) of luminous, unobscured, quasars in SDSS DR7 that
possess SDSS and UKIDSS photometry. The quasar SED model (Maddox et
al. 2012) takes into account luminosity-dependent changes in the
H$\alpha$ equivalent width due to the Baldwin effect \citep{Baldwin:77}.
The base quasar-SED model has already been shown to provide a very
good match to the observed colours of luminous SDSS DR7 quasars over
the redshift range 0.2$<$$z$$<$4.0 (Maddox et al. 2012).  The default
rest-frame equivalent width of the H$\alpha$ line in our $z$=2.3 model
is $\simeq$250\,\AA.

For each of our reddened quasars, we estimate the rest-frame H$\alpha$
equivalent widths directly from the spectra. All equivalent widths are
calculated by integrating the line within $\pm$3$\sigma$, where
$\sigma$ is derived from a single Gaussian fit to the line
profile. The continuum is defined beyond this 3$\sigma$ region out to
$\pm$10000 km/s from the H$\alpha$ line centroid. The exact details of
the equivalent width calculations are not important as measurements
are are made consistently for both the SED model and the observed
spectra. The rest-frame H$\alpha$ equivalent widths for each of our
reddened quasars are given in Table A1 and the median value is
$\simeq$270\,\AA, which is consistent with the value for the
unobscured quasar population.  There is therefore no evidence that the
reddened quasars have unusual H$\alpha$ equivalent widths relative to
their unobscured counterparts.

There are, however, a few examples of reddened quasars with H$\alpha$
equivalent widths of up to $\simeq$500\,\AA. For comparison, the
$K$-band filter in the VISTA survey has an effective width of
3090\,\AA \ or 940\,\AA \ in the rest-frame at $z$=2.3. When deriving
$E(B-V)$-estimates, it is important that we account for the effect of
the H$\alpha$ line on the continuum $K$-band flux. We therefore
proceed as follows. The H$\alpha$ line in the SED model is scaled to
have the same rest-frame equivalent width present in each individual
reddened quasar. As is evident from Table A1 the scaling-factor is
close to unity for the majority of quasars. The $E(B-V)$ values are
then derived by fitting the scaled SED to all the observed photometry
available for each quasar. In this way, the specific contribution to
the $K$-band flux by the H$\alpha$ emission line in each quasar is
taken into account and the $E(B-V)$ estimates are independent of the
emission-line strength.

Our model also includes a luminosity dependent contribution from a host
galaxy. Even allowing for the effects of extinction on the quasar SED,
at the redshifts and $K$-band magnitudes of our sample, the
NIR-colours of the majority of the objects are dominated by the quasar SED. 
Recall that a key element in the sample selection is that the
candidates must be unresolved in the $K$-band and, by selection,
the host-galaxy contribution cannot be large. The form of the
host-galaxy SED and exact contribution to the observed NIR-colours are
thus not significant in the context of reproducing the $JHKW1W2$
photometry. 

The strongest constraints on the reddening for each quasar predominantly
come from the $J-K$ colour, which probes rest-frame wavelengths of
$\sim$3400-6400\AA, below the inflexion in the quasar SEDs at
$\simeq$10\,000\AA \, and above the UV-portion of the spectrum where
different extinction curves exhibit quite different forms. The derived
values of $E(B-V)$ are thus completely insensitive to which extinction
curve (e.g. curves appropriate for the Milky Way, Large or Small
Magellanic Clouds) is incorporated in the model fits. Given the
photometric uncertainties, the estimates of $E(B-V)$ are accurate to
$\simeq\pm$0.1\, mag. Extinction values are summarised in
Table~\ref{tab:demo} and the median $E(B-V)$ is found to be 0.8\,mag, 
consistent with the values reported in B12 and B13. 

As demonstrated in \citet{Banerji:14}, these dust extinction estimates
derived from SED-fitting to the broadband photometry, are in excellent agreement with those
from independent X-ray observations, even for the reddest and
therefore dustiest quasars.  The observed quasar SEDs are de-reddened
using the $E(B-V)$ values, assuming an SMC-like extinction law
\citep{Pei:92}, and the optical luminosity at rest-frame 5100\AA
\ is calculated\footnote{As with the determination of the
$E(B-V)$-values, the derived optical luminosities are insensitive to
which conventional extinction curve is employed.}.  The observed and
de-reddened $i$-band quasar magnitudes, together with the de-reddened $i$-band absolute magnitudes and 
bolometric luminosities are presented in Table \ref{tab:demo}.  Note, these $i$-band absolute magnitudes are related to 
the SDSS $M_{i[z=2]}$ magnitudes via $M_i$=$M_{i[z=2]}+0.596$. The
total luminosities are estimated using a bolometric correction factor
of 7 applied to the rest-frame 5100\AA\@ luminosity
\citep{Netzer:07}\footnote{A bolometric correction factor of closer to
9 has also been used in the literature \citep{Shen:11} and would
increase our estimated bolometric luminosities by $\sim$0.1 dex.}.

\subsection{Black-Hole Masses}

\label{sec:bh}

Single-epoch quasar black-hole masses are normally derived based
on a virial estimator, employing the widths of broad emission lines
and assuming a calibration between the size of the broad-line
region and the quasar luminosity. The calibration has been derived
using reverberation mapping techniques at low redshift
\citep{Kaspi:00, Kaspi:05} and for relatively low-luminosity AGN; the
validity of extrapolating the trends to high redshift and higher
luminosities, still remains uncertain. In addition, virial black-hole
mass estimates can be sensitive to the choice of broad line used, and
the details of the line-fitting. A further complication can result if
there are significant inflows and outflows affecting the gas close to
the accreting black-hole that can artificially broaden the observed
linewidths. If reddened quasars do correspond to the AGN-feedback
phase in galaxy formation, such inflows and outflows would be more
predominant in our sample and indeed several of the H$\alpha$ line
profiles shown in Figure \ref{fig:lines} clearly show evidence for very broad
wings that could result from such flows.

Despite these caveats, single-epoch black-hole masses are still widely used and there is
general consistency between mass estimates obtained from different
broad-lines \citep{Shen:11, Matsuoka:13}. We therefore provide such
mass estimates for our reddened quasar sample. Specifically, we employ
the mass estimator of \citet{Vestergaard:06}, using the calibration
between H$\alpha$ and H$\beta$ linewidths from \citet{Greene:05}:

\begin{equation}
\rm{FWHM}_{H\beta}=(1.07\pm0.07)\times10^{3}\left(\frac{\rm{FWHM}_{H\alpha}}{10^{3}\rm{kms}^{-1}}\right)^{1.03\pm0.03}\rm{kms}^{-1}
\label{eq:HaHb}
\end{equation}

\begin{equation}
M_{\rm{BH}}/M_\odot=10^{6.91}\left(\frac{\rm{FWHM_{H\beta}}}{1000 \rm{kms}^{-1}}\right)^2\left(\frac{L_{5100}}{10^{44}\rm{erg s}^{-1}}\right)^{0.5}
\label{eq:BHmass}
\end{equation}

\noindent where L$_{5100}$ is estimated using the best-fit model SEDs
of each quasar. The H$\alpha$ rest-frame
full-width-half-maximum (FWHM) is calculated from either a single- or
a double-Gaussian fit to the line profile in the velocity range
$\pm$10000\kms. Double Gaussians are only used if they produce a
statistically significant improvement to the line-profile fit corresponding to a lower mean-fit error.  All
Gaussian components are constrained to possess FWHM in the range
1000-10\,000\kms, ensuring they trace broad-line region gas. Errors on 
the black-hole masses are derived by propagating the mean fit error on 
the FWHM measurements. Contributions to the H$\alpha$ emission-line profile from any narrow
component present are small and we verified that including such a
narrow Gaussian component did not alter significantly the FWHM of the
broad single- or double-Gaussian components. Parametrization of any
narrow H$\alpha$ components, which could potentially be tracing star
formation in the quasar host galaxy, is discussed in Alaghband-Zadeh
et al (in preparation).

The Gaussian line fits are included in Figure \ref{fig:lines}, while the
FWHM values and derived black-hole masses are tabulated in Table
\ref{tab:fwhm} and Table \ref{tab:demo} respectively.

\subsection{Comparison to Unobscured Quasars from SDSS}

\begin{figure*}
\begin{center}
\begin{tabular}{cc}
\includegraphics[scale=0.4,angle=0]{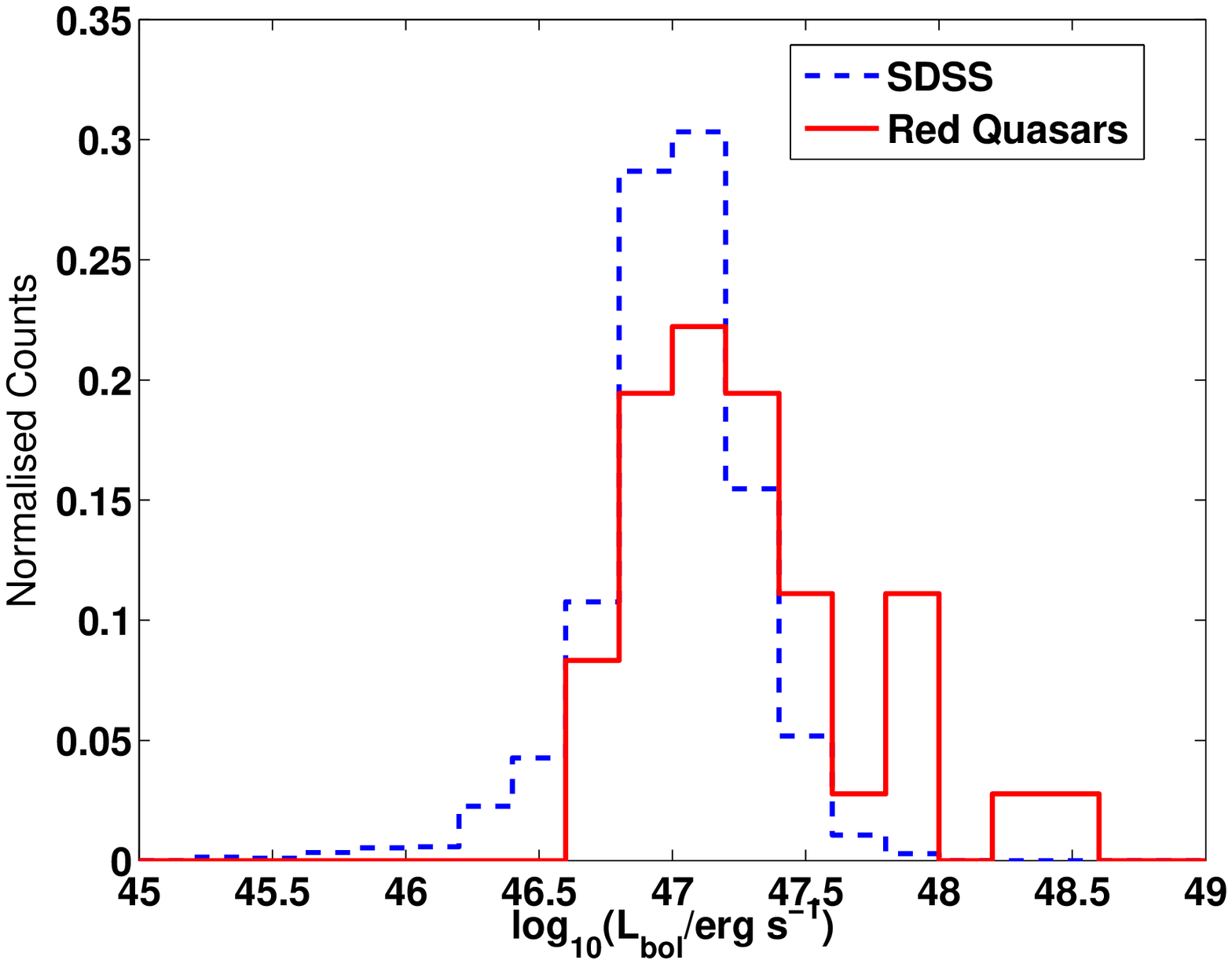} & \includegraphics[scale=0.4,angle=0]{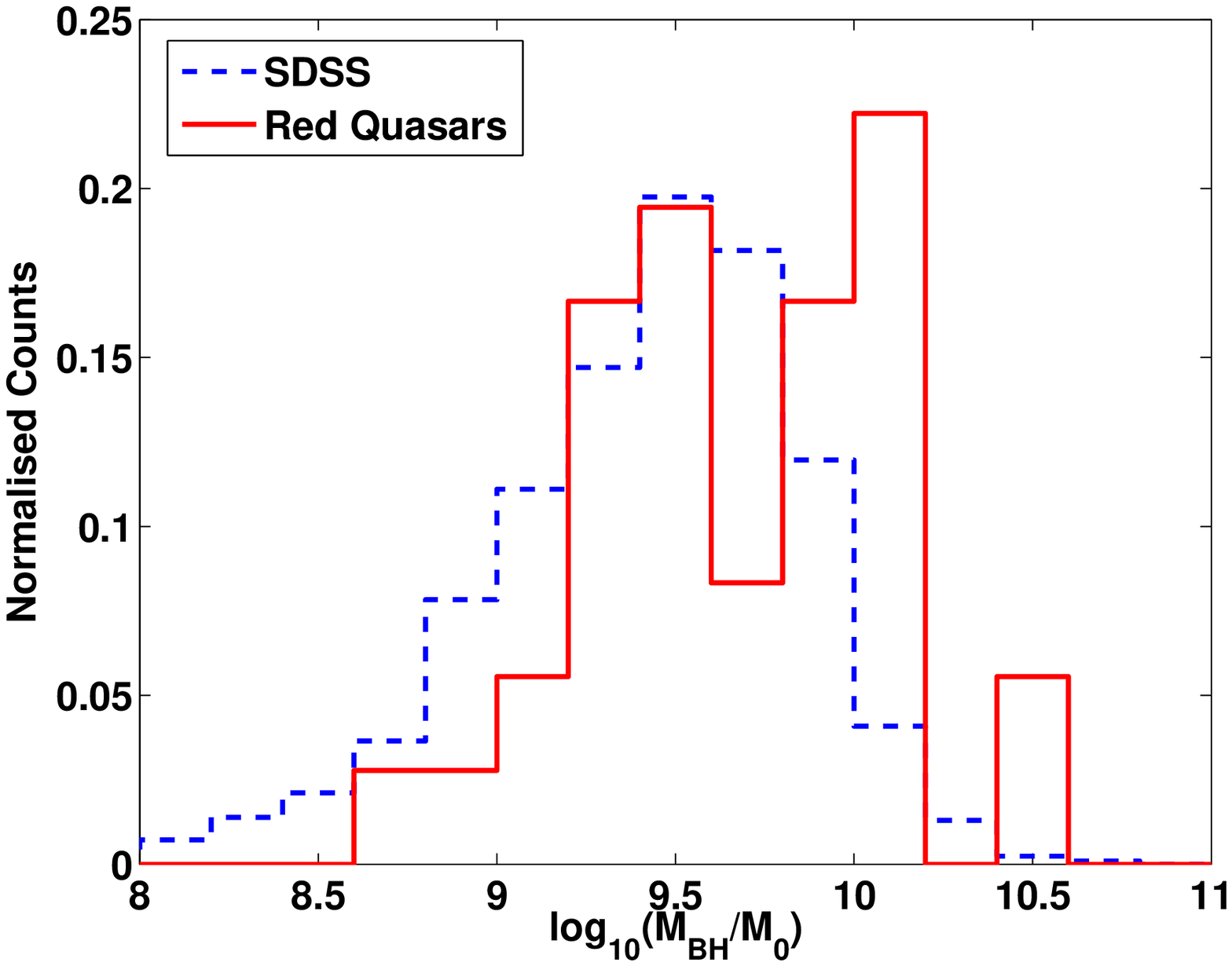} \\ 
\end{tabular}
\caption{Distribution of bolometric luminosity (left) and black-hole mass (right) 
for the red and unobscured quasar populations (see descriptions in Section~4.3).
There is significant overlap between the samples in both luminosity
and mass, although there is a suggestion that the red population
includes a tail of quasars with very high bolometric luminosities and
black-hole masses.}
\label{fig:hists}
\end{center}
\end{figure*}

The black-hole masses and bolometric luminosities can be used to
derive Eddington ratios ($L/L_{\rm Edd}$) which are typically $>$0.1, as
was found in B12 and B13. The reddened quasars therefore appear to be
massive black holes, accreting at a high rate, resulting in their very
high luminosities. Large numbers of unobscured, luminous and highly
accreting black-holes are known, e.g.from SDSS \citep{Schneider:07}
and BOSS \citep{Paris:13}, and the luminosity function and space
density of the obscured and unobscured populations are of potential
interest.  We define a sample of 2081 `unobscured' quasars from SDSS DR7,
with redshifts 2.1$<z<$2.7 and magnitudes $K<$18.9 as well as black-hole
mass and bolometric luminosity estimates from
\citet{Shen:11}. The combined sample of spectroscopically confirmed
reddened quasars from B12, B13 and this paper, over the same redshift
range and brighter than $K<$18.9 represents the `reddened'-quasar
population.  The distribution of black-hole masses and luminosities
for the two samples are shown in Fig.~\ref{fig:hists}. The mean
black-hole masses and bolometric luminosities are
log$_{10}$(M$_{\rm{BH}}$/M$_\odot$)=9.7$\pm$0.4,
log$_{10}$(L$_{\rm{bol}}$/erg s$^{-1}$)=47.1$\pm$0.4 and
log$_{10}$(M$_{\rm{BH}}$/M$_\odot$)=9.3$\pm$1.1,
log$_{10}$(L$_{\rm{bol}}$/erg s$^{-1}$)=47.0$\pm$0.2 for the reddened and
unobscured samples respectively.

The distributions of black-hole mass and luminosity overlap
significantly (Fig.~\ref{fig:hists}), but there is an indication that
the reddened quasars possess a tail out to higher bolometric luminosities
and black-hole masses. Several selection effects and systematic errors
could be responsible for this observed tail. The black hole masses,
derived from UV-emission lines for objects in SDSS and H$\alpha$ for
the reddened population, may not be on exactly the same scale. Progress is
being made via direct comparison of emission-line properties in quasar
samples \citep{Matsuoka:13, Tilton:13} but the uncertainty in black
hole mass estimates from individual emission lines remains significant
\citep[e.g.][]{Shen:11}.  More importantly, inflows and outflows are
likely to be more common in the reddened population, thought to be caught
in the AGN radiative feedback phase. The presence of such flows could
add significant additional spread to the observed linewidths,
increasing black-hole masses by factors of a few (B12). The black-hole
masses should therefore be interpreted with caution.

The bolometric luminosities of the samples may also differ systematically.
Luminosities for the unobscured quasar sample are calculated using 
the 1350\AA\@ rest-frame luminosity and assume a different model
quasar SED  to ours. Most likely, however, is a selection against
the most luminous quasars in the SDSS resulting from the bright
$i$-band flux-limit of 15.0\,mag (c.f. the values for the reddened quasars
in Col. 4 of Table~\ref{tab:demo}).

Overall, notwithstanding the caveats summarised above, we can
conclude that the reddened quasar sample consists of a population
of very high mass black-holes, comparable to the most massive
black-holes included in the SDSS quasar sample.

\begin{table*}
\begin{center}
\caption{Best-fit dust extinction, reddened and de-reddened optical
$i$-band magnitudes, absolute $i$-band magnitudes, bolometric luminosities, black-hole masses and
Eddington ratios of the reddened quasars.}
\label{tab:demo}
\begin{tabular}{lcccccccc}
\hline \hline
Name & E(B-V) & i$_{\rm{AB}}^{\rm{red}}$ & i$_{\rm{AB}}^{\rm{dered}}$ & M$_i$ & log$_{10}$(L$_{\rm{bol}}$/ erg s$^{-1}$) & log$_{10}$(M$_{\rm{BH}}$/M$_\odot$) & lEdd & dm/dt (M$_\odot$yr$^{-1}$) \\
\hline
\multicolumn{9}{c}{\textbf{VHS-DES SPT Region}} \\
\hline
VHSJ2024-5623 & 0.6 & 22.9 & 17.8 & $-27.37$ & 46.7 & 9.8$\pm$0.1 & 0.06 & 9 \\
VHSJ2028-4631 & 0.6 & 22.9 & 17.9 & $-27.49$ & 46.7 & 9.2$\pm$0.2 & 0.2 & 9 \\
VHSJ2028-5740 & 1.2 & 24.3 & 14.6 & $-30.18$ & 47.9 & 10.1$\pm$0.1 & 0.4 & 127 \\
VHSJ2048-4644 &  0.7 & 23.2 & 17.4 & $-27.65$ & 46.8 & 9.2$\pm$0.1 & 0.3 & 11 \\
VHSJ2100-5820 & 0.8 & 23.6 & 16.5 & $-28.49$ & 47.1 & 9.1$\pm$0.1 & 0.8 & 24 \\
VHSJ2101-5943 & 0.8 & 21.8 & 14.9 & $-30.03$ & 47.8 & 10.5$\pm$0.1 & 0.1 & 106 \\
VHSJ2115-5913 & 1.0 & 23.5 & 15.4 & $-29.48$ & 47.5 & 9.3$\pm$0.1 & 1.4 & 61 \\
VHSJ2130-4930 & 0.9 & 24.7 & 16.5 & $-28.75$ & 47.2 & 9.5$\pm$0.2 & 0.4 & 31 \\
VHSJ2141-4816 &  0.8 & 23.8 & 16.1 & $-29.10$ & 47.3 & 9.6$\pm$0.2 & 0.4 & 40 \\
VHSJ2212-4624 &  0.8 & 23.7 & 17.2 & $-27.64$ & 46.9 & 9.9$\pm$0.1 & 0.07 & 14 \\
VHSJ2220-5618 &  0.8 & 21.6 & 14.9 & $-30.09$ & 47.8 & 9.9$\pm$0.1 & 0.5 & 106 \\
VHSJ2227-5203 &  0.9 & 25.0 & 16.3 & $-29.15$ & 47.4 & 10.0$\pm$0.1 & 0.2 & 41 \\
VHSJ2235-5750 &  0.6 & 21.9 & 16.9 & $-28.28$ & 47.1 & 10.1$\pm$0.1 & 0.07 & 20 \\
VHSJ2256-4800 &  0.6 & 21.7 & 16.7 & $-28.49$ & 47.1 & 10.1$\pm$0.1 & 0.08 & 24 \\
VHSJ2257-4700 &  0.7 & 23.0 & 17.3 & $-27.67$ & 46.9 & 9.5$\pm$0.1 & 0.2 & 13 \\
VHSJ2306-5447 & 0.7 & 23.2 & 17.1 & $-28.15$ & 47.0 & 10.0$\pm$0.1 & 0.08 & 18 \\
VHSJ2332-5240 &  0.6 & 23.4 & 18.0 & $-27.39$ & 46.7 & 9.5$\pm$0.1 & 0.1 & 8 \\
\hline
\multicolumn{9}{c}{\textbf{VHS-ATL SGC and NGC Regions}} \\
\hline
VHSJ1556-0835 & 0.7 & 22.7 & 16.9 & $-28.15$ & 47.0 & 8.6$\pm$0.1 & 1.9 & 19 \\
VHSJ2143-0643 & 0.8 & 23.8 & 16.7 & $-28.33$ & 47.1 & 10.0$\pm$0.1 & 0.09 & 22 \\
VHSJ2144-0523 & 0.6 & 22.6 & 17.7 & $-27.33$ & 46.7 & 9.9$\pm$0.1 & 0.04 & 9 \\
\hline
\multicolumn{9}{c}{\textbf{VHS-DES Stripe82 Region}} \\
\hline
VHSJ2109-0026 & 0.7 & 23.3 & 17.2 & $-27.95$ & 46.9 & 9.8$\pm$0.1 & 0.1 & 15 \\
VHSJ2355-0011 & 0.7 & 23.2 & 16.6 & $-28.67$ & 47.2 & 10.1$\pm$0.1 & 0.09 & 27 \\
\hline
\multicolumn{9}{c}{\textbf{UKIDSS-LAS DR10 $(H-K)>1.4$}} \\
\hline
ULASJ0123+1525 & 1.3 & 27.7 & 15.4 & $-30.12$ & 47.8 & 9.7$\pm$0.2 & 0.9 & 114 \\
ULASJ2315+0143 & 1.1 & 26.2 & 16.1 & $-29.43$ & 47.5 & 10.1$\pm$0.2 & 0.2 & 57 \\
\hline
\end{tabular}
\end{center}
\end{table*}

\section{Infrared SEDs from \textit{WISE}}

Comparing the multi-wavelength properties of the reddened quasar
sample to other well-studied galaxy and AGN populations at similar
redshifts helps in understanding the origin of the sample.  The
availability of the \textit{WISE} photometry, probing rest-frame
wavelengths out to $\gtrsim$3\,$\mu$m for objects detected in the
$W3$-band, provides information for a key portion of quasar SEDs.  An
unobscured comparison sample is defined by selecting 1604 SDSS DR7 and DR10
quasars with $2.1<z<2.7$, $K_{AB}<18.9$ and detections in the \textit{WISE} $W123$-bands - hereafter the `SDSS sample'.

\begin{figure}
\begin{center}
\includegraphics[scale=0.4,angle=0]{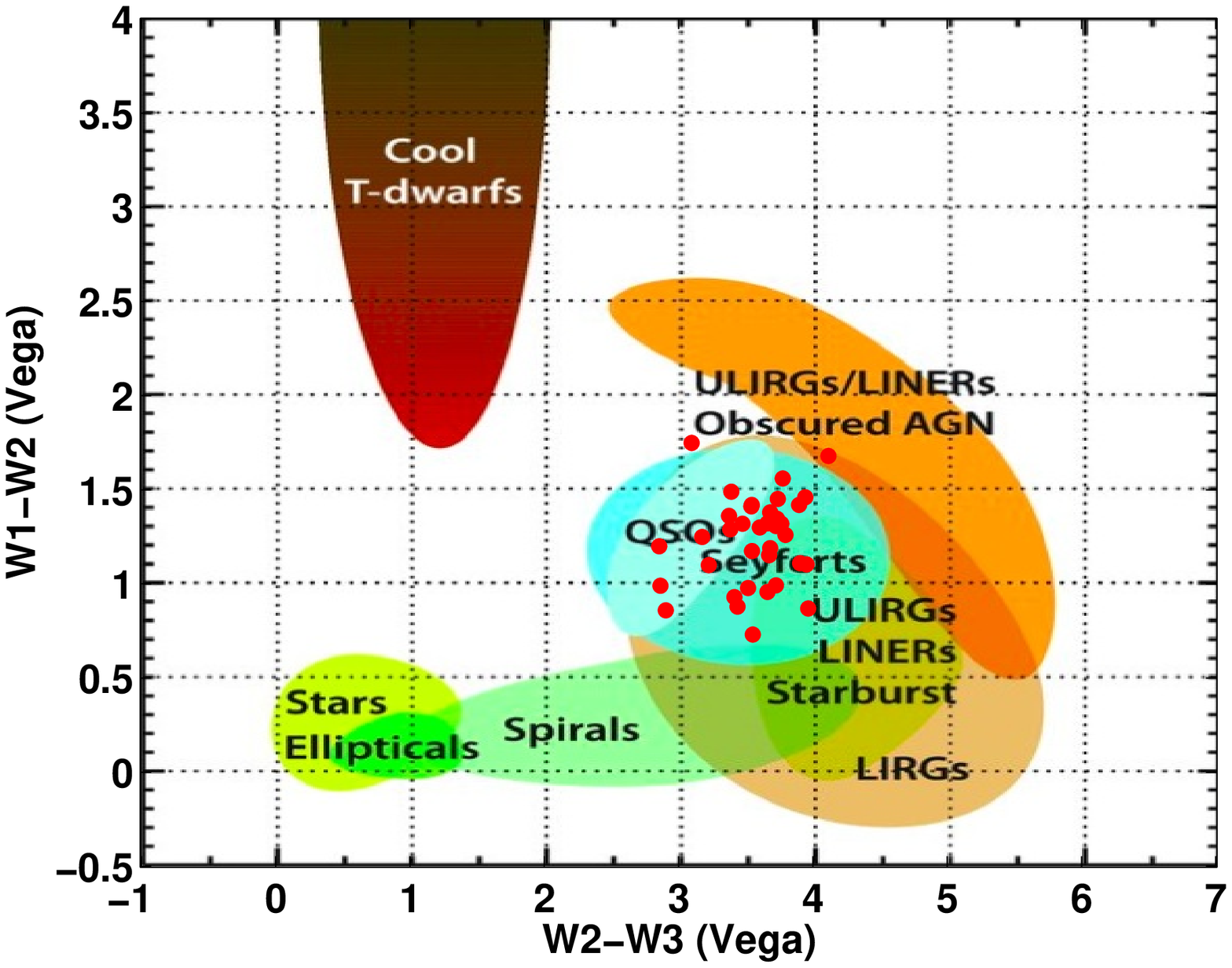} 
\caption{Location of reddened quasars in \textit{WISE} colour-colour
plane taken from \citet{Wright:10} showing that they overlap the
QSO/Seyfert locus.}
\label{fig:wisec}
\end{center}
\end{figure}

The reddened quasar sample was selected using a $(W1-W2)>0.85$ colour cut
as detailed in Section \ref{sec:phot} and are therefore, by
definition, red in the bluer \textit{WISE} bands\footnote{All
\textit{WISE} photometry presented in this section are taken from the
new, improved \textit{WISE} data reductions available through the
\textit{AllWISE} data release. However, the \textit{WISE} main survey
catalogue was used for the target selection as the \textit{AllWISE}
reductions were not available at the time. We find that the
\textit{WISE} and \textit{AllWISE} fluxes are generally consistent
apart from a single quasar: VHSJ2024-5623. The quasar satisfied the
$(W1-W2)>0.85$ colour cut in the original \textit{WISE} data release
with $W1=16.52$ and $W2=15.49$ but has revised magnitudes in
\textit{AllWISE} of $W1=16.17$ and $W2=15.44$.} We show the
\textit{WISE} colours of our spectroscopically confirmed quasars in
Fig.~\ref{fig:wisec} on the colour-colour locus introduced by
\citet{Wright:10}. All the objects are detected at $>3\sigma$ in the
$W3$ 12\,$\mu$m band and the majority have $W1W2W3$-colours placing
them in the QSO/Seyfert locus of \citet{Wright:10}. Just a few of the
reddest sources extend into the Obscured AGN locus.

Fig. \ref{fig:wisec} suggests that, crudely, the reddened quasars are
indistinguishable from unobscured quasars in terms of their \textit{WISE}
colours. We can test this quantitatively by fitting a power-law to the
observed \textit{WISE} fluxes in the $W1$, $W2$ and $W3$ bands of the
form $\lambda F_\lambda \propto \lambda^{\beta_{\rm{NIR}}}$. Emission
from hot dust is expected to dominate the SEDs over the 1--4\,$\mu$m
rest-frame wavelength range probed by the \textit{WISE} passbands.  We
find a mean value of $\beta_{\rm{NIR}}$=0.69 in our sample\footnote{The convention for $\beta_{\rm{NIR}}$ employed here is in wavelength
rather than frequency space compared to our previous work (B13). The two slopes are related
by $\beta_{\rm{NIR}}$=$\beta_{\rm{NIR}}$(B13)+1.}  with a standard
deviation of 0.29. The mean NIR power-law index for the SDSS sample is
0.50$\pm$0.26, consistent with the results from \citet{Wang:13} and
\citet{Hao:11}.

\begin{figure}
\begin{center}
\includegraphics[scale=0.4,angle=0]{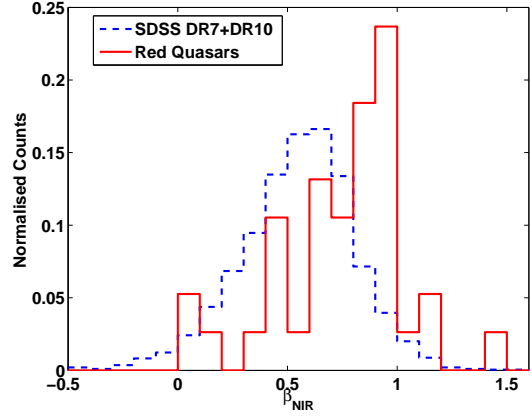} 
\caption{Distribution of NIR power-law indices, $\beta_{\rm{NIR}}$,
fit in the rest-frame wavelength range 1-4$\mu$m both for our reddened
quasar sample and SDSS unobscured quasars over the same redshift range
and down to the same $K$-band flux limit. }
\label{fig:beta}
\end{center}
\end{figure}

The SDSS and reddened-quasar power-law indices agree within 1$\sigma$ of
the population deviation and the steeper mean value for the reddened
quasars is consistent with the most obscured of our quasars showing
some extinction at rest-frame wavelengths of 1--2$\mu$m, which
steepens the slope of the SED.  Fig. \ref{fig:beta} shows the
distribution of $\beta$ for both populations. The high
$\beta_{\rm{NIR}}$ tail is due to the reddest quasars in our sample,
with $E(B-V) \gtrsim 0.8$, as expected. 

Several recent studies have focused on the subset of high redshift AGN
that are extremely bright in the $W4$ 22\,$\mu$m band. \citet{Ross:14}
studied a sample of SDSS and BOSS spectroscopically confirmed quasars
with $W4<=8$\,mag and the \textit{WISE} consortium have selected
hyperluminous infrared galaxies, with $W4 \lesssim 7.7$\,mag, that are
extremely red in their $(W1-W2)$ colours \citep{Eisenhardt:12, Wu:12}.
Our selection for reddened quasars does not use the 22\,$\mu$m fluxes,
although, by selecting objects that are bright in the $K$-band, we are
by definition not sensitive to the \textit{WISE} `W12' drop-outs.
Twenty-five of our 38 reddened quasars are detected at $>3\sigma$ in
the $W4$ band. Of these, 14 have $W4<=8$\,mag and 11 have
$W4<=7.7$\,mag, 37 and 29 per cent of the sample respectively. The
22\,$\mu$m-detected objects are generally the most bolometrically
luminous quasars in our sample, as expected.

\begin{table*}
\begin{center}
\caption{\textit{WISE W3} and $W4$ magnitudes for all reddened quasars
detected at $>3\sigma$ at 22$\mu$m along with the rest-frame 6$\mu$m
luminosity.}
\label{tab:irl}
\begin{tabular}{lccc}
\hline \hline
Name & $W3$ (mags)  & $W4$ (mags) & log$_{10}$(L$_{6\mu m}$ / erg s$^{-1}$) \\
\hline
ULASJ0123+1525 & 8.69$\pm$0.03 & 6.33$\pm$0.06 & 47.8 \\
ULASJ0144-0114 & 10.07$\pm$0.05 & 8.02$\pm$0.21 & 47.1 \\
ULASJ0221-0019 & 10.46$\pm$0.06 & 8.11$\pm$0.17 & 46.9 \\
ULASJ1234+0907 & 9.17$\pm$0.04 & 7.28$\pm$0.17 & 47.4 \\
VHSJ1350-0503 & 10.62$\pm$0.07 & 8.59$\pm$0.26 & 46.7 \\
VHSJ1409-0830 & 10.33$\pm$0.06 & 8.31$\pm$0.23 & 46.9 \\
ULASJ1455+1230 & 10.43$\pm$0.05 & 8.11$\pm$0.15 & 46.3 \\
ULASJ1539+0557 & 9.10$\pm$0.03 & 6.56$\pm$0.06 & 47.8 \\
VHSJ2028-5740 & 9.15$\pm$0.03 & 7.01$\pm$0.08 & 47.3 \\
VHSJ2048-4644 & 10.42$\pm$0.08 & 7.92$\pm$0.18 & 46.9 \\
VHSJ2100-5820 & 10.70$\pm$0.10 & 8.14$\pm$0.24 & 46.9 \\
VHSJ2101-5943 & 9.25$\pm$0.03 & 6.84$\pm$0.07 & 47.4 \\
VHSJ2115-5913 & 9.48$\pm$0.04 & 7.06$\pm$0.08 & 47.2 \\
VHSJ2130-4930 & 10.69$\pm$0.09 & 8.52$\pm$0.33 & 46.9 \\
VHSJ2141-4816 & 10.46$\pm$0.07 & 7.68$\pm$0.16 & 47.3 \\
ULASJ2200+0056 & 10.23$\pm$0.06 & 7.80$\pm$0.17 & 47.2 \\
VHSJ2212-4624 & 11.02$\pm$0.12 & 8.37$\pm$0.27 & 46.7 \\
VHSJ2220-5618 & 8.77$\pm$0.03 & 6.37$\pm$0.05 & 47.6 \\
VHSJ2227-5203 & 10.11$\pm$0.06 & 8.02$\pm$0.22 & 47.2 \\
VHSJ2235-5750 & 9.97$\pm$0.05 & 7.84$\pm$0.17 & 47.0 \\
VHSJ2256-4800 & 9.67$\pm$0.04 & 7.52$\pm$0.13 & 47.1 \\
VHSJ2257-4700 & 10.77$\pm$0.09 & 8.43$\pm$0.28 & 46.7 \\
ULASJ2315+0143 & 9.57$\pm$0.04 & 7.03$\pm$0.10 & 47.5 \\
VHSJ2332-5240 & 11.51$\pm$0.20 & 8.45$\pm$0.31 & 46.9 \\
VHSJ2355-0011 & 9.83$\pm$0.05 & 7.40$\pm$0.14 & 47.3 \\
\hline
\end{tabular}
\end{center}
\end{table*}

\begin{figure}
\begin{center}
\includegraphics[scale=0.4,angle=0]{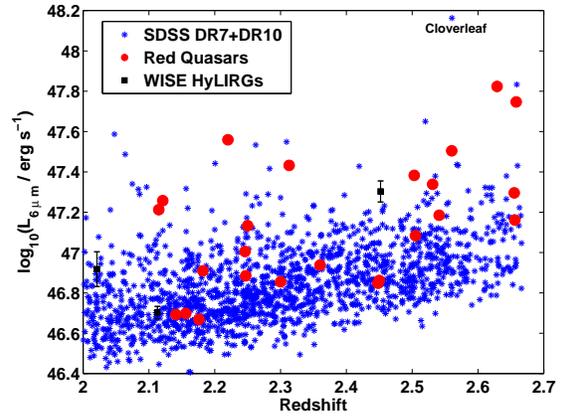} 
\caption{Redshift versus rest-frame $6\mu$m luminosity for our
reddened quasars compared to unobscured quasars from SDSS as well as 3 \textit{WISE} HyLIRGs from
\citet{Stern:14}. The brightest unobscured quasar on this plot is the
lensed Cloverleaf quasar.}
\label{fig:wise_sed}
\end{center}
\end{figure}

For objects detected in the \textit{WISE} $W4$-band, rest-frame
luminosities at 6\,$\mu$m can be estimated using the $W3$ and $W4$
photometry. At the median redshift of $z=$2.3 for our sample, the $W3$
band corresponds to 3.6\,$\mu$m and the \textit{W4} band corresponds
to 6.7\,$\mu$m. The 6\,$\mu$m luminosities are calculated via linear
interpolation, using the rest-frame wavelengths corresponding to the
$W3$ and $W4$ magnitudes, and the results are given in
Table~\ref{tab:irl} and illustrated in Fig. \ref{fig:wise_sed}. The average rest-frame 6\,$\mu$m luminosity is
log$_{10}$(L$_{6\mu\rm{m}}$/erg s$^{-1}$)=47.1$\pm$0.4. For a comparable SDSS DR7+DR10
sample of 1336 2.1$<z<2.7$ quasars with K$<18.9$ and $>$3$\sigma$ detections in the $W4$ band, the average luminosity is log$_{10}$(L$_{6\mu\rm{m}}$/erg
s$^{-1}$)=46.8$\pm$0.2. There are six SDSS quasars with
log$_{10}$(L$_{6\mu\rm{m}}$/erg s$^{-1}$)$>47.5$ (four from SDSS DR7
and two from SDSS DR10).  The brightest of these six is the well-known
gravitationally-lensed Cloverleaf quasar \citep{Magain:88}.
SDSSJ000610.67+121501.2 ($z=2.309$) and SDSSJ155102.79+084401.1
($z=2.520$), the two DR10 objects, were presented in the extremely red
sample studies by \citet{Ross:14}.  By comparison, our reddened quasar
sample, selected using imaging data covering less than a third of the total SDSS imaging area ($\sim$14,555 deg$^2$ from \citealt{Paris:13}), includes four quasars with
log$_{10}$(L$_{6\mu\rm{m}}$/erg s$^{-1}$)$>47.5$, highlighting the
increased prevelance of such luminous quasars among the obscured
sample.  Compared to \textit{WISE}-selected HyLIRGs
where the most luminous object at 6\,$\mu$m, WISEJ1814+3412
\citep{Eisenhardt:12}, has log$_{10}$(L$_{6\mu\rm{m}}$/erg
s$^{-1}$)=47.30$\pm$0.05 \citep{Stern:14}, the reddened quasar sample is also extreme, with
seven objects more luminous than WISEJ1814+3412.

Finally, we compare the 6\,$\mu$m luminosities to the bolometric
luminosities in Table \ref{tab:demo}, and find a typical bolometric
correction factor of $\sim$2, consistent with a model in which a large
fraction of the total luminosity is re-radiated at mid infrared
wavelengths. This bolometric correction from the mid infrared is also broadly consistent
with that derived for the most luminous essentially unobscured quasars in SDSS and \textit{WISE} \citep{Weedman:12}. 
As the 6\,$\mu$m rest-frame luminosities are inferred
directly from the observed photometry without any assumptions
regarding the dust extinction law and the form of the quasar SED, this
analysis corroborates our conclusion in Section \ref{sec:sed}, that
our reddened quasars correspond to some of the most intrinsically luminous
quasars known.

\section{Obscured Black-Hole Growth at the Highest 
Quasar Luminosities}

With a sample of 38 reddened quasars ($E(B-V)>=0.5$), of
which 36 lie at $z>2$, now in place, a direct comparison of
their space density to that of unobscured quasars can be made. 
As deep X-ray and mid-infrared surveys have focussed on studying 
much fainter, dusty, AGN, the sample allows us to address the key
question of how many very luminous quasars are
dust-obscured at redshifts $z\sim2-3$. We begin by considering the 
selection function and completeness of our reddened quasar sample.

\subsection{Sample Completeness}

\label{sec:complete}

\subsubsection{Morphology Cut}\label{sec:morph}

The sample selection (Section~\ref{sec:phot}) includes a requirement
that candidates are unresolved in the $K$-band. While eliminating
contamination by low-redshift quasars, where host galaxies can
contribute significantly to the observed colours (B12), it is
necessary to quantify the fraction of high-redshift quasars that may
also be excluded. To assess such incompleteness, we select a sample of
unobscured quasars from the SDSS DR7 \citep{Schneider:10} and SDSS
DR10 \citep{Paris:13} quasar catalogues, with redshifts $2.1<z<2.7$
and magnitudes $K<19.3$. There are a total of 7871 quasars that satisfy these criteria. The cumulative fraction of unobscured
quasars classified as unresolved as a function of $K$-band magnitude is
shown in Fig. \ref{fig:complete}.  At the bright-end ($K<18$), almost
90 per cent of the quasars are unresolved, with the fraction
decreasing to just under 70 per cent at the faint limit. Only four
of the 48 reddened quasar candidates are in the faintest magnitude bin
and, using the distribution of candidate $K$-magnitudes, we calculate the
incompleteness due to the image morphology constraint to be
20-25\,per cent.

\subsubsection{$(J-K)$ Colour Cut}

\begin{figure}
\begin{center}
\includegraphics[scale=0.45,angle=0]{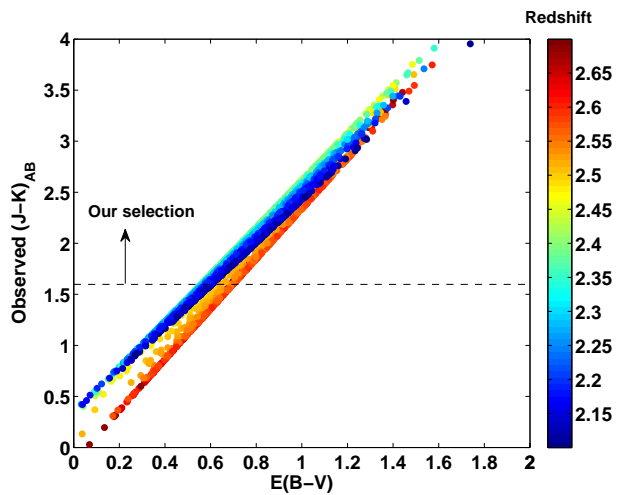} 
\caption{Observed, reddened $(J-K)$ colour as a function of extinction
$E(B-V)$ for a set of simulated quasars assuming the base SED model
described in Section \ref{sec:sed} and an SMC-like extinction
law. Points are colour-coded by the redshift of the quasar. The dashed
horizontal line shows our colour selection for the reddened quasars.}
\label{fig:JKebv}
\end{center}
\end{figure}

The NIR colour-cut of $(J-K)>1.6$ is designed to isolate the most
highly reddened Type 1 broad-line quasars. Fig. \ref{fig:JKebv} shows
model quasar $(J-K)$ colours, calculated using our quasar SED model
(Section \ref{sec:sed}), over the redshift range of our sample, as a
function of the extinction parameter $E(B-V)$. Some dependence of
$(J-K)$ colour on redshift is present; higher redshift quasars have
bluer $(J-K)$ colours due to the H$\alpha$ emission line moving out of
the $K$-band. However, there is a tight correspondence between
extinction and colour, with the $(J-K)>1.6$ threshold predicted to
define a quasar sample with $E(B-V) \gtrsim 0.5$.

A potential concern is that quasars with very large H$\alpha$
equivalent widths and somewhat lower $E(B-V)$ values form a
significant fraction of the sample. The direct measurements of the
H$\alpha$ equivalent widths for the reddened quasars
(Table~\ref{tab:fwhm}), however, demonstrate that the presence of such
objects is not a factor. The highest equivalent width quasar in our sample has an equivalent width
of only a factor of $\sim$2 larger than the average equivalent width in unobscured quasars of similar luminosity. 
The average H$\alpha$ equivalent widths however are very similar to those seen in unobscured quasars, confirming that the reddened quasars we have selected
do not have unusual equivalent widths relative to the unreddened population. 

The systematic dependence of the threshold
$E(B-V)$ as a function of redshift to satisfy the $(J-K)>1.6$
selection, evident in Fig.~\ref{fig:JKebv}, is directly taken into
account when estimating the space-density
(Section~\ref{sec:vvmax}). The S/N of the $K$-band photometry used in the sample selection is
high and any apparent increase in the observed number of objects due
to Eddington bias \citep{Eddington:1913} is small. The S/N of the $J$-band
photometry is somewhat lower and the $E(B-V)$ selection at a
given redshift, resulting from the $(J-K)>1.6$ colour cut, is thus
blurred, with some objects lost from the sample and others
included due to photometric errors. This could be a concern if the $E(B-V)$ distribution was changing
significantly at the boundary of our selection box ($E(B-V)\sim0.5$), in which case unequal numbers of objects 
may be scattered out of the sample relative to the numbers of objects that are scattered in. However, estimates of the form of the $E(B-V)$ distribution from SDSS 
\citep{Hopkins:01} and near-IR selected samples \citep{Maddox:12}
strongly suggest that the $E(B-V)$ distribution is not changing
significantly and our sample should thus represent an accurate
estimate of number of objects with $E(B-V)$ exceeding the limits
calculated. 

\subsubsection{WISE Colour Cut}

We employ the same SDSS sample of unobscured quasars used in
Section~\ref{sec:morph} to determine the incompleteness of the
reddened quasar sample due to the application of the restriction that
\textit{WISE} $(W1-W2)>=0.85$. The \textit{WISE} colours of the
reddened quasar sample are themselves affected by the presence of
dust. Specifically, at the median redshift of the sample ($z=2.3$),
the $W1$ and $W2$ magnitudes of a quasar possessing an $E(B-V)$ of 
0.5\,mag are 0.75 and 0.53\,mag fainter respectively. The requirement
that $(W1-W2)>=0.85$ for reddened quasars with $E(B-V) \gtrsim 0.5$
therefore corresponds to $(W1-W2)>=0.63$ for the unobscured SDSS sample.

\begin{figure}
\begin{center}
\includegraphics[scale=0.45,angle=0]{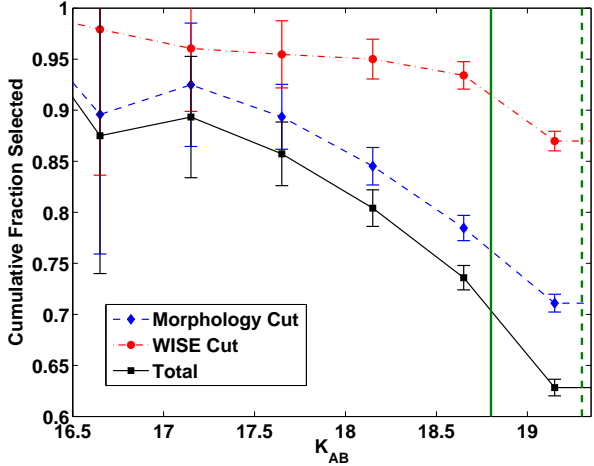} 
\caption{Cumulative fraction of SDSS DR7 and DR10 quasars at
$2.1<z<2.7$ and $K<19.3$, that are also classified as point sources in
the $K$-band, as a function of $K$-band magnitude
(dashed line). The dot-dashed line shows the cumulative fraction of
SDSS quasars that have $(W1-W2)>0.63$ (corresponding to $(W1-W2)>0.85$
for $E(B-V)>0.5$). The solid line shows the cumulative fraction of
SDSS high redshift quasars that satisfy both the morphology and
\textit{WISE} colour selection. The solid vertical line marks the magnitude
limit used for the majority of our reddened quasar sample while the dashed
vertical line shows the magnitude limit used in the 83 sq-deg VHS-DES
SPT deep field. We can see that $>$70 per cent of true high redshift
quasars would satisfy all our selection criteria at $K$=18.9 and this
completeness falls to just under 65 per cent at $K=19.3$.}
\label{fig:complete}
\end{center}
\end{figure}

Figure~\ref{fig:complete} also shows the cumulative fraction of the
unobscured SDSS sample with $(W1-W2)>=0.63$, W1$_{\rm{S/N}}>5$ and
W2$_{\rm{S/N}}>5$ (dot-dashed line). This fraction is $>$90 per cent for most of the
magnitude range but falls to $\simeq$85 per cent in the faintest
magnitude bin, used in the 83 deg$^2$ VHS-DES SPT Deep Field survey.

The cumulative fraction of quasars, with redshifts $2.1<z<2.7$, that
satisfy both the $(J-K)$ and $(W1-W2)$ colour cuts, plus the
unresolved morphology constraint, i.e. the combined sample photometric
completeness, is also shown in Fig.~\ref{fig:complete}. More than
70\,per cent of quasars satisfy all our selection criteria at $K=18.9$
with the completeness falling to just under 65 per cent at $K=19.3$.

\subsection{Redshift Completeness}

\begin{figure}
\begin{center}
\includegraphics[scale=0.45,angle=0]{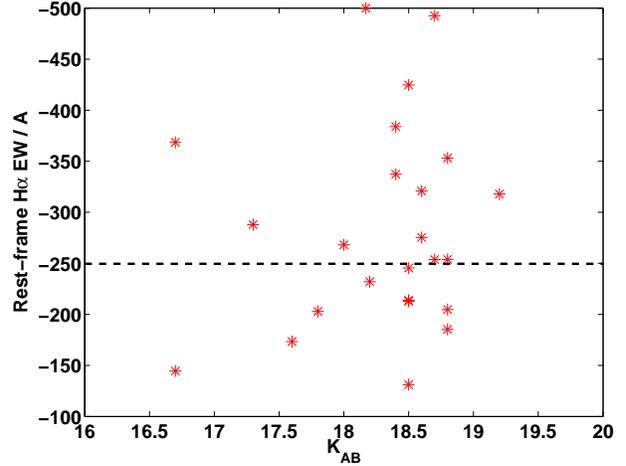} 
\caption{Rest-frame equivalent width of the H$\alpha$ line in our reddened quasars as a function of the $K$-band magnitude. The horizontal dashed line represents the average H$\alpha$ equivalent widths of unobscured quasars at the same redshift and luminosity. The H$\alpha$ lines have a range of equivalent widths between $\sim0.5-2$ times the equivalent widths in unobscured quasars. Particularly at the faint-end of the sample, a range of H$\alpha$ equivalent widths is evident with no strong correlation between the $K$-band magnitude and the equivalent width.}
\label{fig:ew_snr}
\end{center}
\end{figure}

In addition to the candidate photometric completeness, the
effectiveness of the spectroscopic observations for determining
redshifts and hence reliable object classifications must be
considered.  Fig.~\ref{fig:spectra} shows that the high
equivalent-width, broad, H$\alpha$ emission lines are clearly detected
at high S/N, even at the faint $K$-band limit (e.g. VHSJ2332-5240) or
when the line is at the edge of the wavelength interval
(e.g. VHSJ2227-5203). All H$\alpha$ lines have S/N$>4$. 
The observational strategy
(Section~\ref{sec:specobs}), where integration times
(Table~\ref{tab:sample}) were adjusted to achieve a minimum S/N for
each target sufficient to allow detection of broad emission lines,
thereby ensures that the redshift completeness is close to 100 per
cent within the redshift interval $2.1<z<2.7$. Fig.~\ref{fig:ew_snr} shows the rest-frame equivalent width of the H$\alpha$ line as a
function of the quasar K-band magnitude. No significant correlation
between the two quantities is present and, particularly for the
faintest quasars, a range of equivalent widths is evident. There is no
evidence to suggest that, for the faintest sources, our survey is
preferentially detecting quasars with large H$\alpha$ equivalent widths. 

In the space-density analysis that follows, we first consider a
`Bright Sample', from the VHS-DES SPT, VHS-DES S82-W and VHS-ATLAS SGC
regions, as summarised in Table~\ref{tab:area}. These three regions
cover a total area of 1115\,deg$^2$ and all 29 candidates with
$K_{AB}<18.9$ possess spectra. Twenty one of the 29 candidates have
redshifts. In addition, We also consider a `Faint Sample', based on
the $K_{AB}<19.3$ sample, covering the 83\,deg$^2$ SPT Deep Field,
where, again, all candidates possess spectra.

\subsection{The Reddened Quasar Space Density}\label{sec:vvmax}

We calculate the $V/V_{\rm{max}}$ weighted \citep{Schmidt:68, Avni:80} estimate of the reddened
quasar space density from the 21 reddened quasars in the Bright
Sample.  Using the best-fit SED for each reddened quasar, we can
calculate its absolute magnitude as well as the redshift range and therefore total volume $V$, within
which it would have satisfied our survey selection criteria. Our
space density calculations assume the quasar SED model from Section
\ref{sec:sed} in order to derive k-corrections and
therefore absolute magnitudes and bolometric luminosities. As our SED
model is slightly different from the one used in SDSS \citep{Ross:13} and in
order to allow comparison with investigations by other groups.  we
provide a table with the $i$-band and $K$-band k-correction from our
model in Table \ref{tab:kcorr}. However, we note that differences in k-corrections derived from our template and that of \citet{Ross:13} are expected to be small given that both templates are able to reproduce the observed colours of SDSS quasars \citep{Maddox:12,Ross:13}. The maximum available survey volume, $V_{\rm{max}}$ is the co-moving volume
within $2.1<z<2.7$. The cumulative volume and completeness-corrected number density in each
absolute magnitude bin is then given by:

\begin{equation}
N(M<M_i)_{\rm{corr}} / \rm{Mpc}^3=\frac{\rm{f_{sky}}}{V_{max}} \Sigma_{j=1}^{Nbin} \frac{1}{c(K) \times V_j/V_{\rm{max}}}
\label{eq:VVmax}
\end{equation}

\begin{figure*}
\begin{center}
\includegraphics[scale=0.6,angle=0]{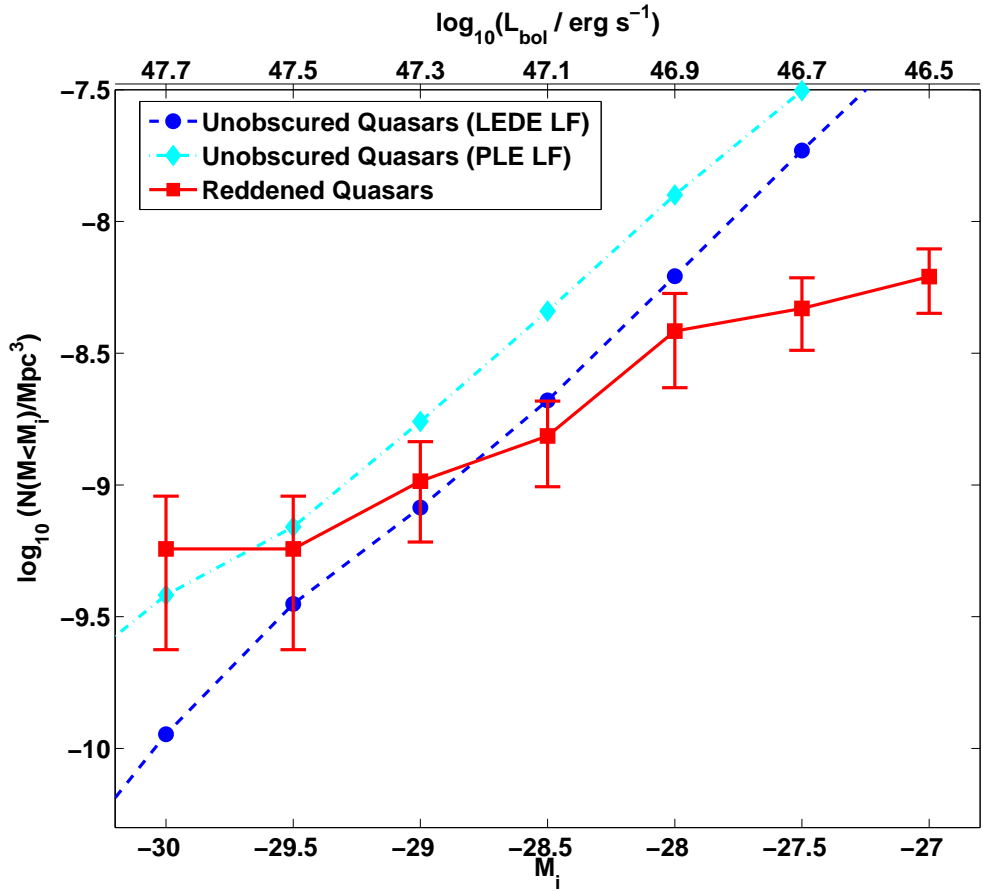}
\caption{Completeness-corrected, $V/V_{max}$ estimates of the cumulative number density of reddened quasars (red solid) compared to the 
equivalent cumulative number-densities of unobscured quasars derived over the same redshift and luminosity range assuming the SDSS+BOSS LEDE 
LF model from \citet{Ross:13} (blue dashed) and the SDSS+BOSS PLE LF from \citet{Ross:13} (cyan dot-dashed). Reddened quasars appear to dominate
the counts at high intrinsic luminosities but their number counts flatten off relative to unobscured quasars as we approach more typical quasar luminosities.}
\label{fig:spacedensity}
\end{center}
\end{figure*}

\noindent The error on these cumulative number counts is:

\begin{equation}
\sigma_{N(M<M_i)}=\frac{\rm{f_{sky}}}{V_{max}} \Sigma_{j=1}^{\rm{Nbin}} \frac{1}{(\rm{c(K)} \times \rm{V}_j/\rm{V}_{\rm{max}})^2}
\label{eq:VVmax2}
\end{equation}

\noindent where $f_{sky}$ denotes the fraction of area on the sky encompassed by our survey and $c(K)$ is the $K$-band magnitude dependent photometric completeness of each 
quasar calculated using the completeness curve in Fig.~\ref{fig:complete}. The resulting cumulative space density is shown in Fig.~\ref{fig:spacedensity} together with the equivalent space density for unobscured quasars derived from the SDSS+BOSS luminosity function (LF; \citealt{Ross:13}) over
exactly the same redshift and absolute magnitude range as the reddened sample. As our luminosity function for reddened quasars extends to brighter absolute magnitudes than the 
SDSS+BOSS quasars, we assume the parametric double power-law form of the unobscured quasar luminosity function and integrate this parametric luminosity function over the appropriate
redshift and luminosity ranges to compare to the reddened quasars in Fig.~\ref{fig:spacedensity}. We consider two best-fit forms for this parametric luminosity function for unobscured quasars
from \citet{Ross:13}. Firstly, a luminosity and density evolution (LEDE) model and secondly a pure luminosity evolution model (PLE). We note that the SDSS+BOSS luminosity
function is constrained in the redshift range $2.2 < z < 3.5$
and only a very modest extrapolation of the luminosity function
and redshift-evolution is necessary to compare with our new data.
In \citet{Ross:13}, a pure luminosity evolution (PLE) model has
instead been fit to the data at $z < 2.2$ but the effect on the
cumulative LEDE model based number counts of the unobscured quasars plotted in Fig.~\ref{fig:spacedensity}
is small. We also note that the systematic differences in k-corrections between our models and those used by BOSS, are expected to be smaller than the bin size in Fig.~\ref{fig:spacedensity}.

We find that the LF of the reddened quasars is significantly
flatter than that of the SDSS+BOSS quasars over the entire magnitude
range. At bright absolute magnitudes
($M_i < -29$), the reddened quasars possess larger space densities than predicted by the SDSS+BOSS LEDE LF,
but the more steeply rising SDSS+BOSS quasar LF means the unobscured
quasars become significantly more common at magnitudes $M_i >
-27.5$. The much flatter increase in the number of quasars as a
function of decreasing luminosity leads to very different
interpretations regarding the fraction of the total energy liberated
by the Type 1 AGN population in an `obscured' phase, compared to
previous investigations which probed only the most luminous quasars (e.g. \citealt{White:03, Glikman:07}). Our 
survey now covers enough dynamic range in luminosity to demonstrate that while the numbers of obscured ($0.5 \lesssim E(B-V) \lesssim 1.5$)
quasars are significant at the highest luminosities, their space density shows a much shallower
rise relative to unobscured quasars at these redshifts. There is no evidence from our data that the reddened quasars have significantly different
H$\alpha$ equivalent width distributions relative to unobscured quasars so we are unlikely to be missing a population of very low equivalent width objects but 
with similar levels of reddening as in our sample. As such, these reddened broad-line quasars are likely to be sub-dominant relative to 
the unobscured quasars in terms of their contribution to the total energy liberated during the luminous quasar phase.  For reference, our $K=$18.9 magnitude limit corresponds to $M_i
\sim -27.5$ at $z=2.3$ for a quasar with $E(B-V)=0.75$. This is $\simeq$1.3\,mag brighter than $M^*$ at these redshifts, where the bulk of the luminosity is produced. A larger survey to fainter $K$-band limits is required to determine
the exact fraction of reddened quasars down to the knee of the intrinsic
quasar luminosity function and hence quantify what fraction of the
accretion luminosity for the Type 1 AGN-population is generated in an
obscured phase. 

Our result, that reddened quasars are more common at bright absolute
magnitudes, contrasts with observations of intrinsically less luminous
AGN-populations, where the obscured fraction shows the opposite trend
and decreases with increasing luminosity \citep{Ueda:03}. On the other
hand, \citet{Assef:14} have recently studied the space density of the
highly obscured population of \textit{WISE}-selected HyLIRGs, which
overlap our sample in terms of both luminosity and redshift, but have
higher extinctions. They also find that the fraction of obscured
luminous AGN is significant at very high luminosities.

Our sample is not large enough to merit fitting a parametric form for
the evolution of the reddened quasar LF as a function of redshift.  Nevertheless, 
the differences in the cumulative number counts of unobscured quasars
seen from the LEDE and PLE LFs in this redshift range, can help shed 
light on the kind of evolutionary model that may be appropriate. Starting with
the SDSS+BOSS luminosity functions in Fig.~\ref{fig:spacedensity} and assuming that (i) the extinction is independent of 
luminosity and (ii) that every quasar possesses a reddening between 0.5$<$$E(B-V)$$<$2.0
with reddening values drawn randomly according to a flat probability distribution in this range, we can estimate how many reddened quasars we would have observed in our survey area
employing the completeness function shown in Fig.~\ref{fig:complete}. These predicted counts, when compared to our observed counts, can 
then help determine whether the underlying luminosity distribution of reddened quasars, is indeed the same as that of unobscured quasars.
Based on the normalisation provided by the BOSS LEDE LF, we predict 15$\pm$4 reddened quasars would be detected in the
Bright Sample, which is lower than the 21 quasars we have observed. Adopting instead the $E(B-V)$ distribution
inferred from SDSS quasars by \citet{Hopkins:01} rather than a flat $E(B-V)$ distribution in the range 0.5$<$$E(B-V)$$<$2.0, results in 
0.35$\pm$0.59 quasars predicted in our survey. The dramatic
difference in the number of objects provides a direct illustration of
the need to incorporate near-IR photometric selection in order to
identify quasars with reddenings of $E(B-V)$$\simeq$0.5-1.0. Finally, adopting the BOSS PLE
LF model instead, which has a brighter break-luminosity ($M^*$) than the LEDE model, 
results in a larger space density for the most luminous quasars as evident in Fig.~\ref{fig:spacedensity}. As a
consequence, using the normalisation from the PLE LF-evolution model
would predict a larger number of reddened quasars in our Bright Sample - 34$\pm$4. Neither model, together 
with the assumption that the extinction is independent of luminosity, provides a good match to our observed number
counts of reddened quasars . The SDSS+BOSS LEDE LF produces too few reddened quasars while the PLE LF produces too many. The results
demonstrate that reddened quasars with $-27.0 < M_i < -30.0$ do not
possess the same intrinsic [unreddened] LF as unobscured quasars
unless the intrinsic luminosity and the $E(B-V)$ distribution are
related. Qualitatively, the probability that a quasar possesses a significant
$E(B-V)$ must be greater for intrinsically more luminous objects in order for the SDSS+BOSS luminosity distributions to be consistent with our data. The
quasar luminosity and reddening are thus not independent.  A model in which the probability of a quasar having high $E(B-V)$ increases with increasing
luminosity would be viable. However, in the conventional AGN unification
model, incorporating a receding torus, the dependence is predicted to
take the opposite sign. Alternatively, for an explanation related to 
some element of the AGN-fueling or life-cycle, individual quasars could
experience episodes where both the luminosity and the $E(B-V)$
increase, causing them to migrate brightward in the quasar LF for a
period determined by the lifetime of the fueling cycle.

While the number of quasars in our Faint Sample is modest,
if we calculate the observed number density of reddened quasars down
to $K=$18.9 (Bright) and 19.3 (Faint) we obtain surface densities of
$0.04 \pm 0.01$ deg$^{-2}$ and $0.04 \pm 0.03$ deg$^{-2}$
respectively. The Poisson error associated with the Faint Sample is
large but the result provides additional evidence that the LF of the
reddened quasar populations continues to increase only slowly down to
fainter absolute magnitudes.

\citet{Assef:14} came to the same conclusion using an independently
selected sample of more highly obscured AGN from \textit{WISE}. Our
sample covers a larger dynamic range in terms of intrinsic quasar
luminosity as well as selecting quasars that fall at the lower
boundary of their $E(B-V)$ distribution. In combination, the two
investigations provide evidence for the existence of a significant
population of optically-obscured, hyperluminous quasars with
$E(B-V)>0.5$, with our results probing smaller reddenings and
fainter intrinsic luminosities than those in \citet{Assef:14}. Quasars with $E(B-V )$$\gtrsim$1.5 are not included in our sample
as their observed $K$-band magnitudes put them below the
flux limit of our survey. At the bright-end however, \citet{Assef:14} have confirmed the existence
of a population of quasars with such extinctions. 
The space-densities shown in Fig.~\ref{fig:spacedensity} therefore represent a lower
limit to the numbers of obscured quasars that may be missing from optical surveys.

\section{CONCLUSIONS}

We have presented a search for heavily dust-obscured luminous quasars
at $z>2$ selected from the new wide-field infrared photometry provided
by the UKIDSS Large Area Survey, the ESO VISTA Hemisphere Survey and
the \textit{WISE} All Sky Survey. From 48 candidates, we used VLT-SINFONI NIR spectroscopy to confirm 24
new $z>2$, heavily reddened broad-line quasars with $0.5<E(B-V)<1.5$. Three of these
new quasars (VHSJ2028$-$5740, ULASJ2315$+$0143 and ULASJ0123$+$1525) have $E(B-V)>1$ and correspond to the dustiest broad-line
Type 1 quasars known. Combining the new quasars with objects presented
in B12 \& B13, the full sample consists of 38 spectroscopically
confirmed, heavily reddened quasars, 36 of which are at $z>2$.  There is no evidence for these near infrared selected reddened quasars
having systematically different H$\alpha$ equivalent widths relative to their unobscured counterparts.
We present a space density estimate for the reddened quasar
population at $z\sim2-3$ enabling detailed comparisons to the
space-density of unobscured quasars over the same redshift and
luminosity range. In particular, we reach the following conclusions:

\begin{itemize}

\item{The reddened quasars have very high bolometric luminosities and
black-hole masses that are comparable to the brightest unobscured
quasars at these redshifts. The median bolometric luminosity
of the new sample presented here is log$_{10}$(L$_{\rm{bol}}$/erg
s$^{-1}$)=47.1$\pm$0.4 and the median black-hole mass is
log$_{10}$(M$_{\rm{BH}}$/M$_\odot$)=9.7$\pm$0.4, consistent with our
previous samples of reddened quasars. There is some evidence for the
presence of a tail of quasars out to very high bolometric luminosities 
of $\sim10^{48}$ erg/s. Around 70 per cent of the reddened quasars have Eddington ratios of $L/L_{\rm Edd}>0.1$
confirming the presence of massive black-holes accreting at a high
rate. Apart from the high dust extinctions along the
line-of-sight, the properties of the reddened quasars appear to be similar to
luminous unobscured quasars.}

\item{We compile \textit{WISE} photometry out to 22\,$\mu$m for all
our reddened quasars and using a simple power-law fit to the
rest-frame 1-4\,$\mu$m SED, we show that the power law indices in the
reddened quasars are consistent with those measured for unobscured
quasars from the SDSS over the same redshift range and down to the
same $K$-band magnitude limit. The mean power-law index ($\lambda$ F$_\lambda \propto \lambda^{\beta_{\rm{NIR}}}$)
at rest-frame wavelengths of 1-4$\mu$m is $\beta_{\rm{NIR}}$=0.69$\pm$0.29 for the reddened quasars
compared to 0.50$\pm$0.26 for the unobscured quasars. The reddened
quasars have slightly steeper power-law indices on average due to the
non-negligible extinction at rest-frame wavelengths of $1-2$\,$\mu$m
for some of our reddest objects.}

\item{We use the subset of 25 22\,$\mu$m-detected reddened quasars to
calculate their rest-frame mid infrared 6\,$\mu$m luminosity. We find
an average value of log$_{10}$(L$_{6\mu\rm{m}}$/erg
s$^{-1}$)=47.1$\pm$0.4 compared to log$_{10}$(L$_{6\mu\rm{m}}$/erg
s$^{-1}$)=46.8$\pm$0.2 for equivalent SDSS 22\,$\mu$m-detected
quasars. The most luminous \textit{WISE} HyLIRG, WISEJ1814+3412 has
log$_{10}$(L$_{6\mu\rm{m}}$/erg s$^{-1}$)=47.30$\pm$0.05 and we find
seven of our sample are more luminous at 6\,$\mu$m than
WISEJ1814+3412. The
6\,$\mu$m rest-frame luminosities are robust, estimated purely from
the observed \textit{WISE} photometry without assumptions regarding
the quasar SED or the extinction law, corroborating our conclusion
that these reddened quasars constitute some of the most luminous
quasars known at these epochs.}

\item{We present the first estimate of the space density of
reddened ($0.5<E(B-V)<1.5$) quasars at $z\sim2-3$. We find that the
most luminous reddened quasars, with de-reddened $M_i < -29$, are at least as
numerous at these epochs as unobscured quasars. We also find however
that the reddened quasar luminosity function has a much flatter slope
and the number density of unobscured quasars, due to the rapidly rising
luminosity function, greatly exceeds that of the reddened population by
$M_i \sim -27.5$. The slower
increase in the number of reddened quasars as a function of decreasing
luminosity suggests that as a whole, luminous reddened Type 1 
quasars likely do not account for a significant fraction of the total energy liberated by the Type 1 AGN population in an
`obscured' phase.}

\item{The observed evolution of of the reddened quasar space-density
with luminosity is also opposite to that seen in the much less luminous Type 2 AGN samples
and to the predictions from the receding torus model, where the obscured
fraction is predicted to increase with decreasing luminosity. Our population of reddened quasars therefore probe a very
different region of the luminosity-extinction parameter space compared to these Type 2 AGN samples, which have previously
been selected through deep, small-area surveys in the X-ray and mid infra-red.}

\item{We test whether different assumptions for the reddened quasar luminosity function, based on the luminosity function of unobscured quasars, 
are able to reproduce our observations. A model where the luminosity and density are
evolving with redshift (LEDE model), and the normalisation is set by
the space-density of unobscured quasars, produces too few reddened quasars if the extinction is independent of luminosity.  A pure
luminosity evolution (PLE) model with a brighter break luminosity
reproduces better the observed counts of reddened quasars in the
brightest absolute magnitudes in our survey, but in general results in too many reddened quasars being predicted in our survey. 
Neither model together with the assumption that extinction is independent of luminosity, is able to reproduce the observed number
counts of reddened quasars which demonstrates that either reddened quasars do not have the same intrinsic luminosity distribution as unobscured quasars, or
that the extinction must depend on luminosity. A more elaborate model,
involving an increase in the probability of finding high extinction
quasars at high luminosities, would be consistent with the
observations. Qualitatively, this behaviour could be related to some
element of the AGN fuelling cycle where both the extinction and
luminosity increase for a period, causing the quasars to migrate
brightward in the luminosity function.}

\end{itemize}

We conclude that our sample of high redshift ($z\sim2.5$), heavily
reddened (0.5$\lesssim E(B-V) \lesssim 2$), hyperluminous
(L$_{\rm{bol}}\gtrsim10^{13}$L$_\odot$) quasars effectively bridge the gap
between the optically-luminous, unobscured, quasars known at these
redshifts and the much more heavily obscured AGN
($E(B-V) \gtrsim 5$) now emerging from the \textit{WISE} All-Sky
Survey. We demonstrate that sensitive near-infrared observations such as those provided by the UKIDSS Large Area Survey (in the northern hemisphere) and the 
VISTA Hemisphere Survey (in the southern hemisphere), are critical
for identifying the population of quasars with intermediate
extinctions at high redshift. The space density of such obscured quasars dominates the total quasar
counts at very bright luminosities but they are unlikely to account for a
significant fraction of the total energy liberated by the luminous, Type 1 AGN
population. Reddened quasars likely constitute
a brief evolutionary phase in galaxy formation associated with high luminosities and moderate extinctions towards the quasar line-of-sight. Discriminating between different
models for the evolution of these AGN, as they experience changes in
their black hole growth rate, $L/L_{\rm Edd}$, degree of obscuration
and outflow properties, will require multi-wavelength investigations
of carefully selected sub-samples of objects and is now becoming
possible with sensitive NIR integral field units (e.g. SINFONI on the
VLT) and ALMA.

\section*{Acknowledgements}

MB is grateful for the assistance provided by the staff at the ESO
Paranal Observatory, in particular Jonathan Smoker and John Carter,
during the Visitor Mode observations that were conducted for this
analysis. Based on observations obtained as part of the VISTA
Hemisphere Survey, ESO Program, 179.A-2010 (PI: McMahon) and the ESO
Program, 091.A-0341 (PI: Banerji).

\bibliography{}

\appendix
\section{Spectra \& Spectral Line Fits}

We present the SINFONI $K$-band spectra for all our spectroscopically
confirmed reddened quasars along with the best-fit SED models
described in Section \ref{sec:sed} in Fig. \ref{fig:spectra}. We note that the SED
models include a scaling of the H$\alpha$ line to match the observed equivalent
width in each quasar. However, no effort has been made to match the velocity widths
of the lines as these do not affect the reddening estimates derived from this SED-fitting.
The H$\alpha$ line-fits are shown in Fig. \ref{fig:lines} in the velocity
range $\pm$10000 km/s. Where double Gaussian fits were possible as
described in Section \ref{sec:bh}, the fits show these double
Gaussians. For the remaining objects, the fits correspond to a single
broad Gaussian. The wavelength dependent noise spectrum is shown as
the dashed line in the upper panel while the lower panel shows the fit
residuals. The rest-frame full-width-half-maxima for both the single
Gaussian and double Gaussian fits, are given in Table \ref{tab:fwhm}.

\begin{table*}
\begin{center}
\caption{Rest-frame FWHM from both single and double Gaussian fits to
the H$\alpha$ line as well as the rest-frame H$\alpha$ equivalent width (EW) for each of our reddened quasars. The double gaussian fit is only considered if the
individual components in this Gaussian have FWHM between 1000 and
10000 km/s.}
\label{tab:fwhm}
\begin{tabular}{lcccc}
Name & FWHM$_{\rm{single}}$/km s$^{-1}$ & FWHM$_{\rm{double 1}}$/ km s$^{-1}$ & FWHM$_{\rm{double 2}}$ / km s$^{-1}$ & Rest-frame H$\alpha$ EW / \AA \\
\hline 
\hline
VHSJ2024-5623 & 7000 & 1600 & 8700 & $-350$ \\
VHSJ2028-4631 & 4300 & -- & -- & $-200$ \\
VHSJ2028-5740 & 6100 & -- & -- & $-290$ \\
VHSJ2048-4644 & 4000 & -- & -- & $-250$ \\
VHSJ2100-5820 & 3100 & -- & -- & $-200$ \\
VHSJ2101-5943 & 5800 & 3300 & 9600 & $-360$ \\
VHSJ2115-5913 & 3000 & -- & -- & $-170$ \\
VHSJ2130-4930 & 4400 & -- & -- & $-130$ \\
VHSJ2141-4816 & 3900 & 1100 & 4600 & $-210$ \\
VHSJ2212-4624 & 5400 & 3100 & 7900 & $-490$ \\
VHSJ2220-5618 & 3800 & 1600 & 5200 & $-140$ \\
VHSJ2227-5203 & 7300 & -- & -- & $-190$ \\
VHSJ2235-5750 & 4900 & 2100 & 9000 & $-260$ \\
VHSJ2256-4800 & 3900 & 2600 & 8800 & $-200$ \\
VHSJ2257-4700 & 4000 & 2100 & 5200 & $-420$ \\
VHSJ2306-5447 & 7900 & 1500 & 8600 & $-500$ \\
VHSJ2332-5240 & 5900 & -- & -- & $-300$ \\
VHSJ1556-0835 & 1900 & -- & -- & $-230$ \\
VHSJ2143-0643 & 3300 & 2000 & 8300 & $-250$ \\
VHSJ2144-0523 & 3200 & 2300 & 9600 & $-250$ \\
VHSJ2109-0026 & 3500 & 1900 & 6900 & $-320$ \\
VHSJ2355-0011 & 6000 & 2400 & 9000 & $-330$ \\
ULASJ0123+1525 & 3500 & 1000 & 4100 & $-380$ \\
ULASJ2315+0143 & 4700 & 3000 & 7000 & $-270$ \\
\hline
\end{tabular}
\end{center}
\end{table*}

\begin{figure*}
\begin{center}
\centering
\begin{tabular}{cc}
\includegraphics[scale=0.35,angle=0]{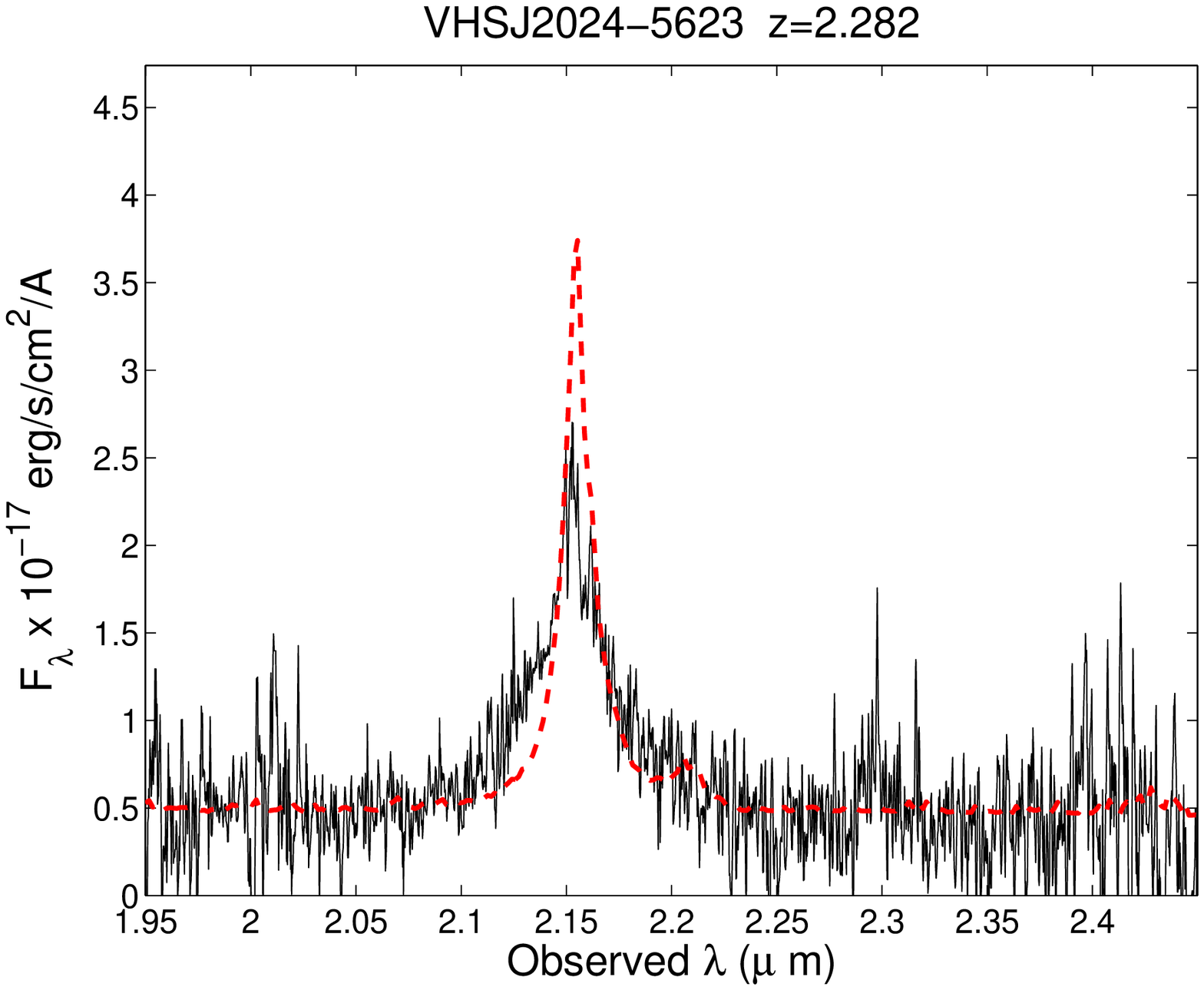} & \includegraphics[scale=0.35,angle=0]{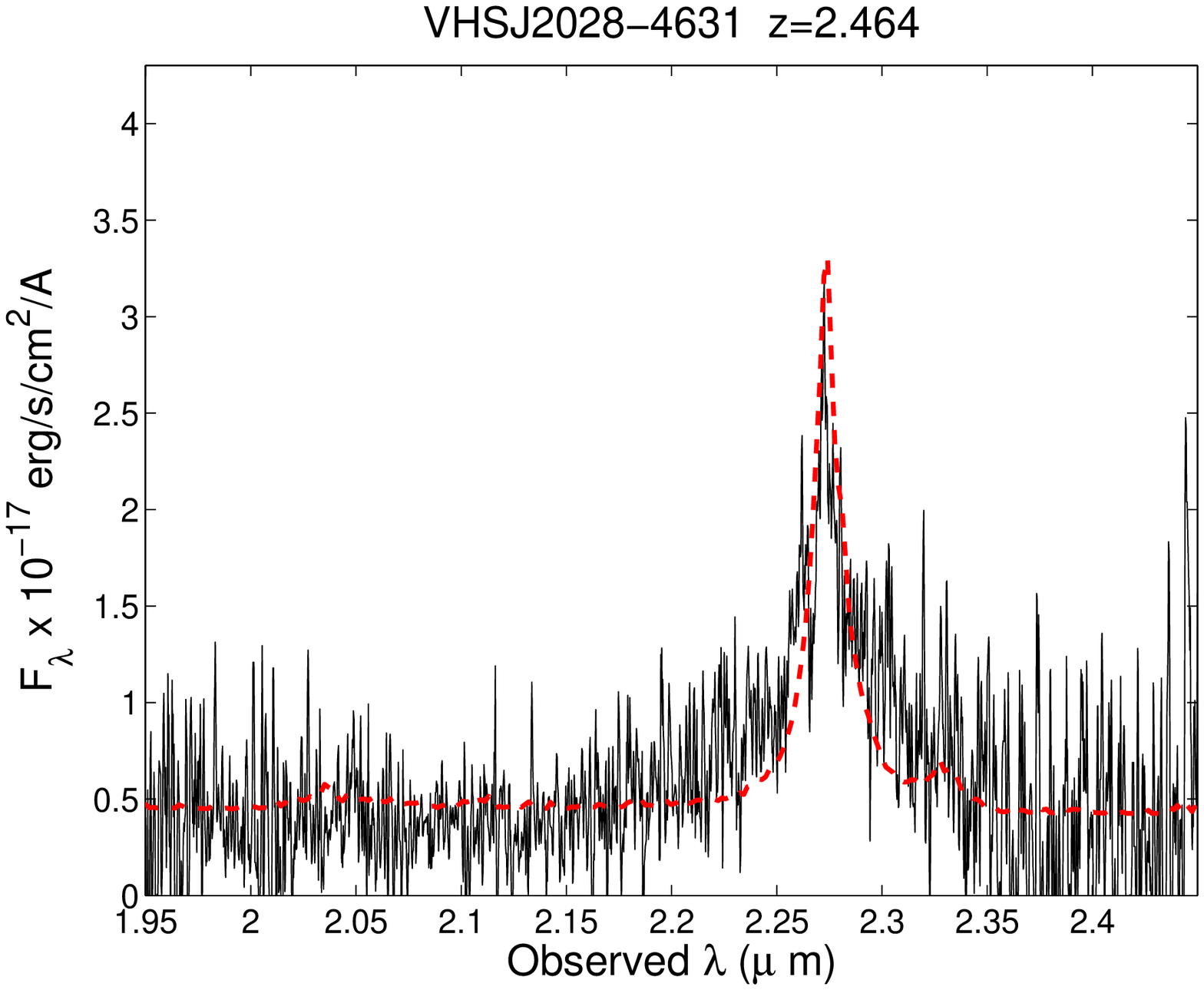} \\
\includegraphics[scale=0.35,angle=0]{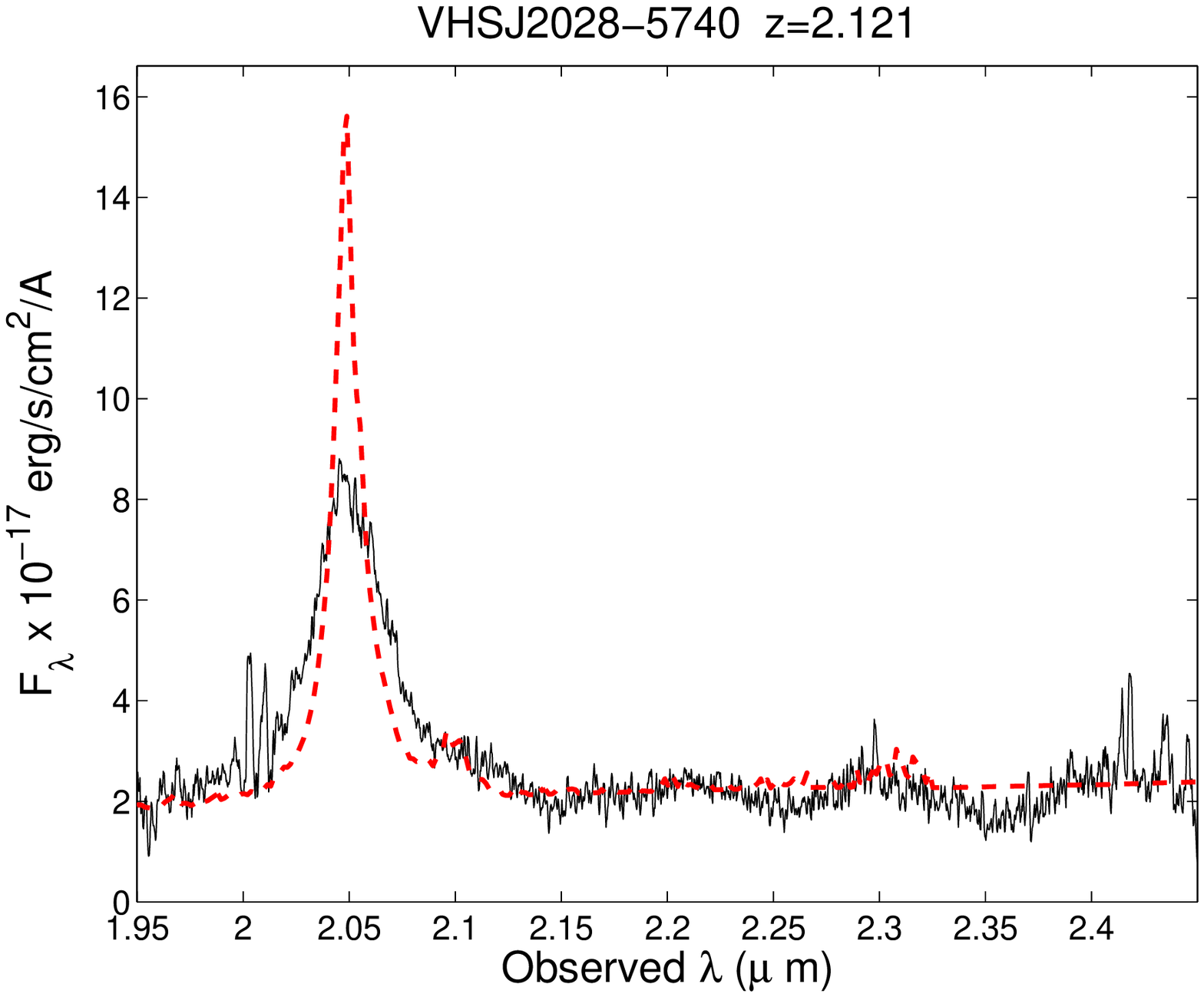} & \includegraphics[scale=0.35,angle=0]{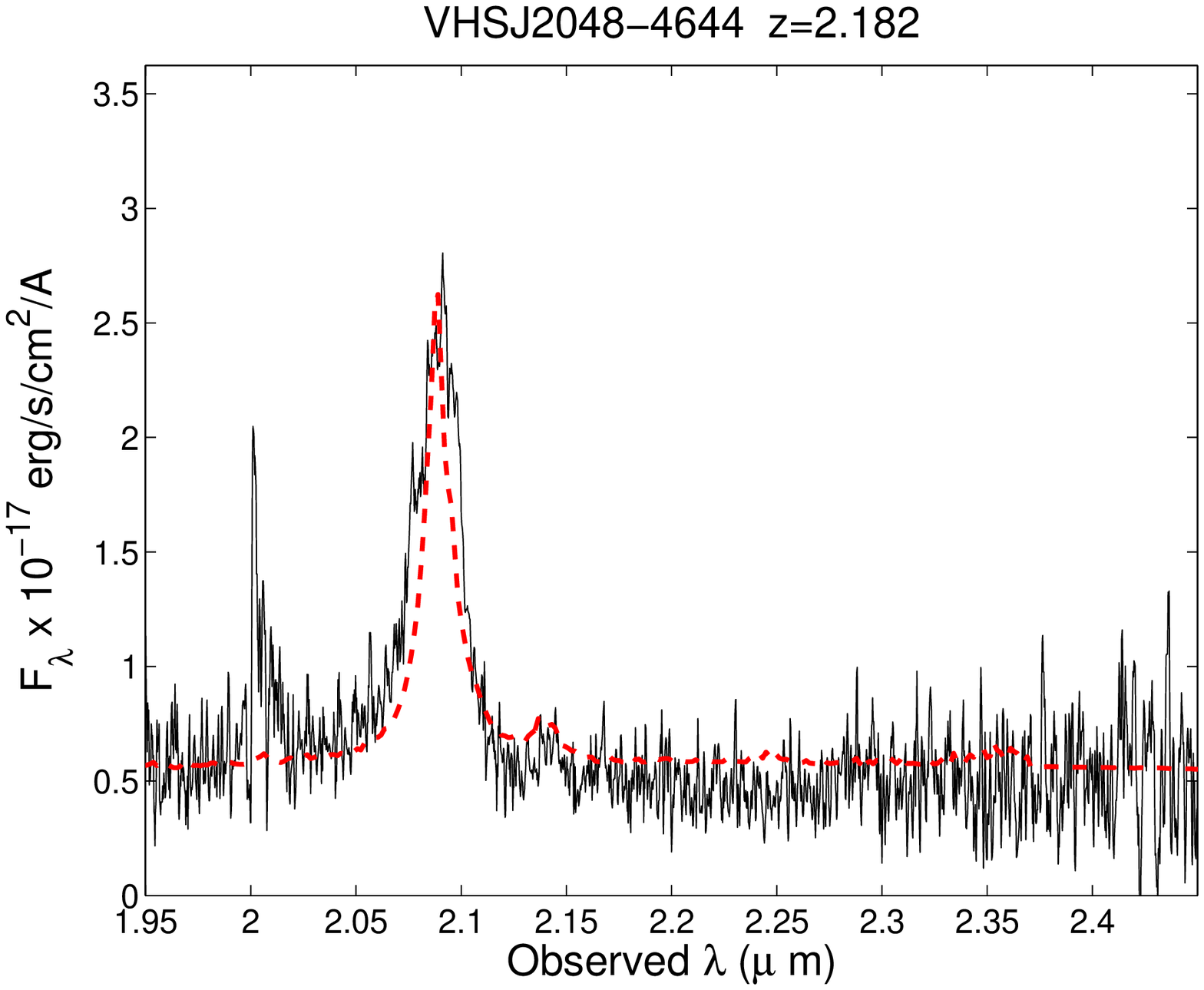} \\
\includegraphics[scale=0.35,angle=0]{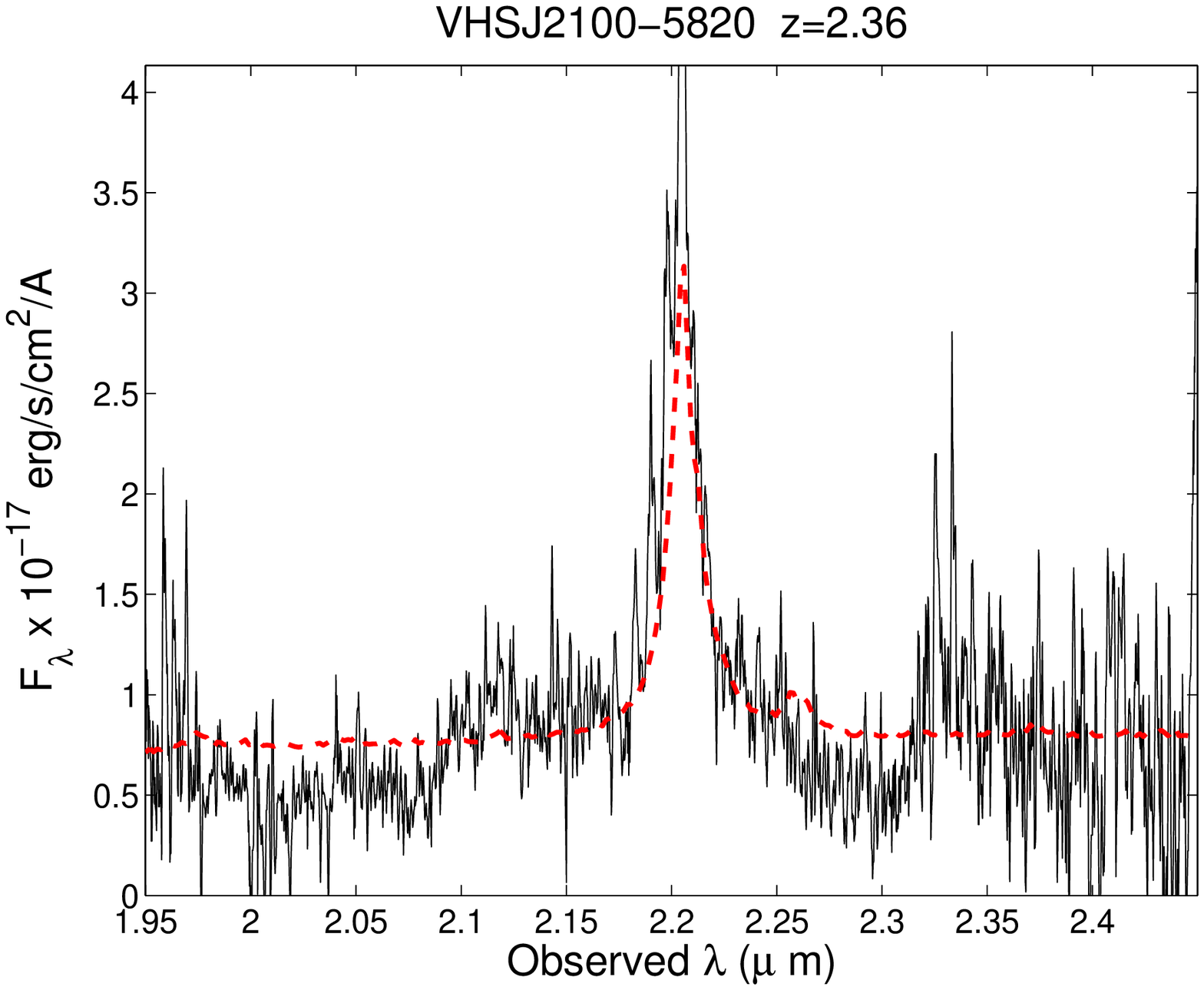} & \includegraphics[scale=0.35,angle=0]{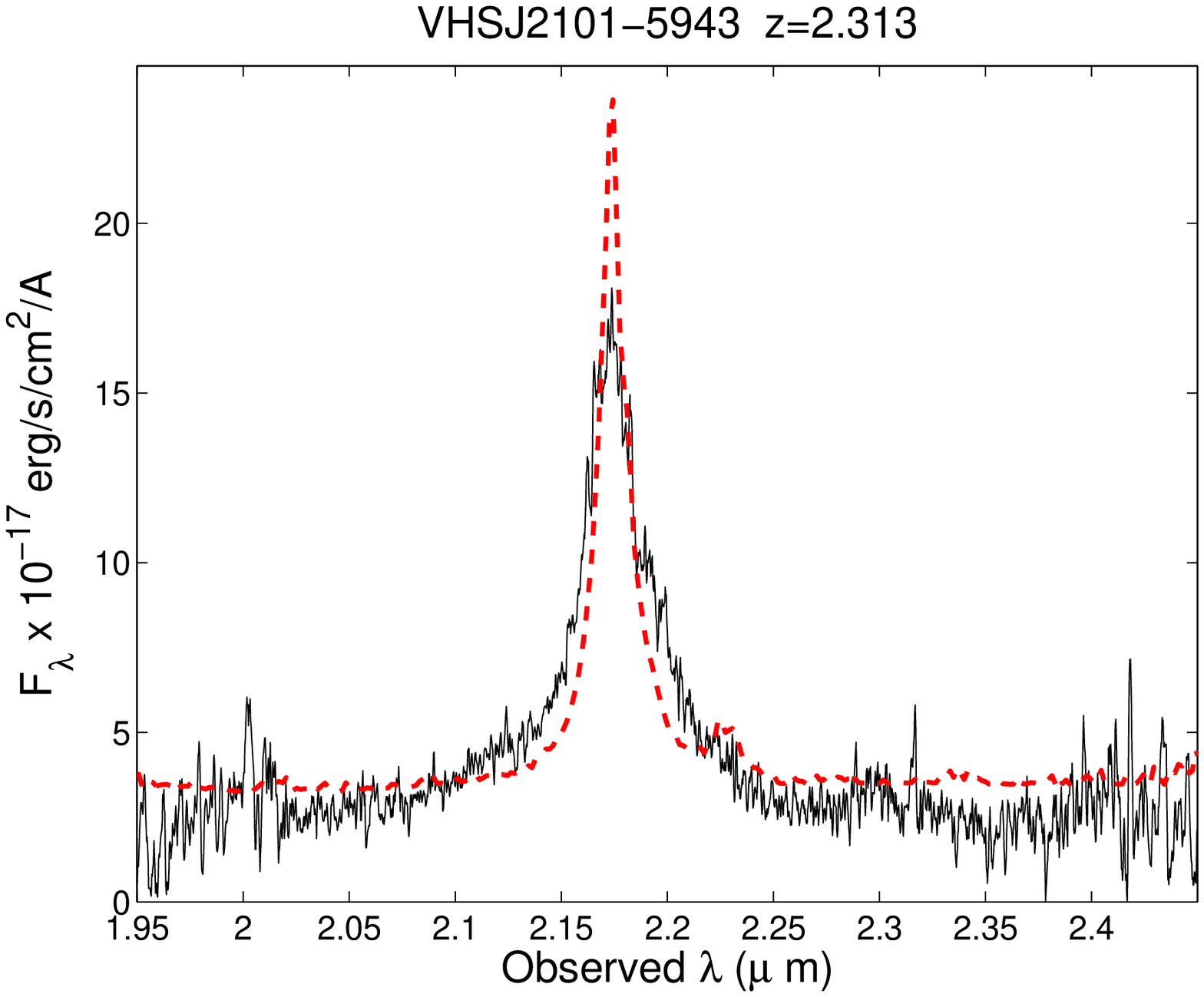} \\
\includegraphics[scale=0.35,angle=0]{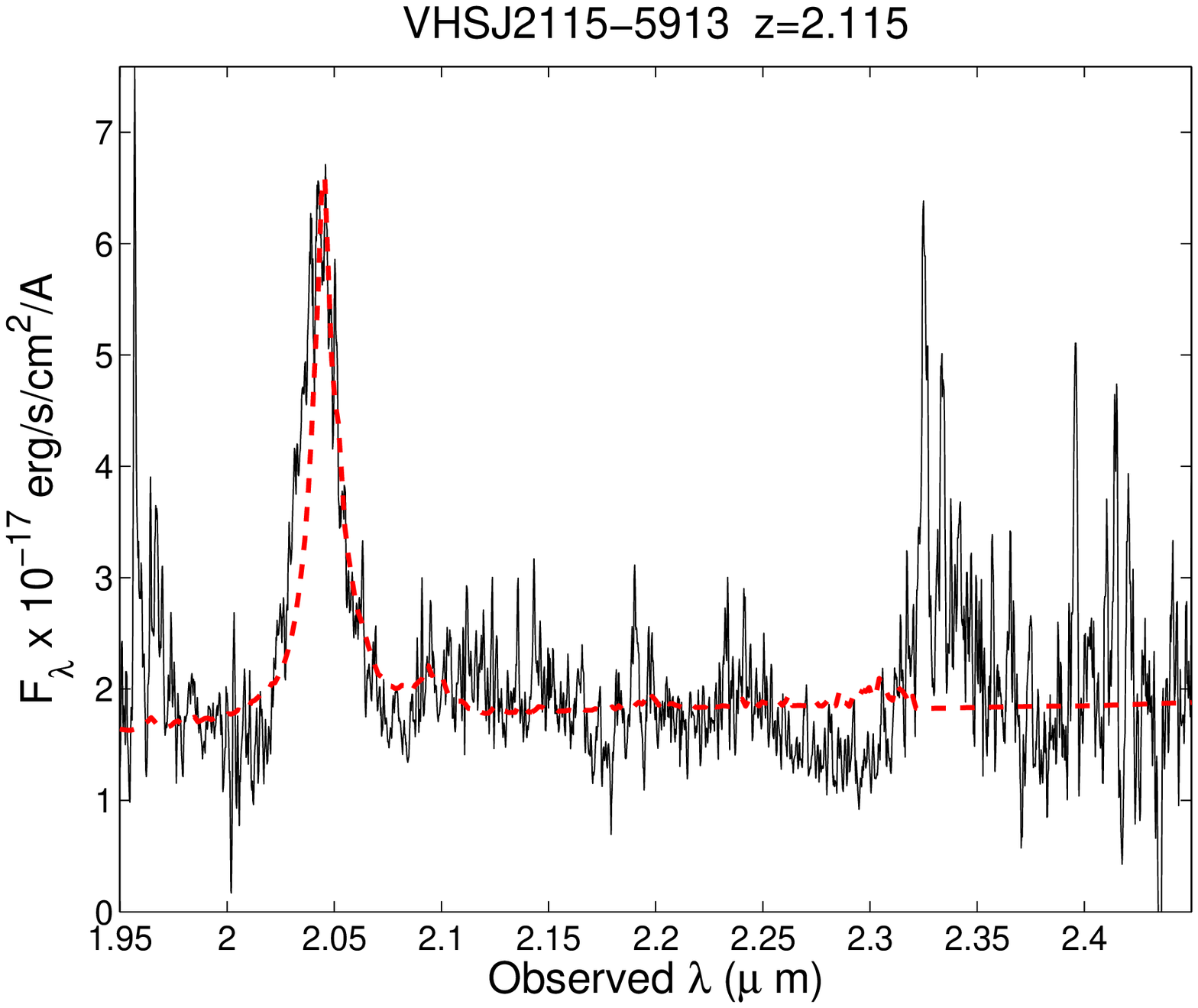} & \includegraphics[scale=0.35,angle=0]{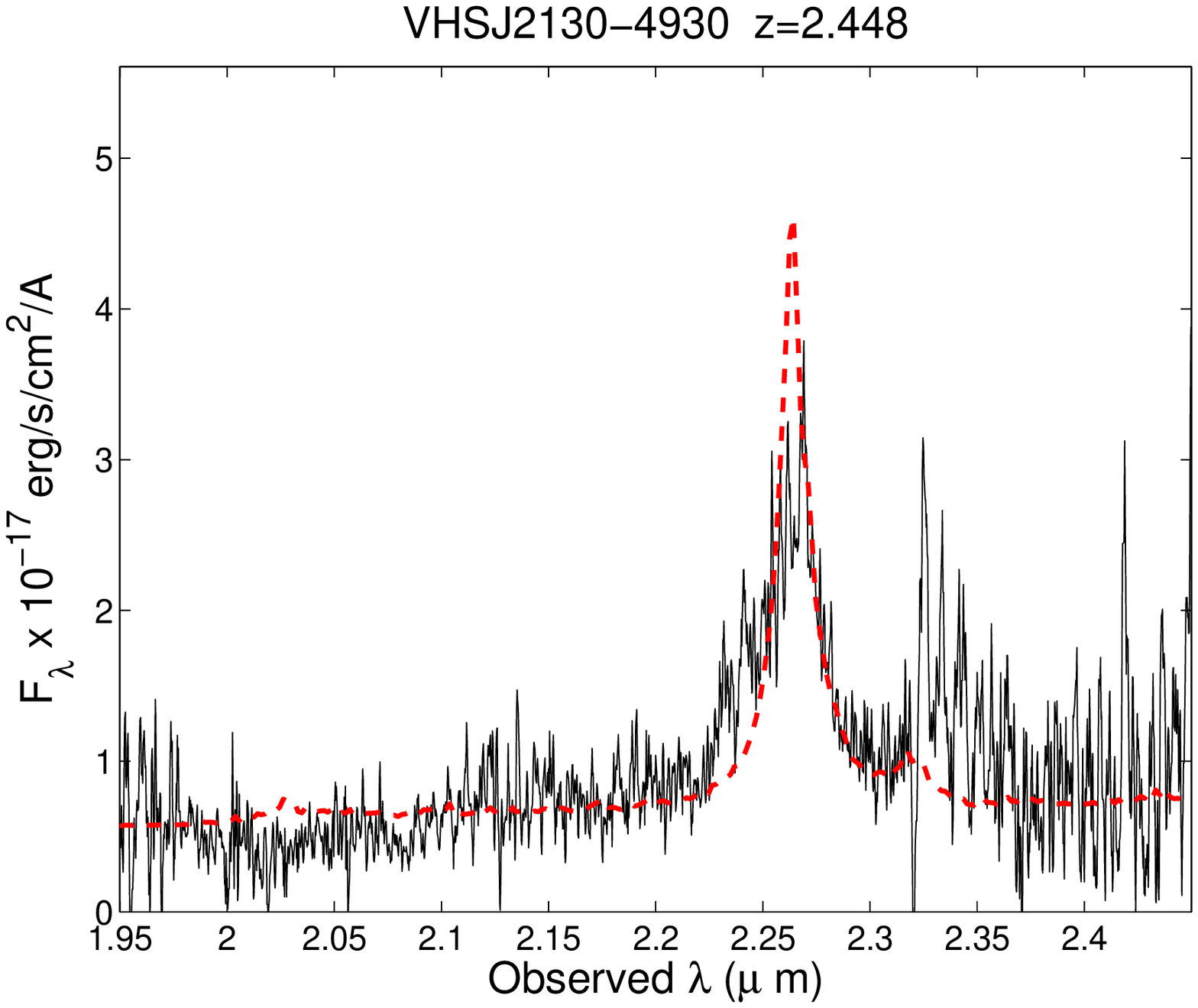} \\
\end{tabular}
\caption{VLT-SINFONI $K$-band spectra of spectroscopically confirmed reddened quasars (solid line) together with our best-fit SED model (dashed line). Note that the equivalent width of the H$\alpha$ line in the SED model has been adjusted to match the observed equivalent width of each quasar. However, all the SED models assume the same velocity width for the H$\alpha$ line and no effort has been made to match the velocity widths as these do not affect the reddening estimates derived from the SED-fitting.}
\label{fig:spectra}
\end{center}
\end{figure*}

\clearpage
\setcounter{figure}{1}
\begin{figure*}
\contcaption{}
\begin{center}
\centering
\begin{tabular}{cc}
\includegraphics[scale=0.35,angle=0]{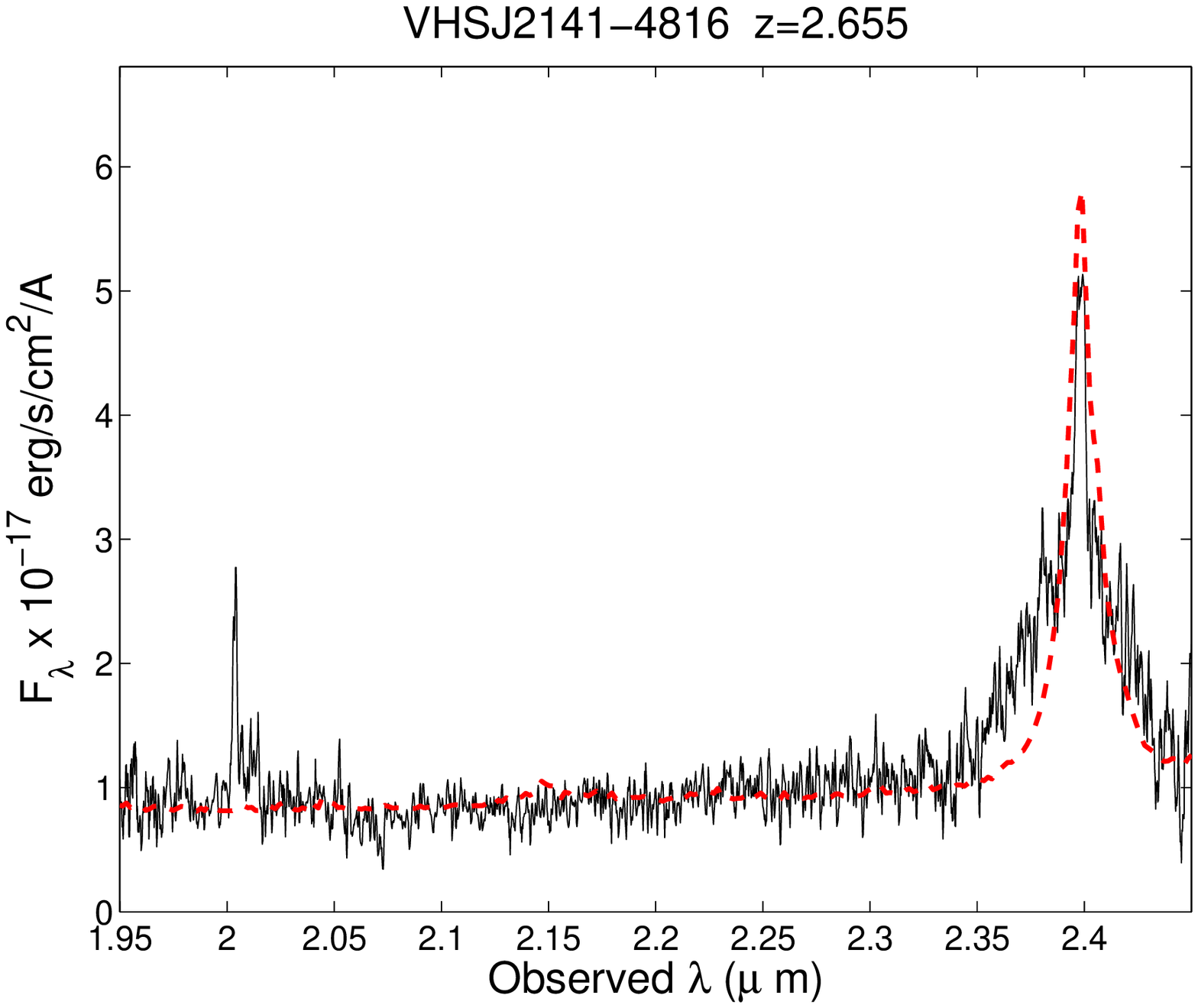} & \includegraphics[scale=0.35,angle=0]{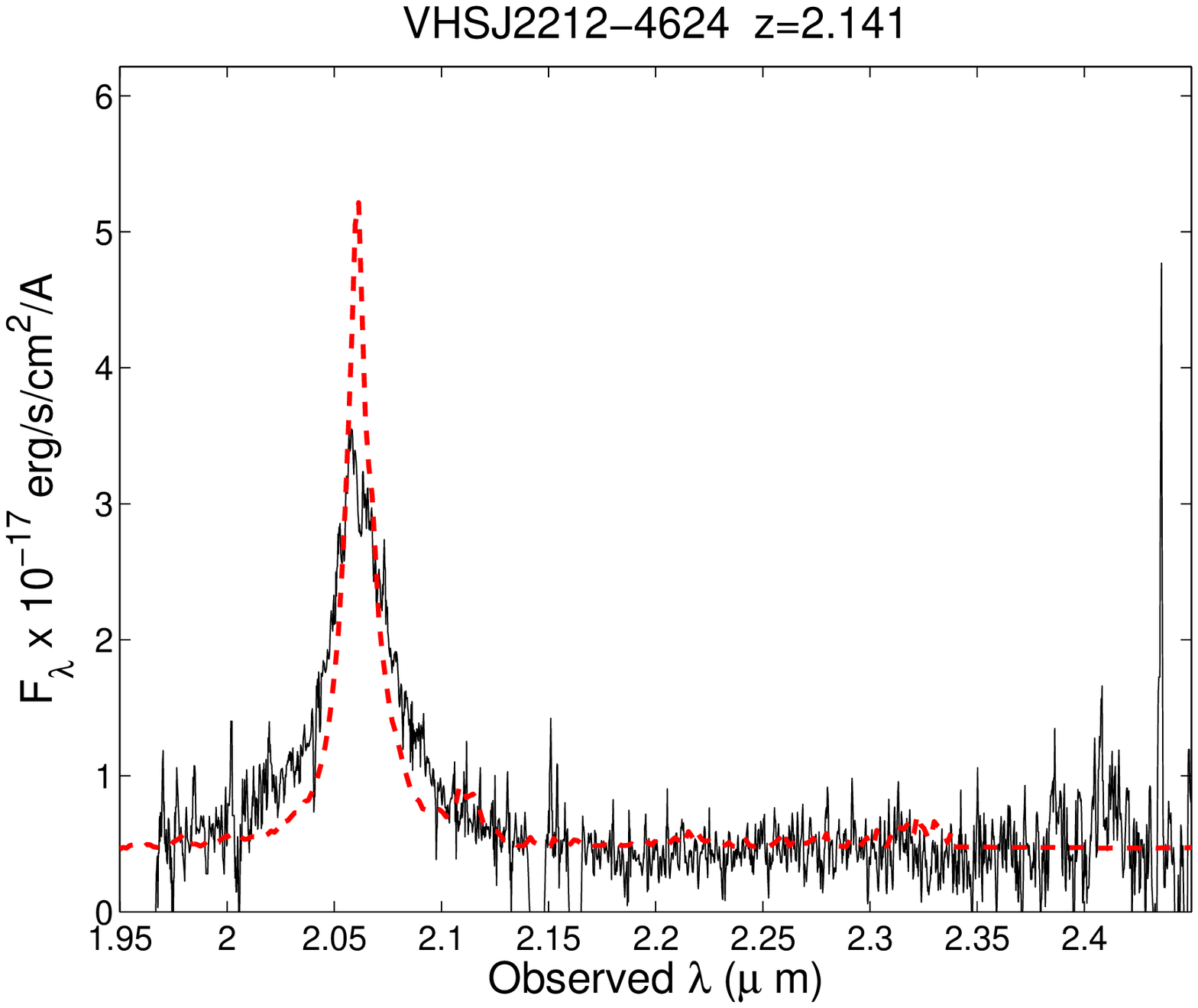} \\
\includegraphics[scale=0.35,angle=0]{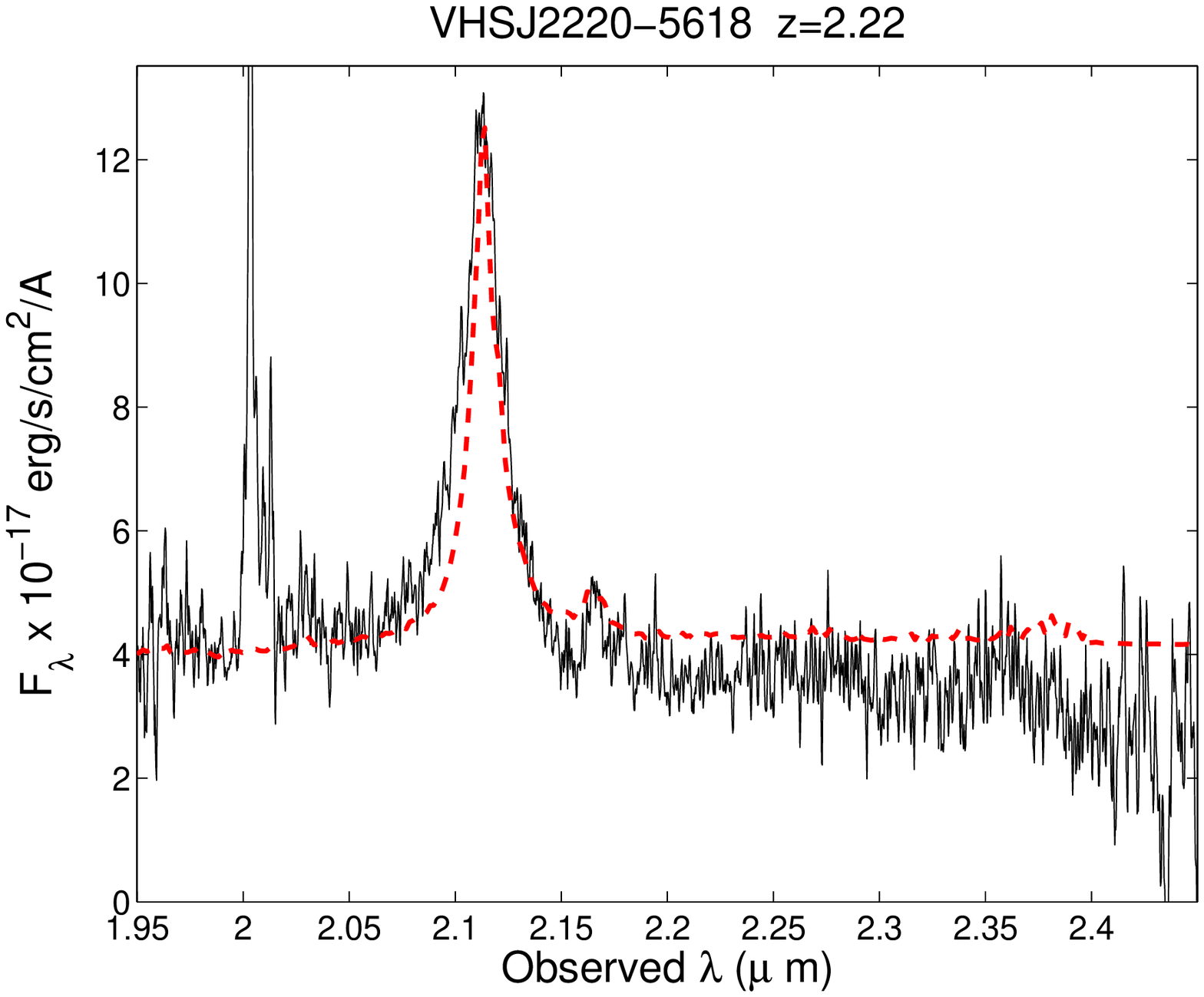} & \includegraphics[scale=0.35,angle=0]{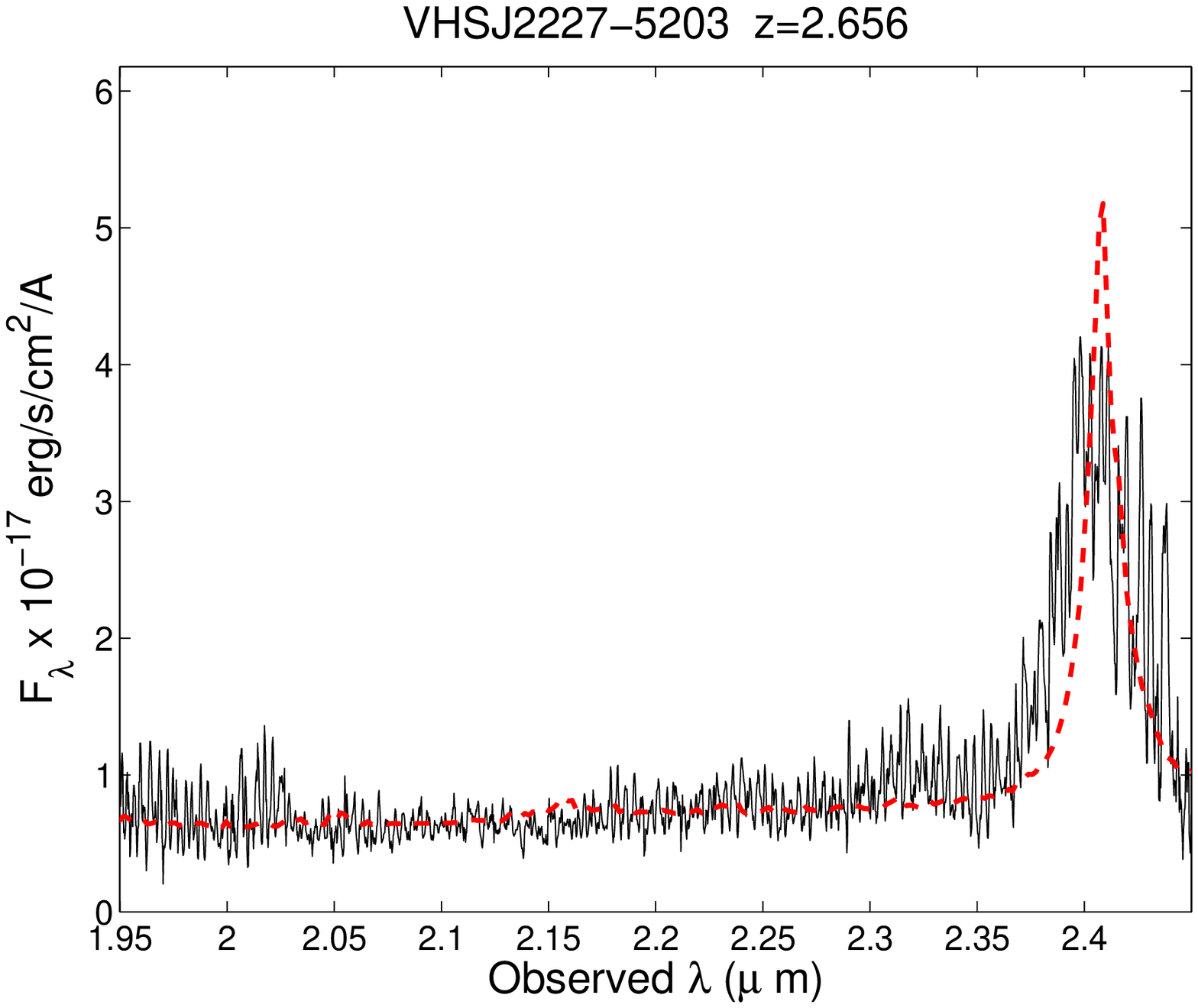} \\
\includegraphics[scale=0.35,angle=0]{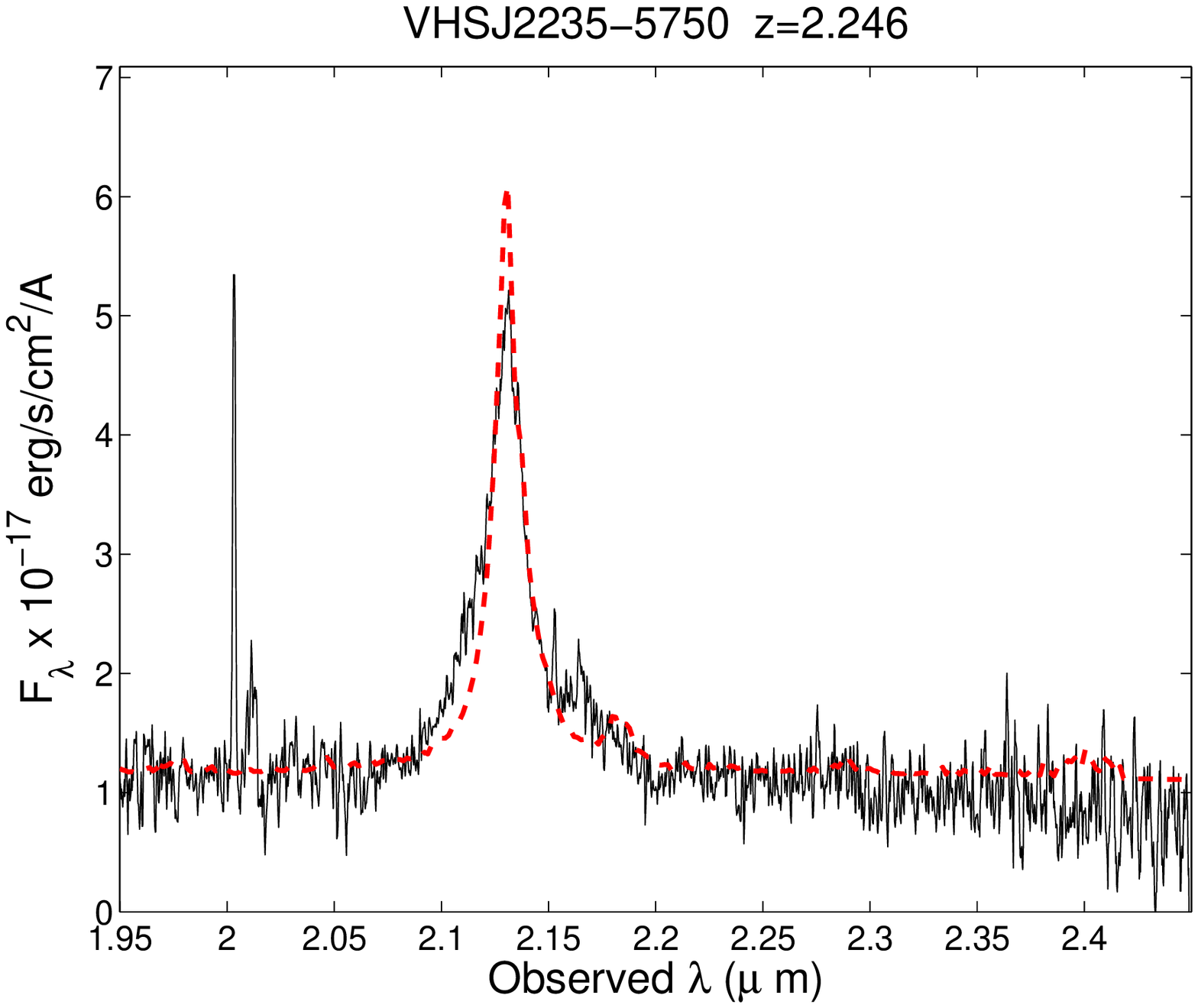} & \includegraphics[scale=0.35,angle=0]{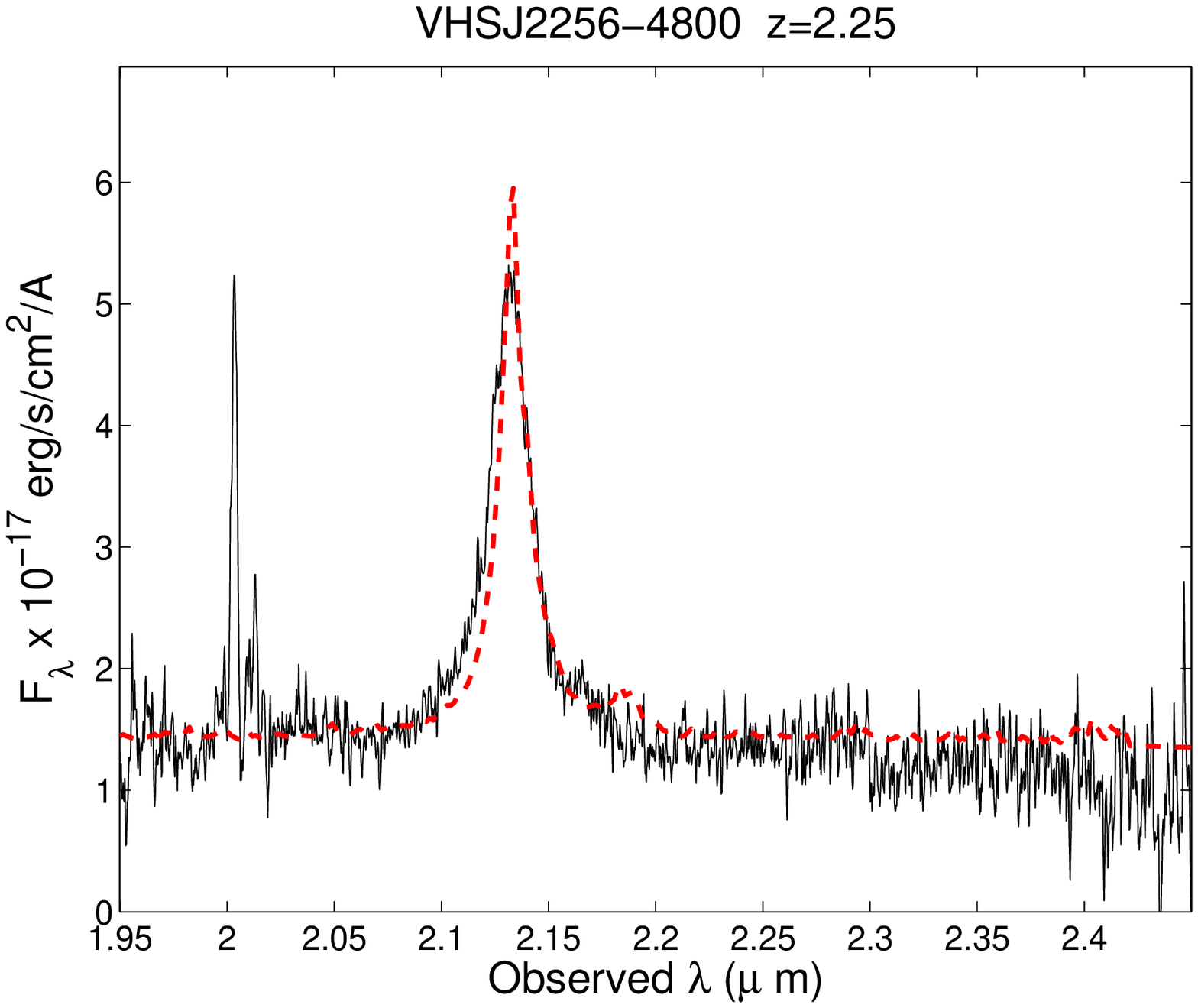} \\
\includegraphics[scale=0.35,angle=0]{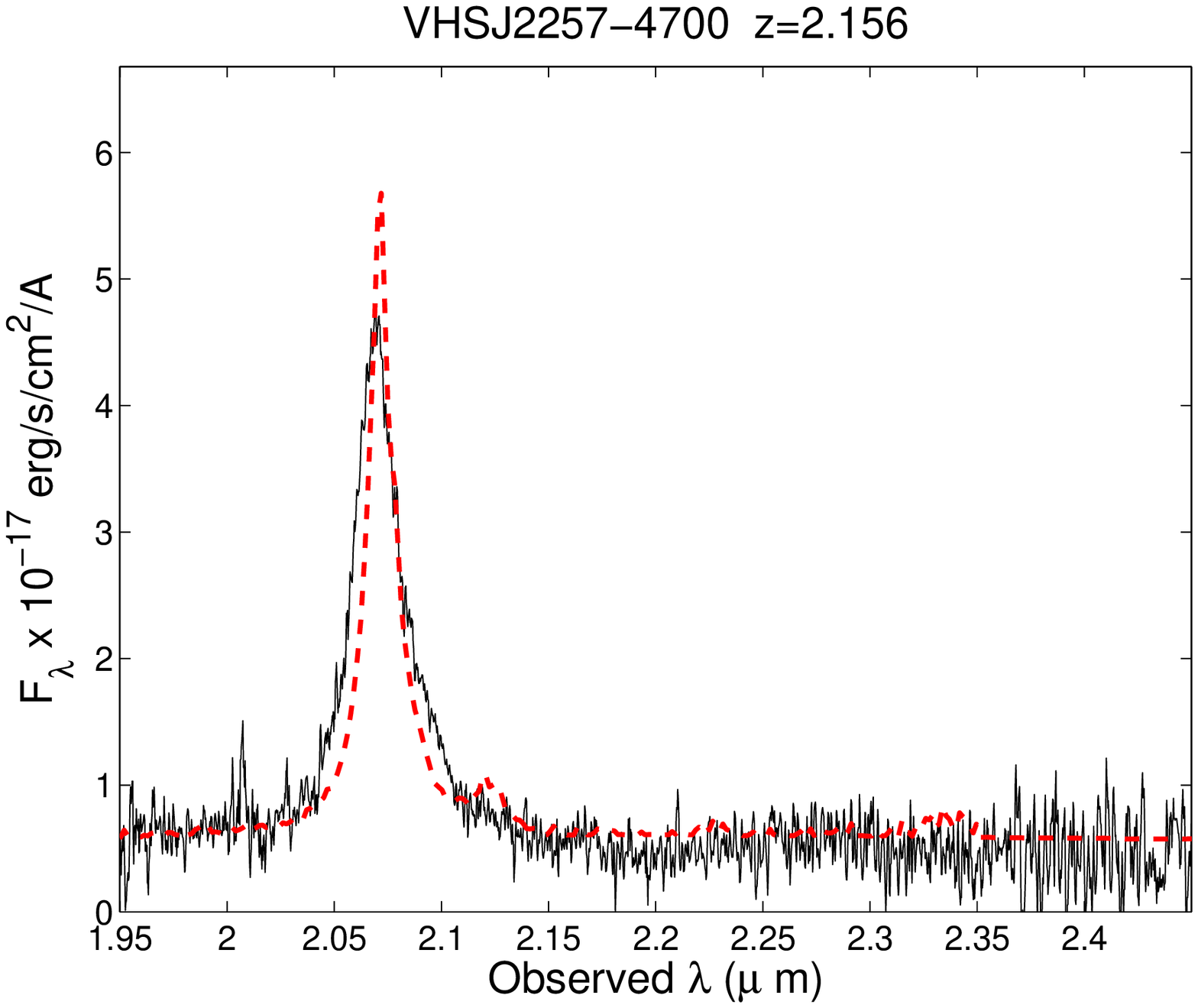} & \includegraphics[scale=0.35,angle=0]{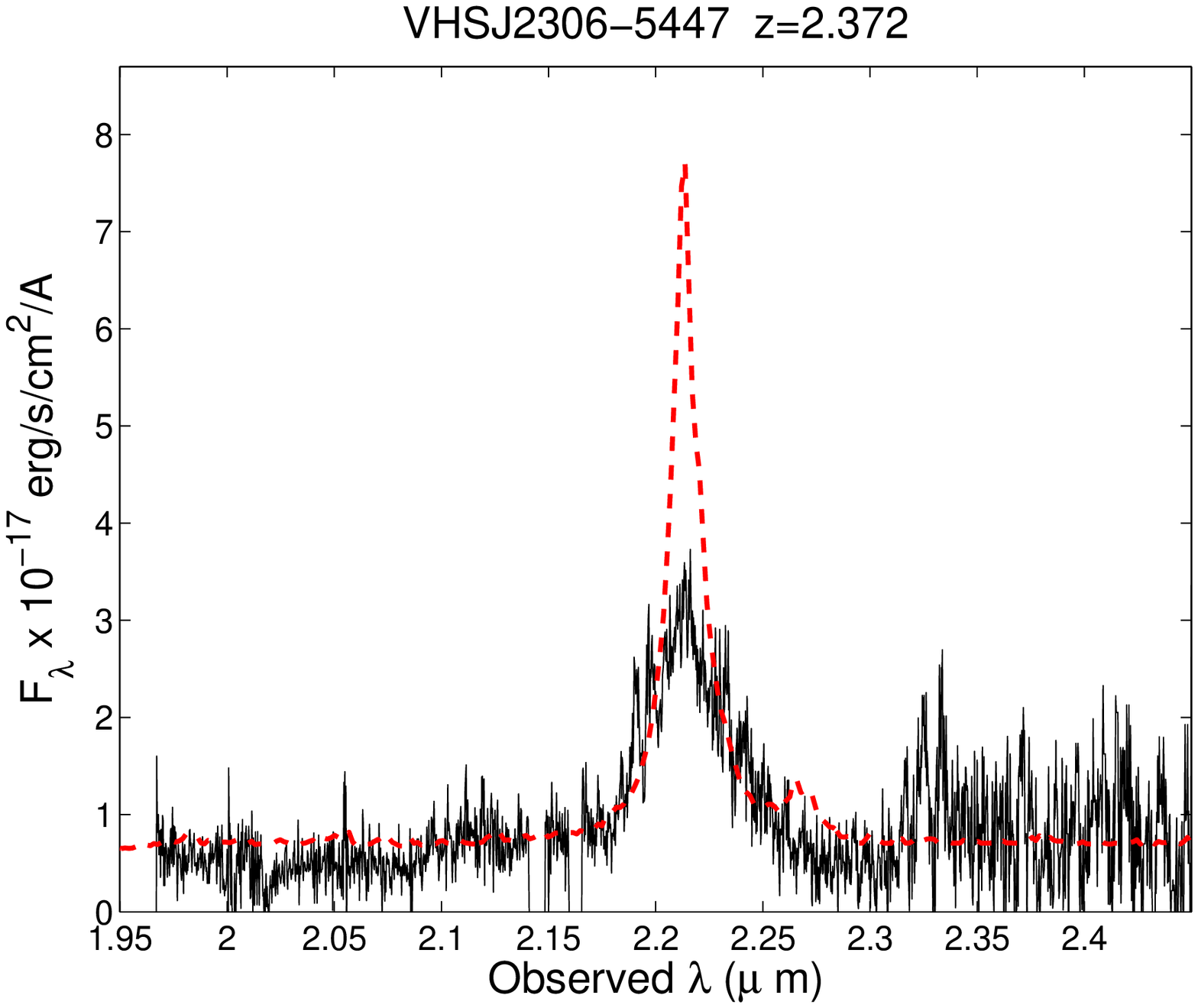} \\
\end{tabular}
\end{center}
\end{figure*}

\clearpage
\setcounter{figure}{1}
\begin{figure*}
\contcaption{}
\begin{center}
\centering
\begin{tabular}{cc}
\includegraphics[scale=0.35,angle=0]{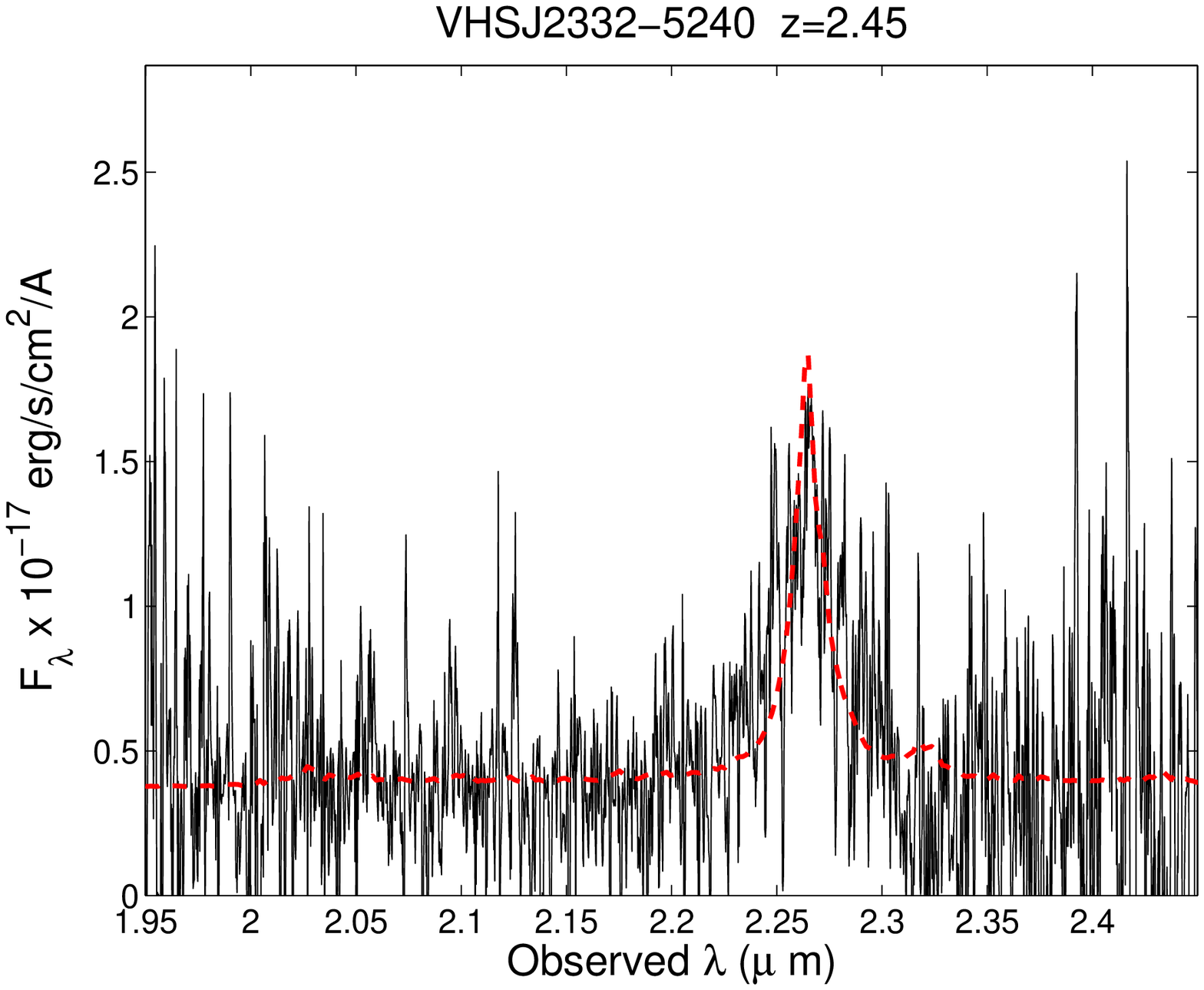}  & \includegraphics[scale=0.35,angle=0]{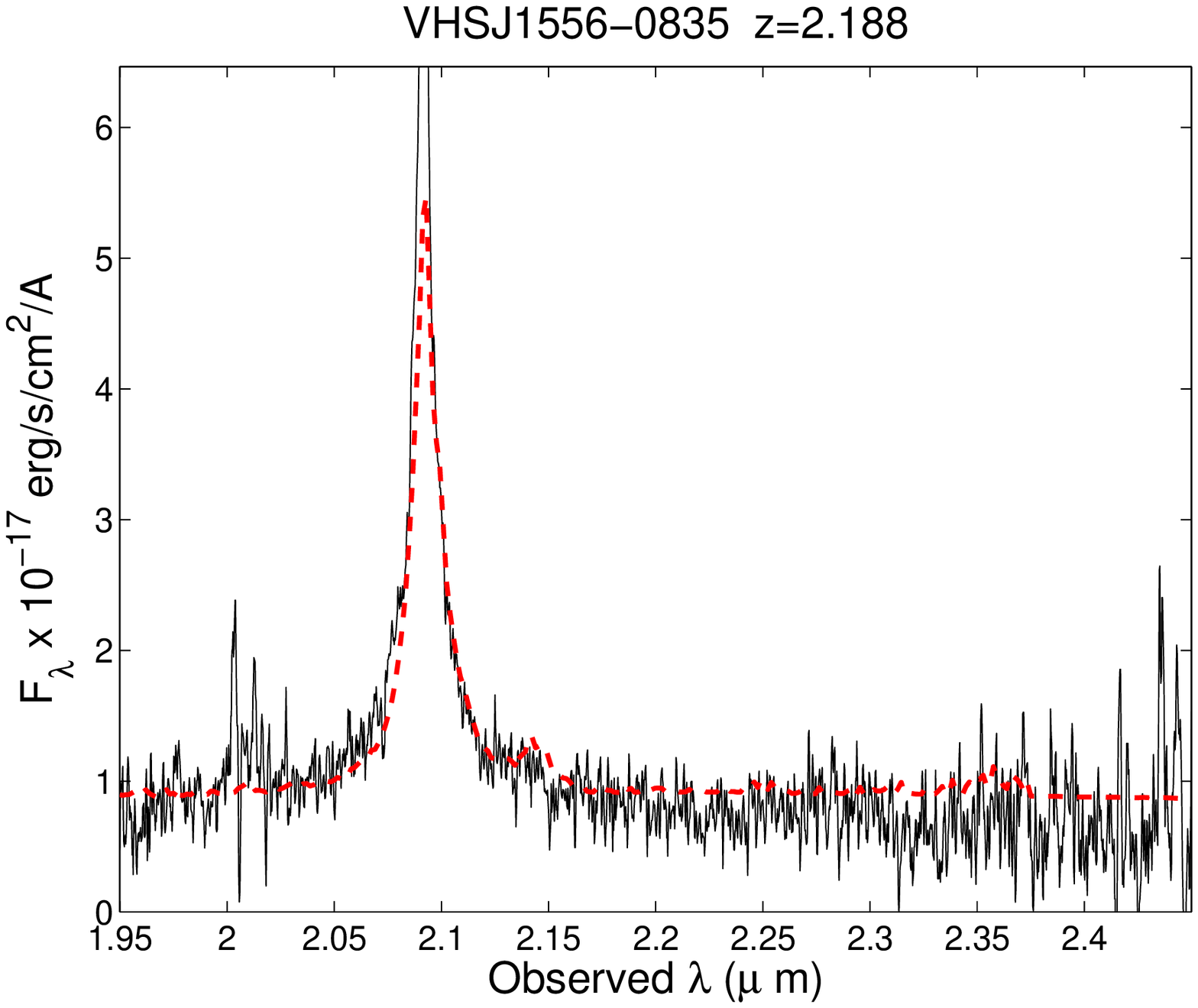} \\
\includegraphics[scale=0.35,angle=0]{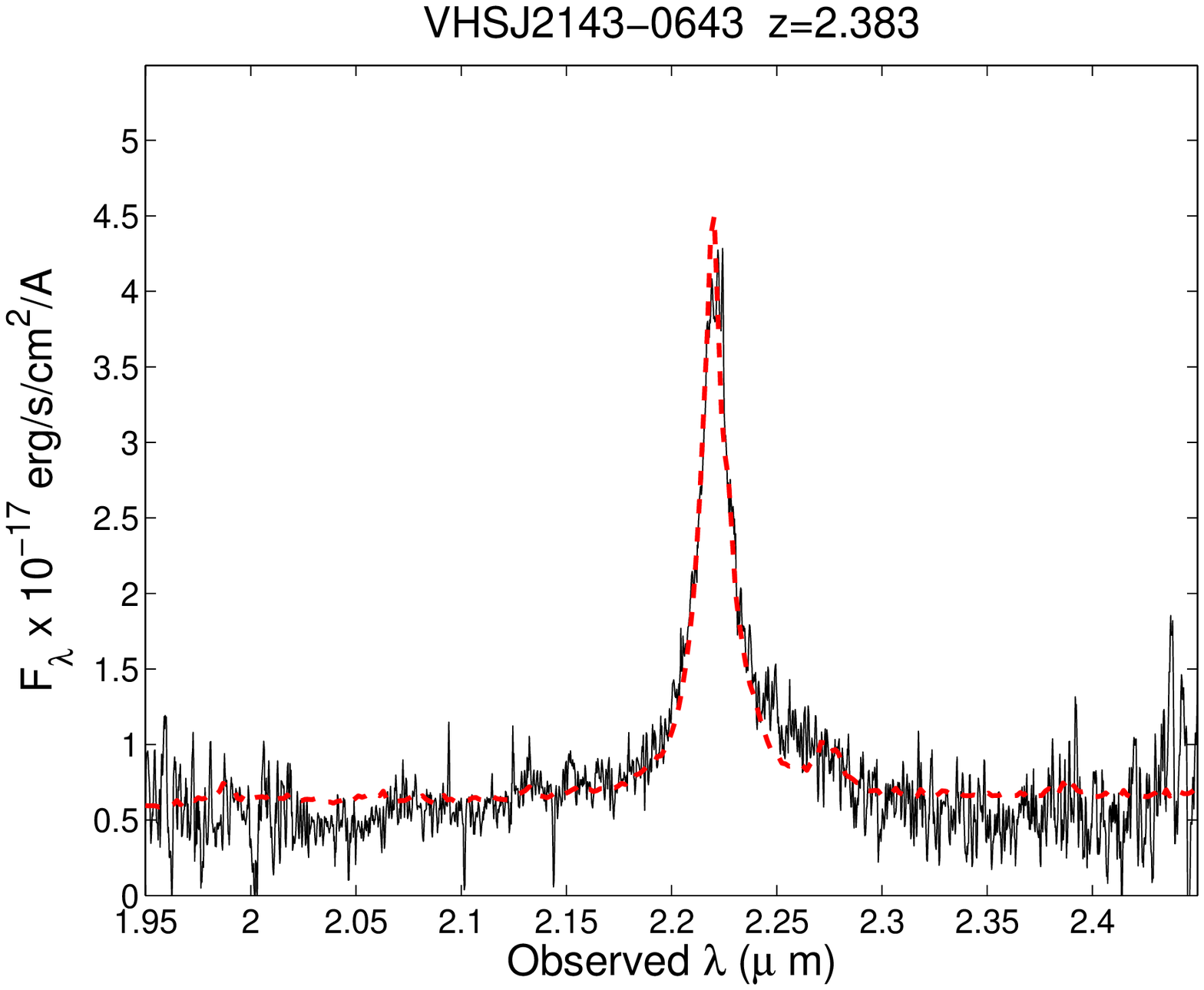} & \includegraphics[scale=0.35,angle=0]{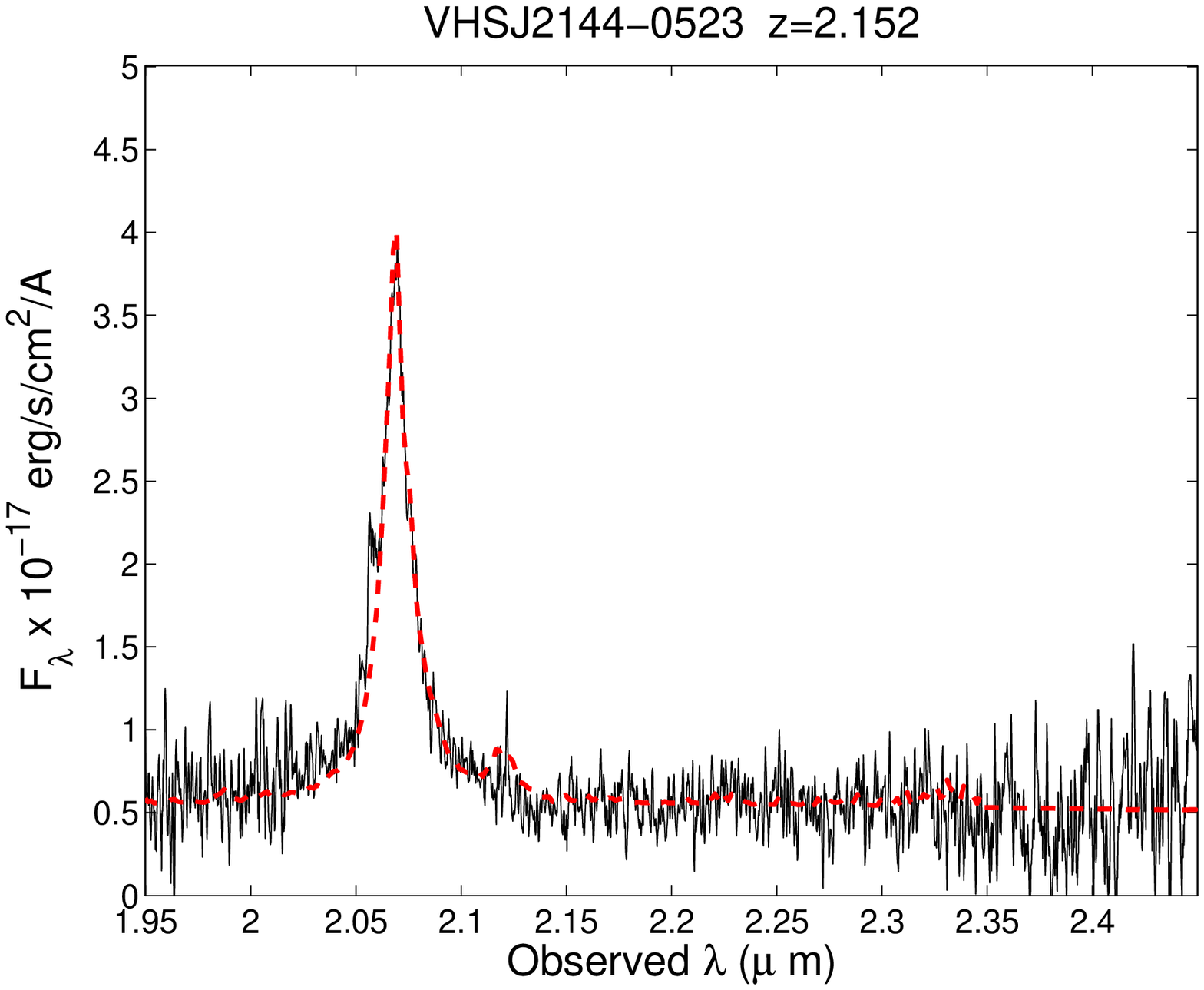} \\
\includegraphics[scale=0.35,angle=0]{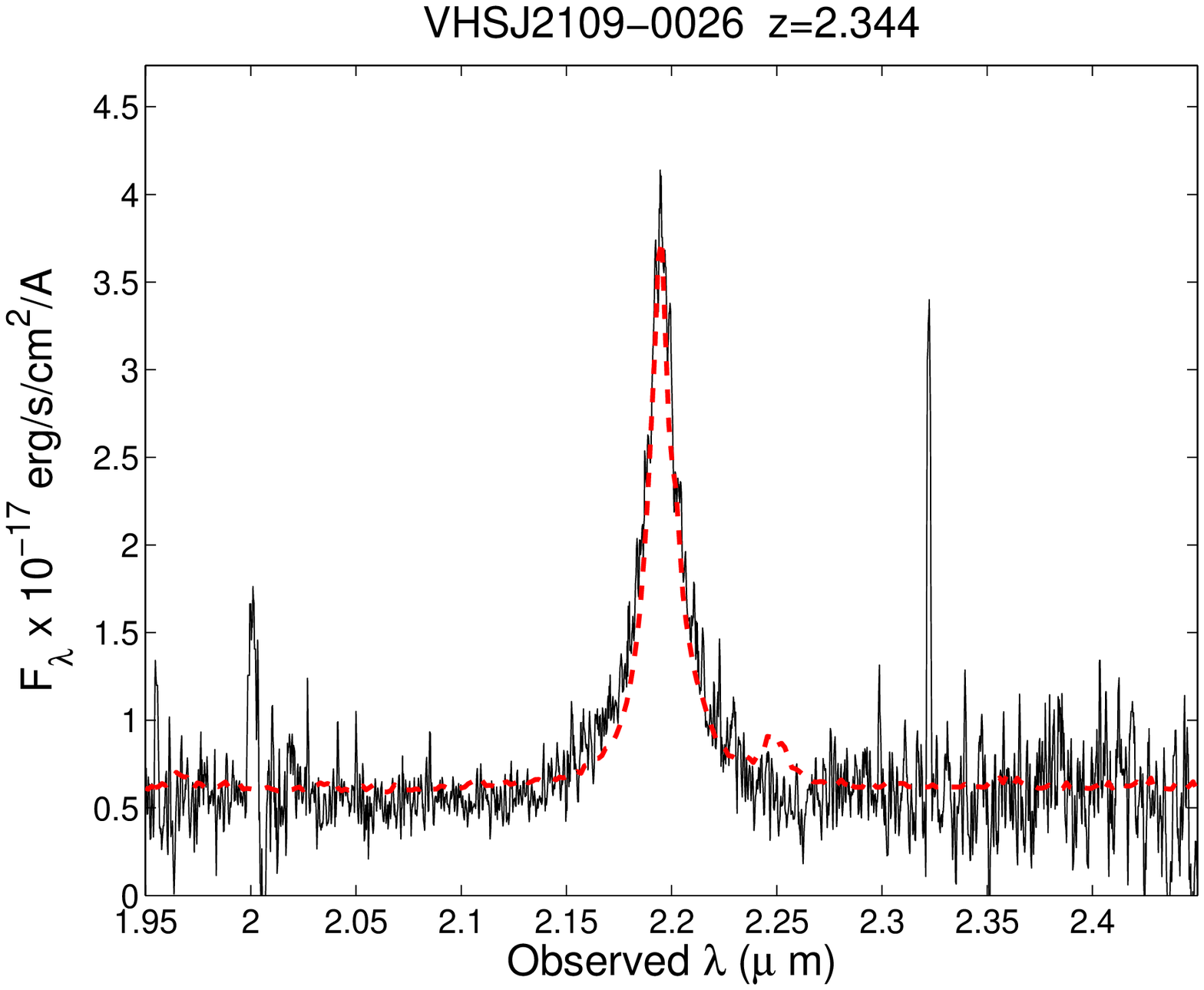} & \includegraphics[scale=0.35,angle=0]{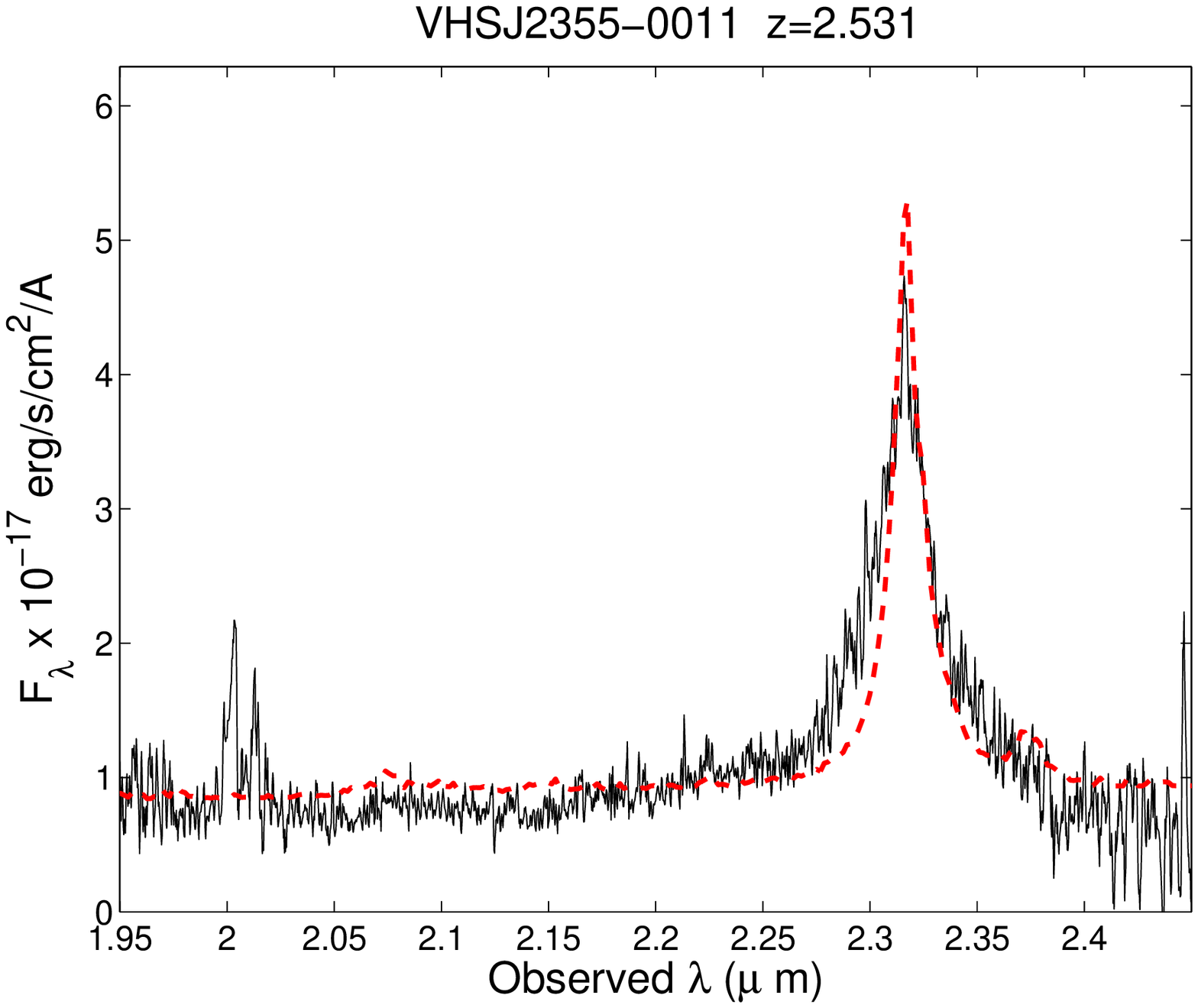} \\
\includegraphics[scale=0.35,angle=0]{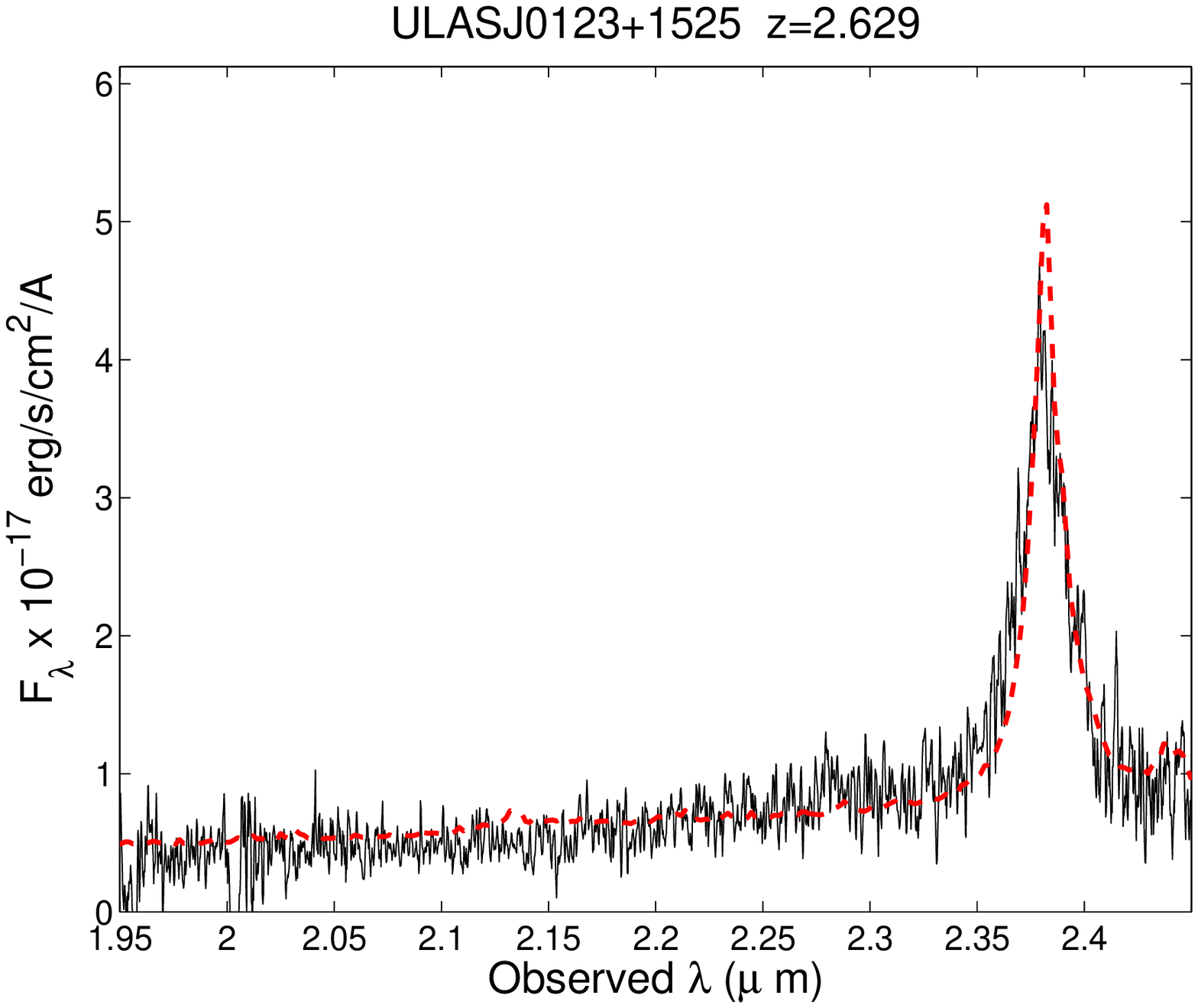} & \includegraphics[scale=0.35,angle=0]{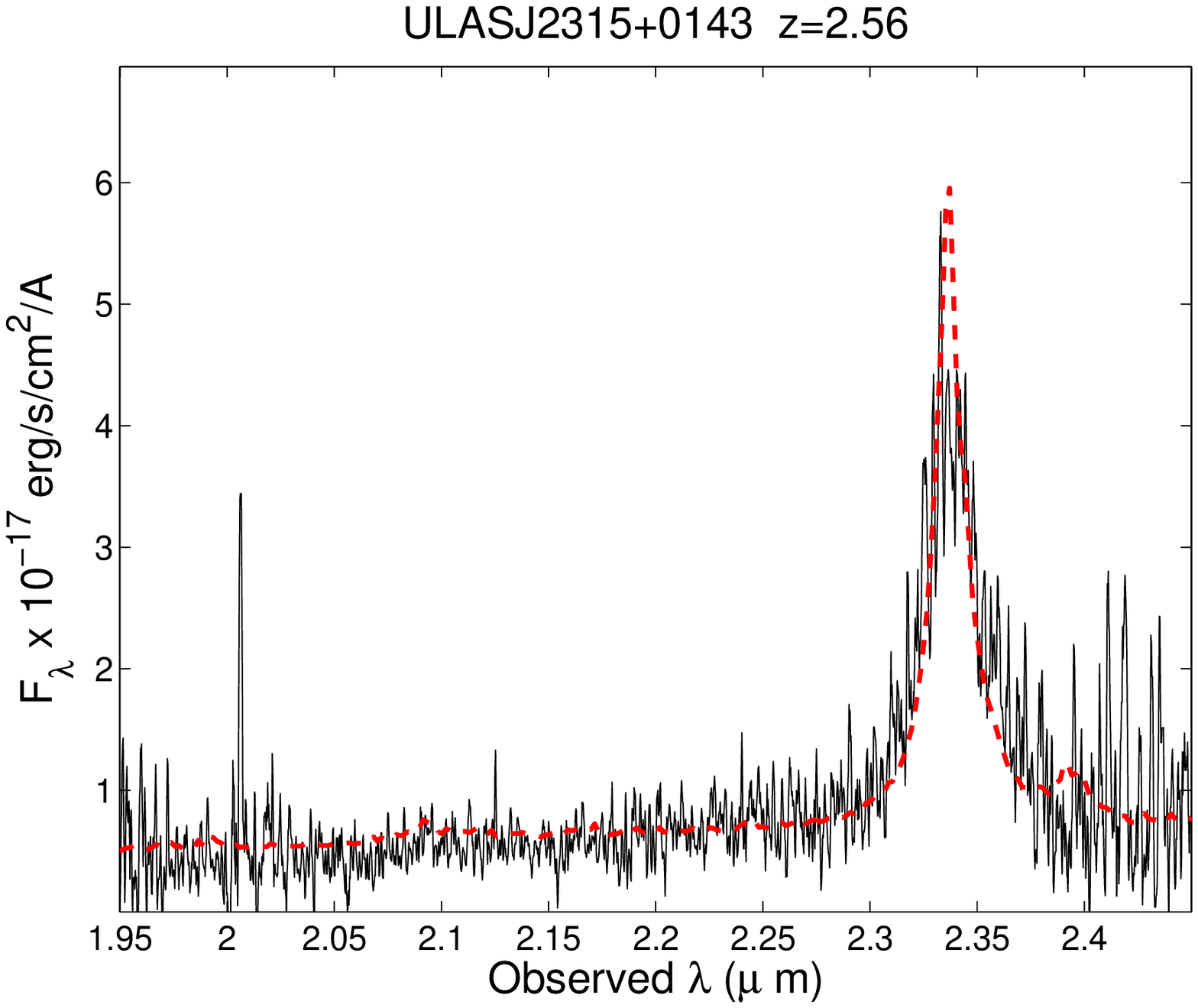} \\
\end{tabular}
\end{center}
\end{figure*}

\begin{figure*}
\begin{center}
\centering
\begin{tabular}{cc}
\includegraphics[scale=0.35,angle=0]{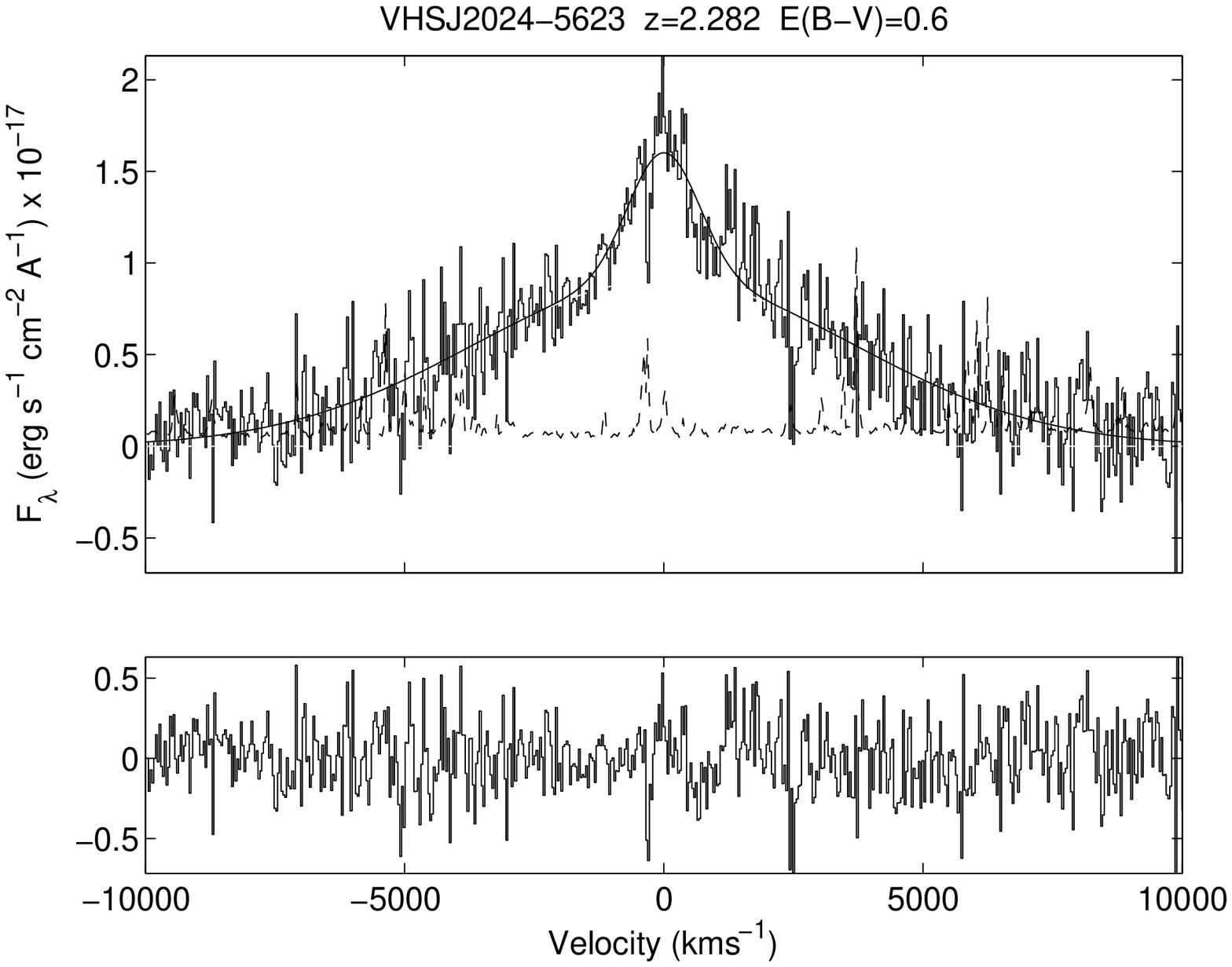} & \includegraphics[scale=0.35,angle=0]{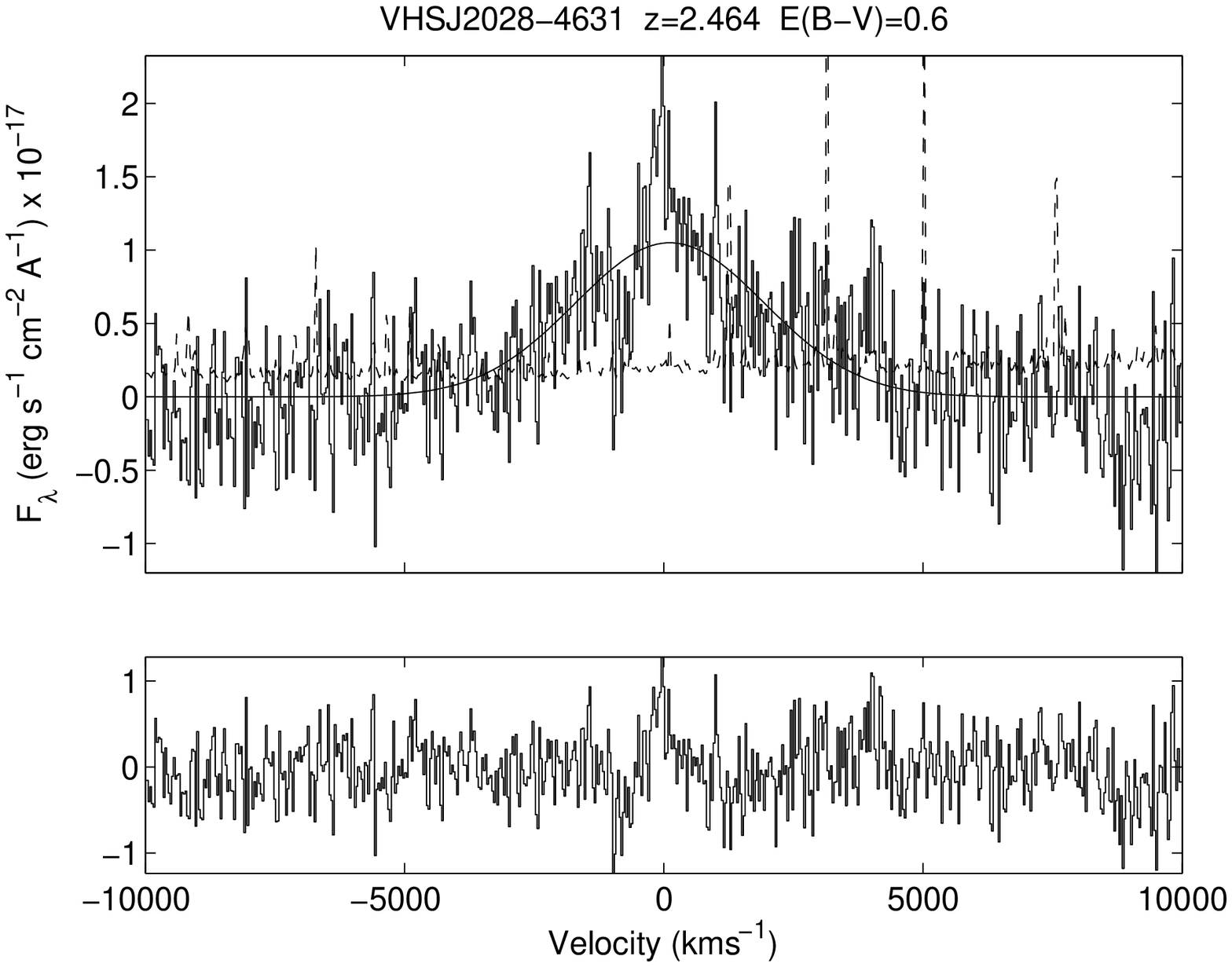} \\
\includegraphics[scale=0.35,angle=0]{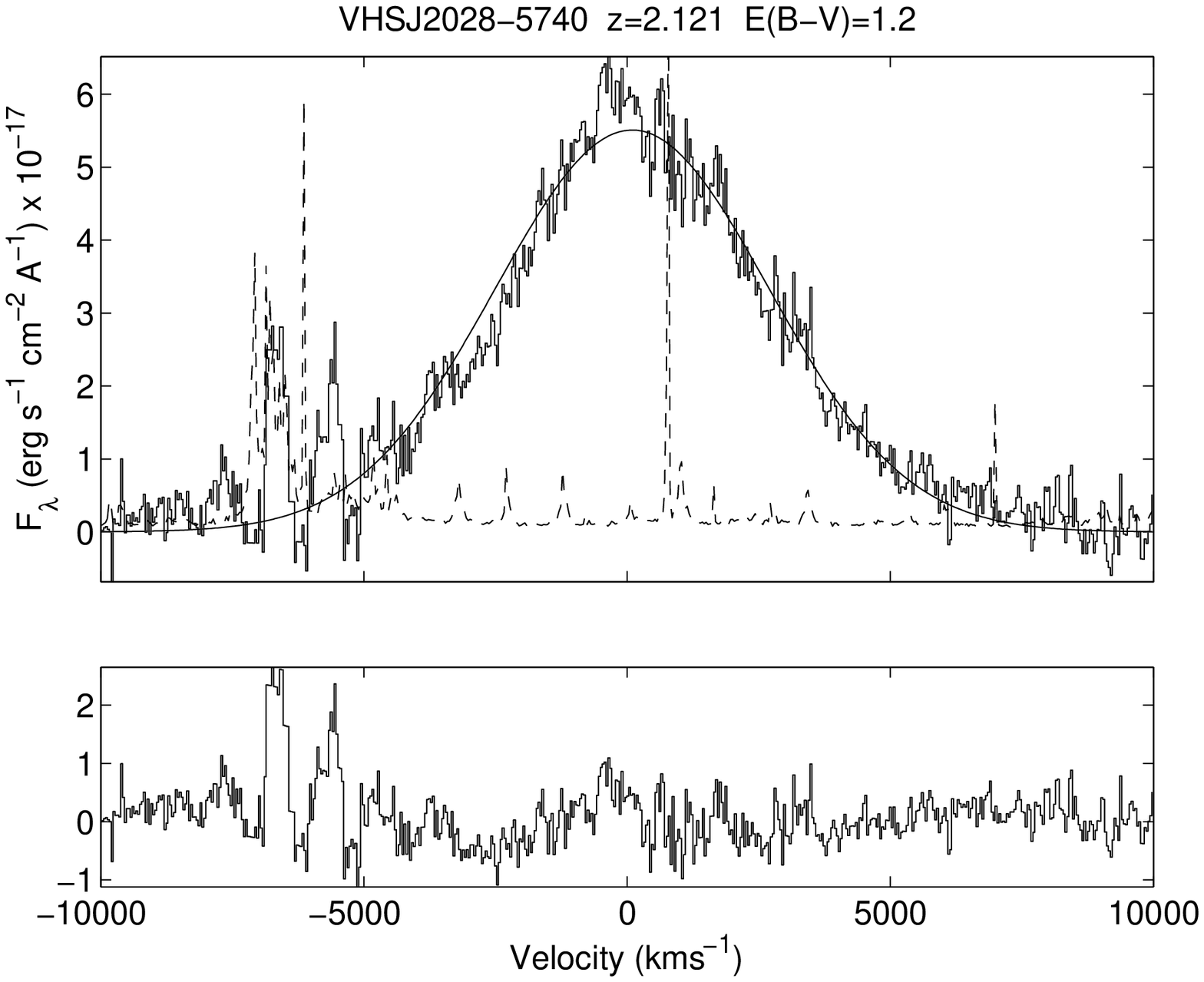} & \includegraphics[scale=0.35,angle=0]{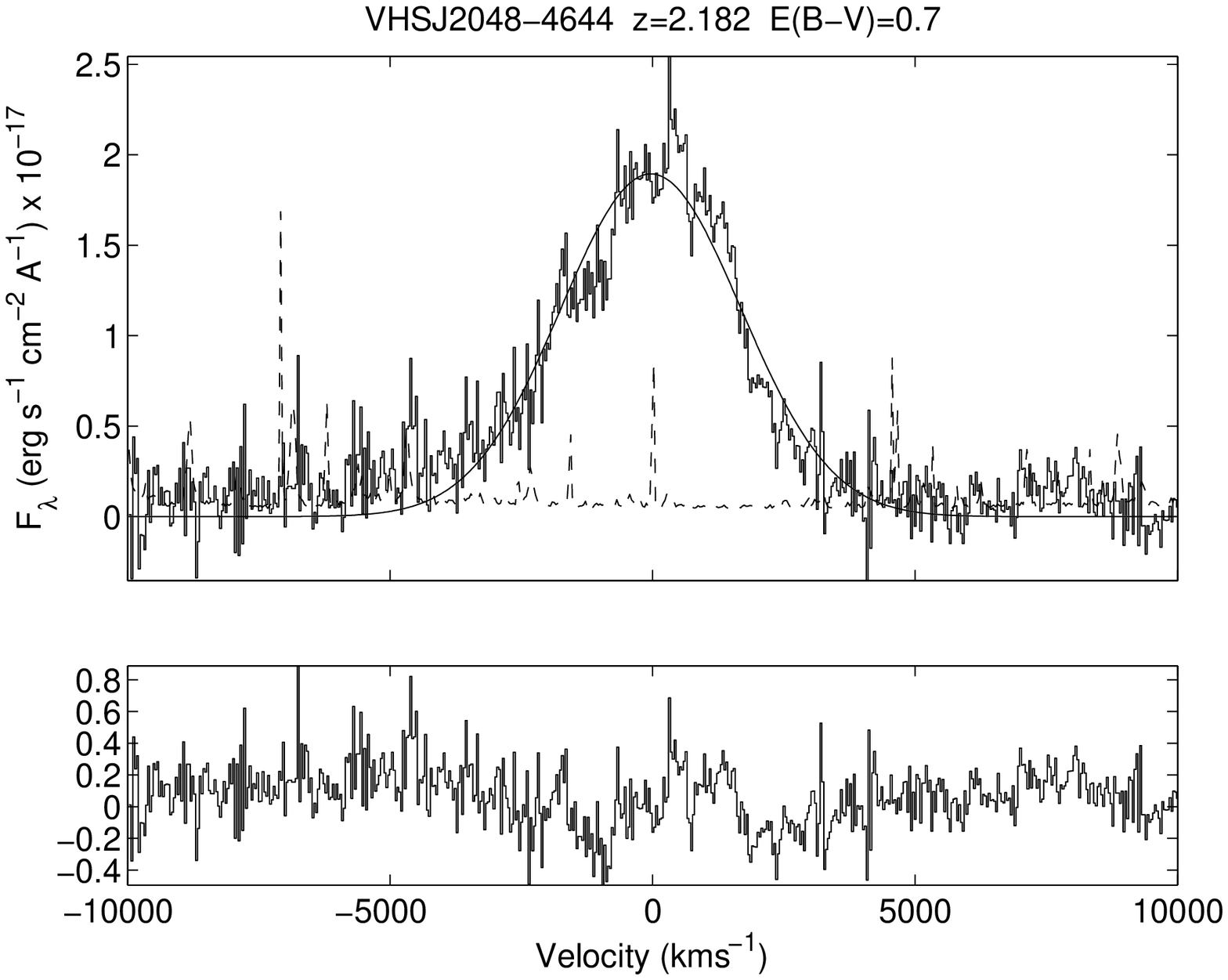} \\
\includegraphics[scale=0.35,angle=0]{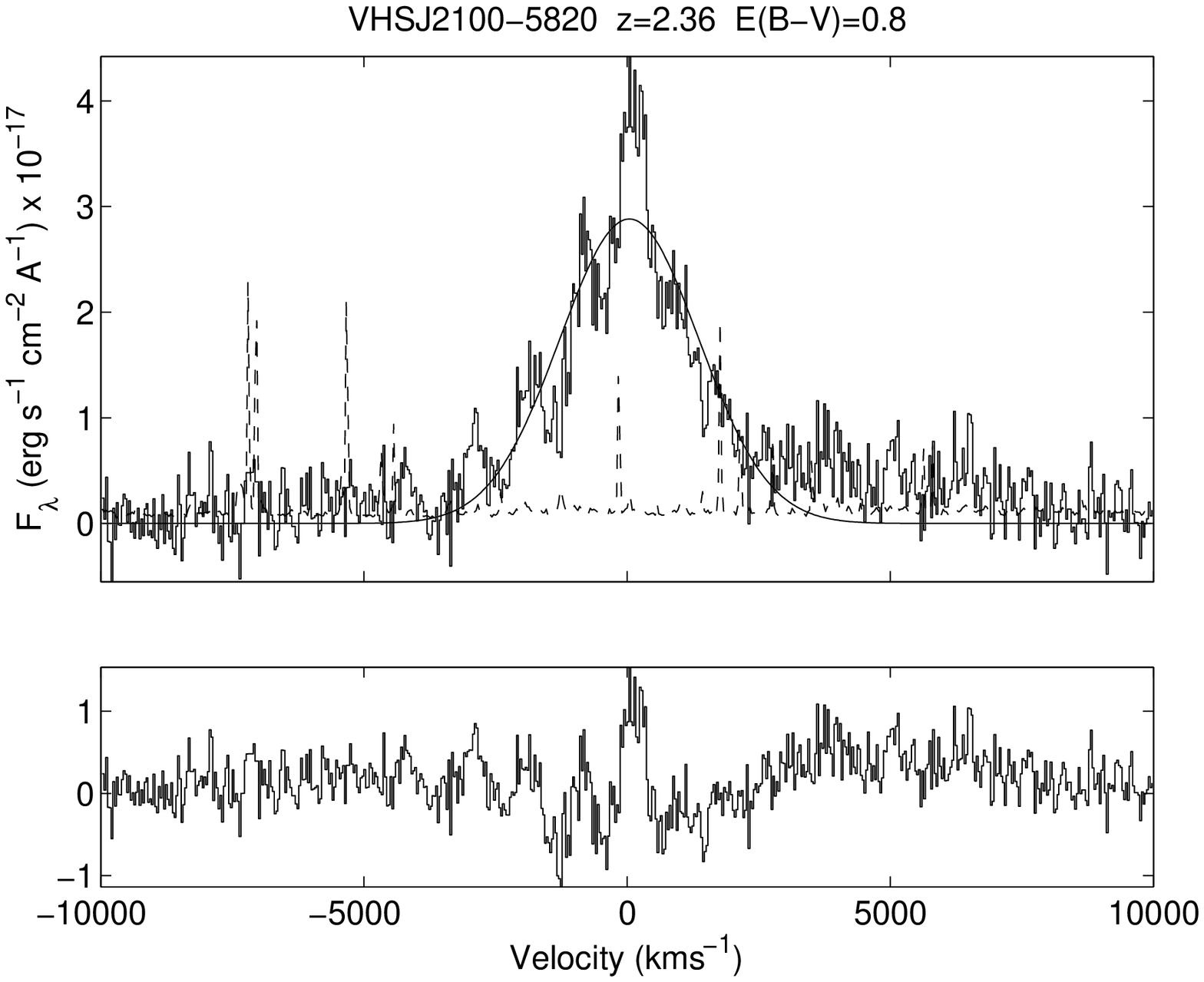} & \includegraphics[scale=0.35,angle=0]{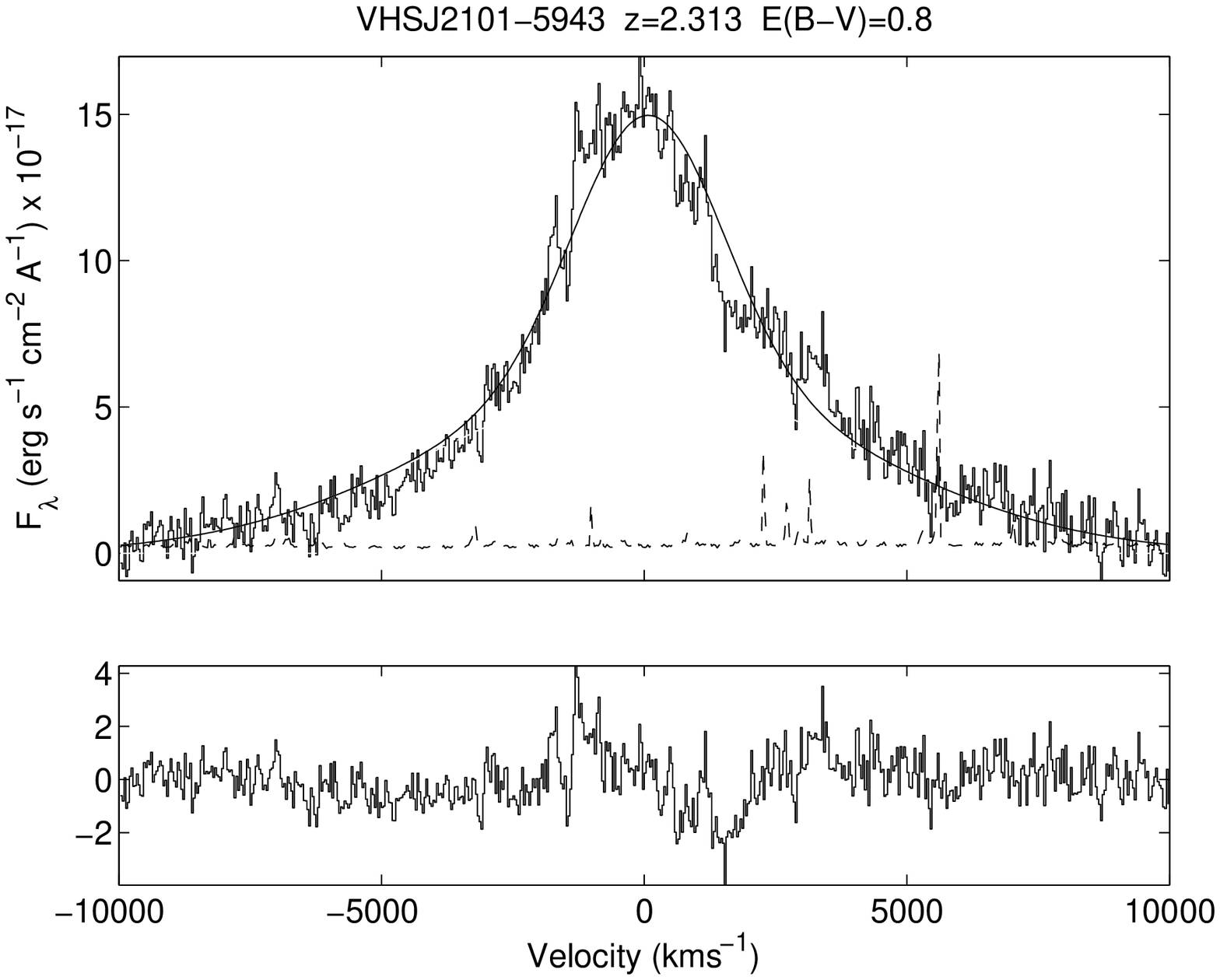} \\
\includegraphics[scale=0.35,angle=0]{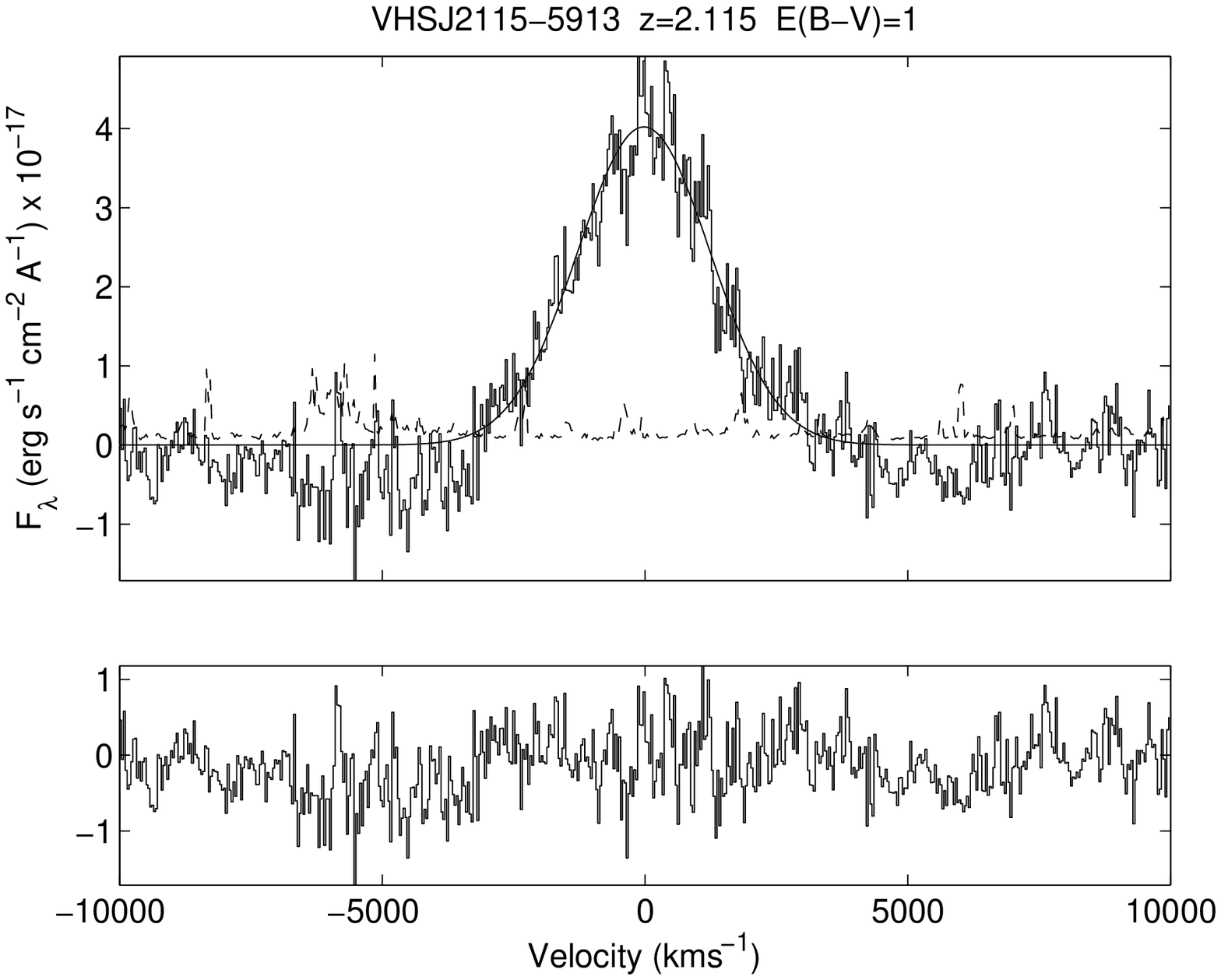}  & \includegraphics[scale=0.35,angle=0]{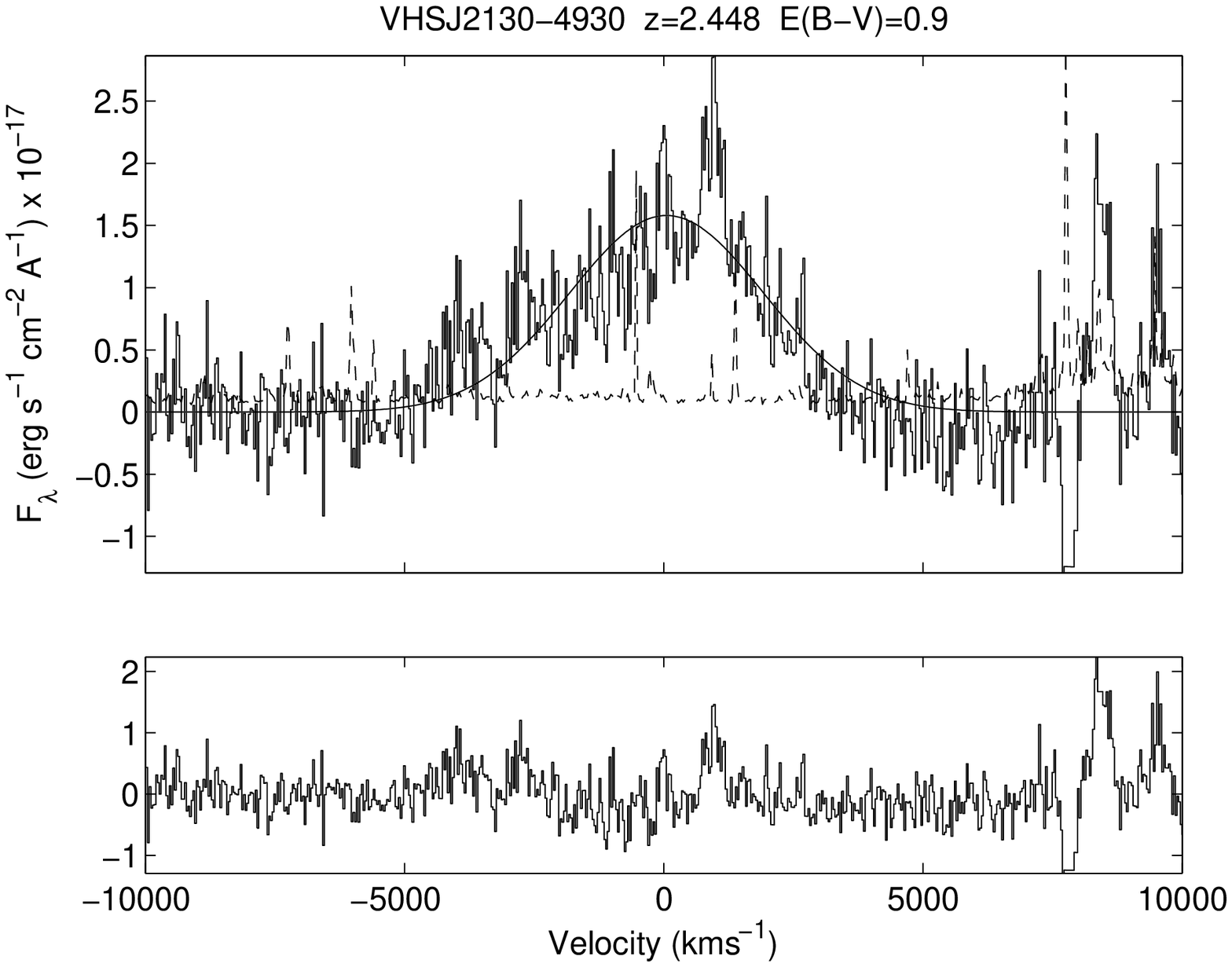} \\
\end{tabular}
\caption{H$\alpha$ line profiles and Gaussian line fits. Bottom panel
shows the residuals and the dashed line is the wavelength dependent
noise spectrum.}
\label{fig:lines}
\end{center}
\end{figure*}

\clearpage
\setcounter{figure}{2}
\begin{figure*}
\contcaption{}
\begin{center}
\centering
\begin{tabular}{cc}
\includegraphics[scale=0.35,angle=0]{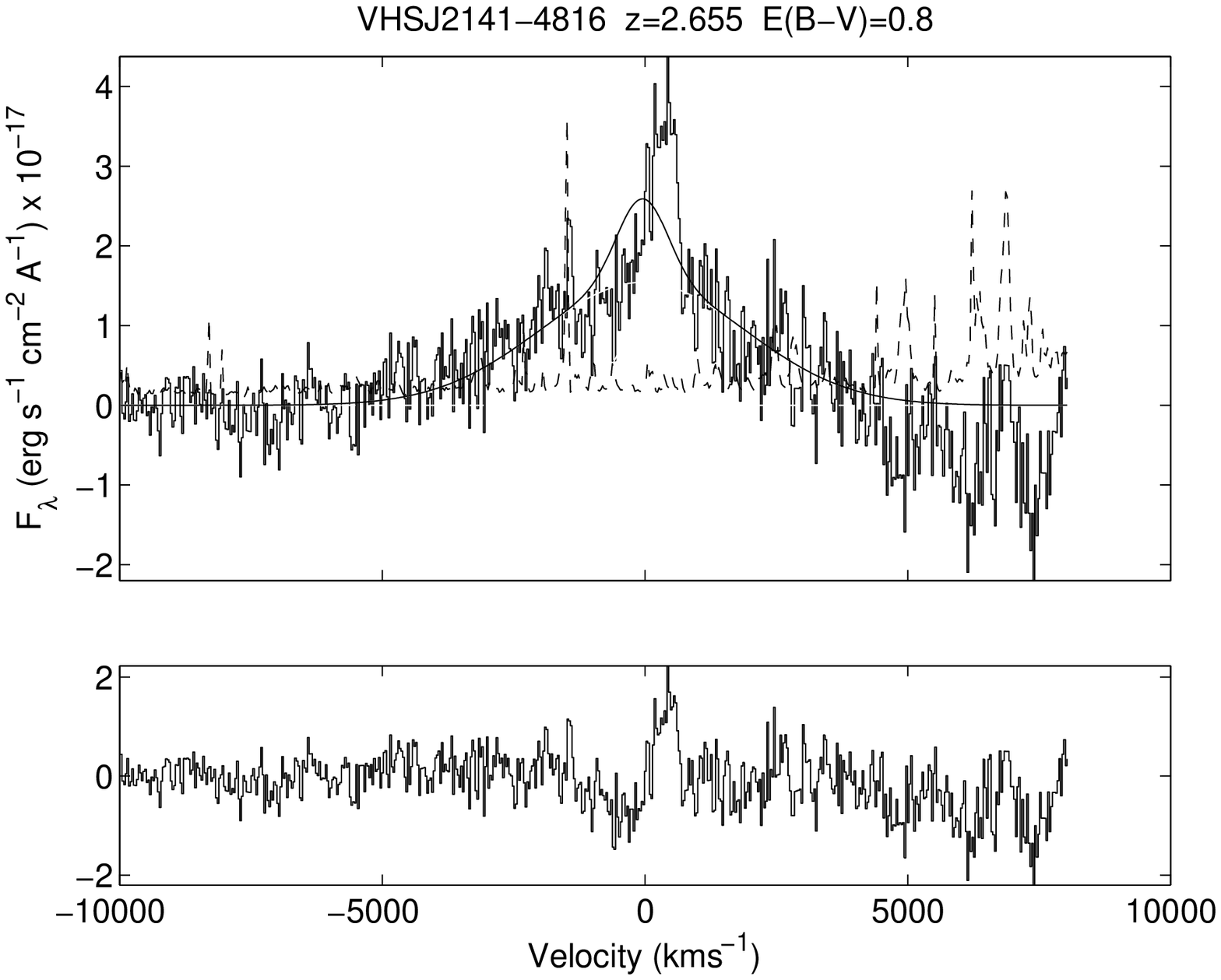} & \includegraphics[scale=0.35,angle=0]{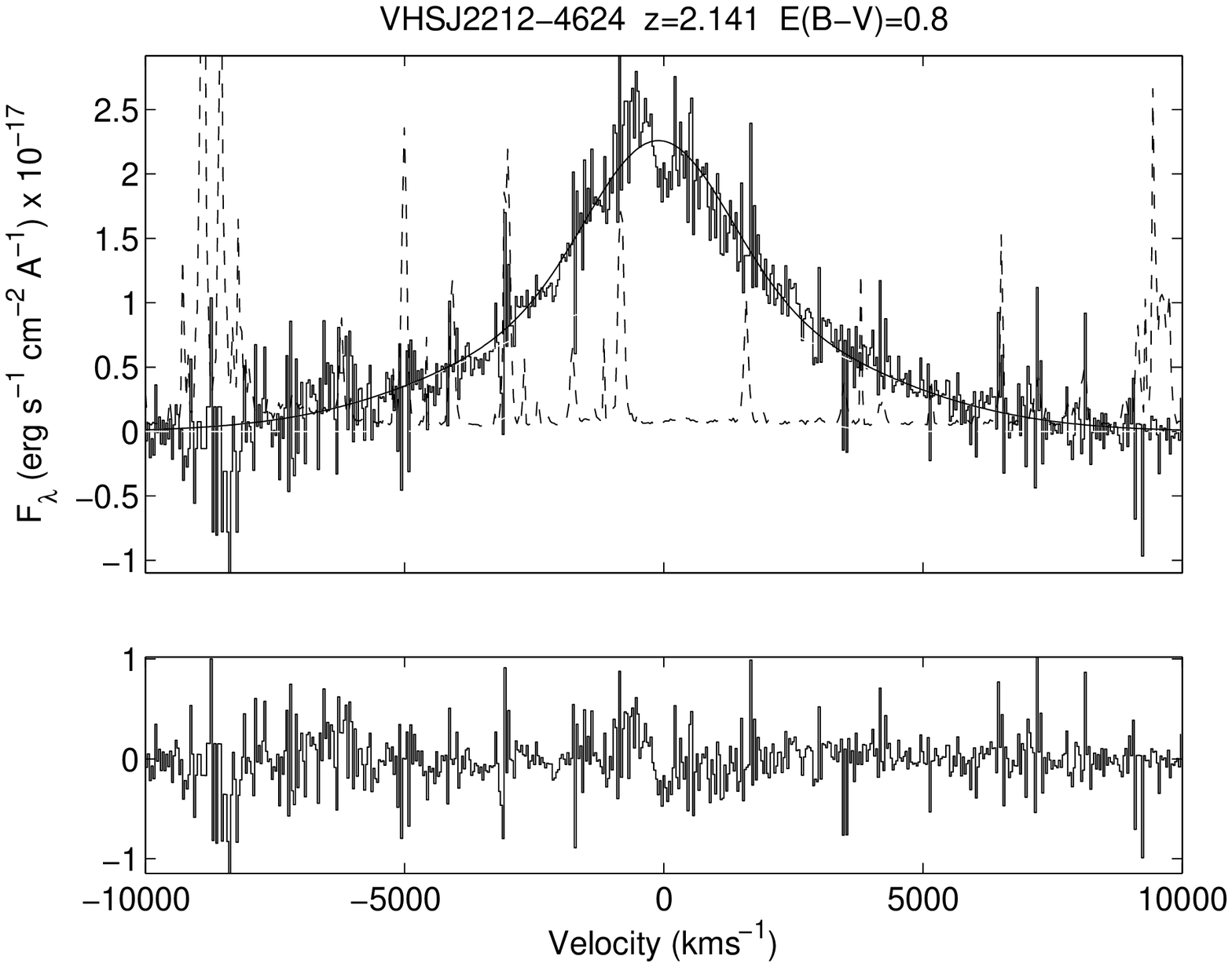} \\
\includegraphics[scale=0.35,angle=0]{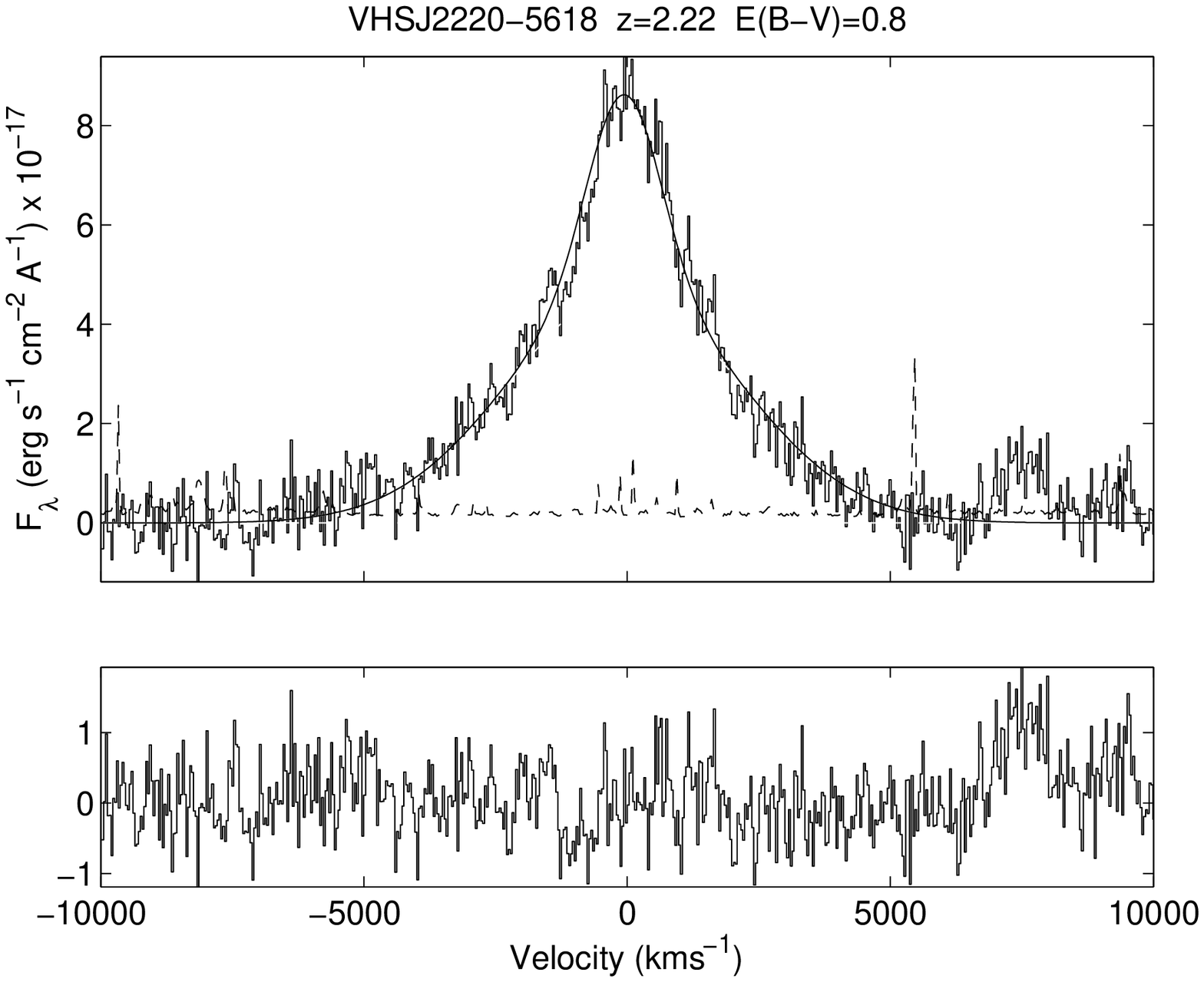} & \includegraphics[scale=0.35,angle=0]{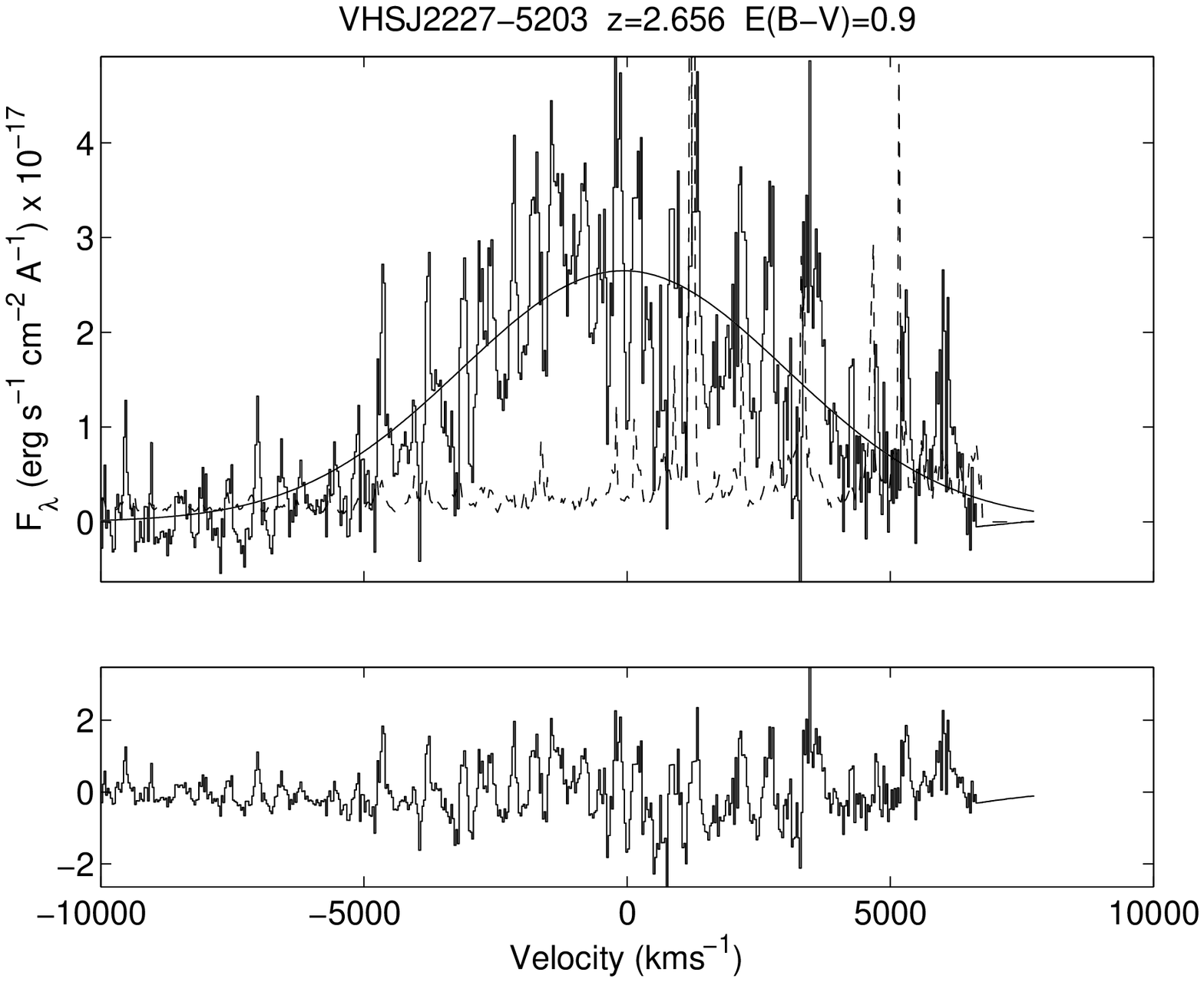} \\
 \includegraphics[scale=0.35,angle=0]{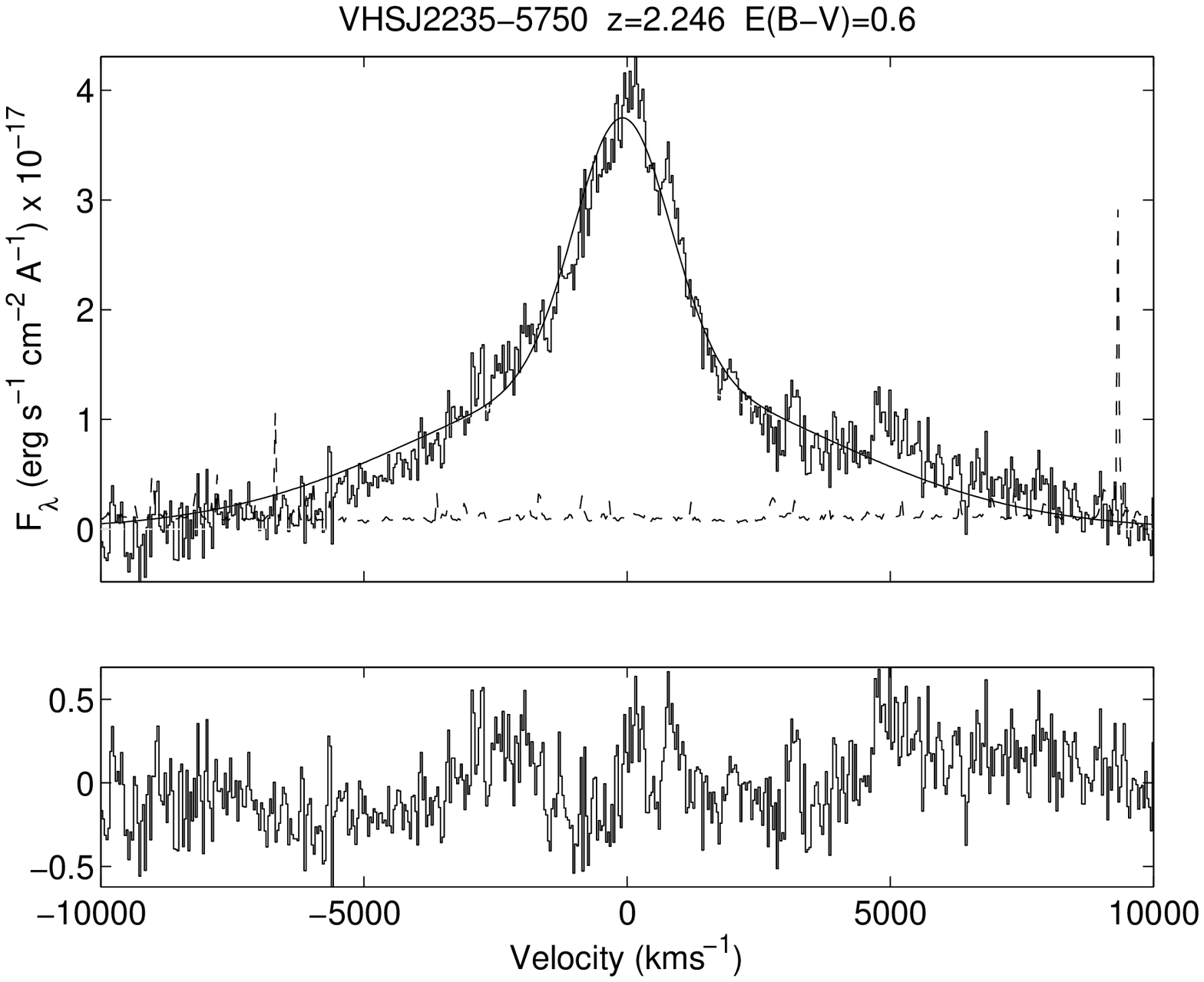} & \includegraphics[scale=0.35,angle=0]{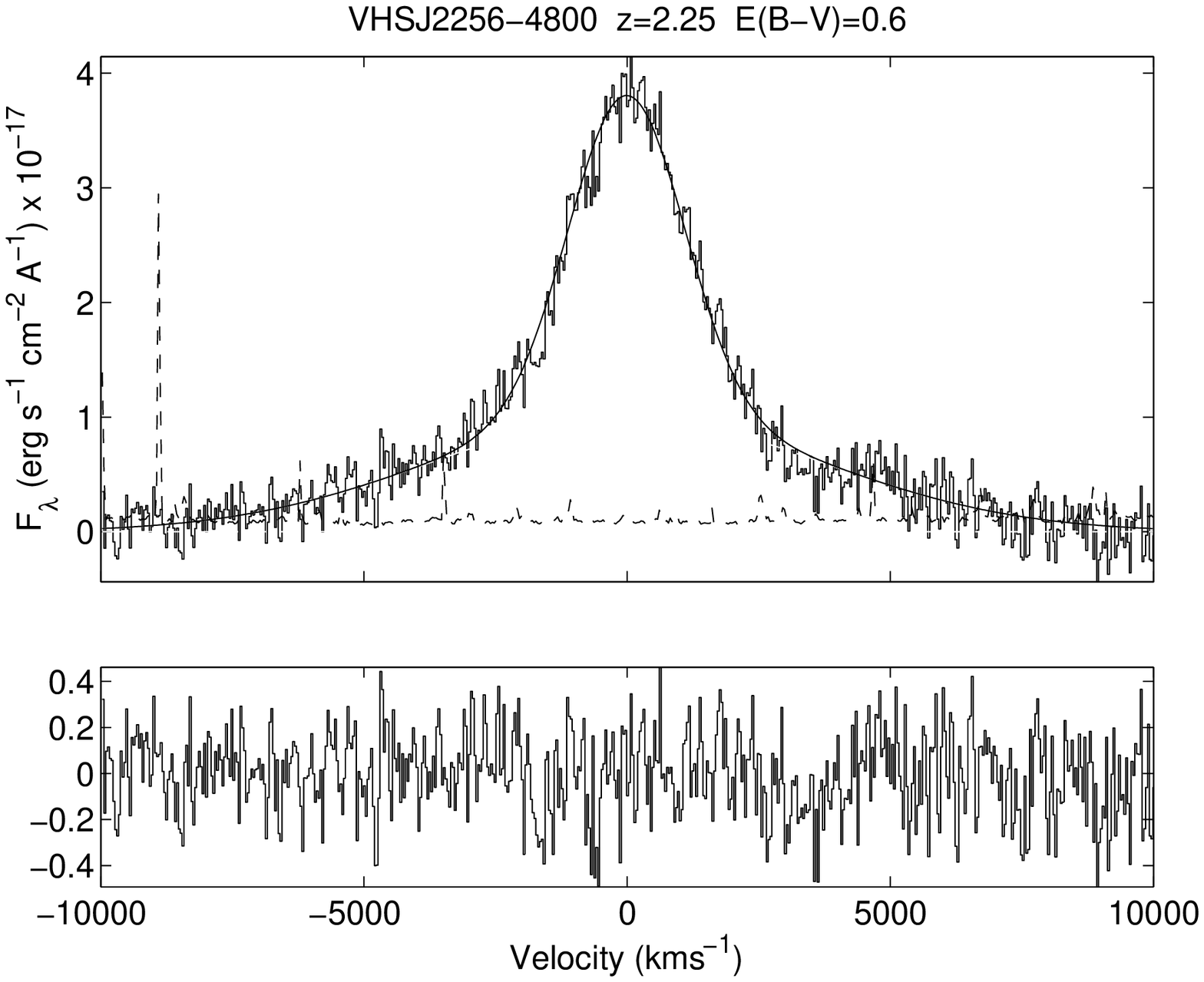} \\
 \includegraphics[scale=0.35,angle=0]{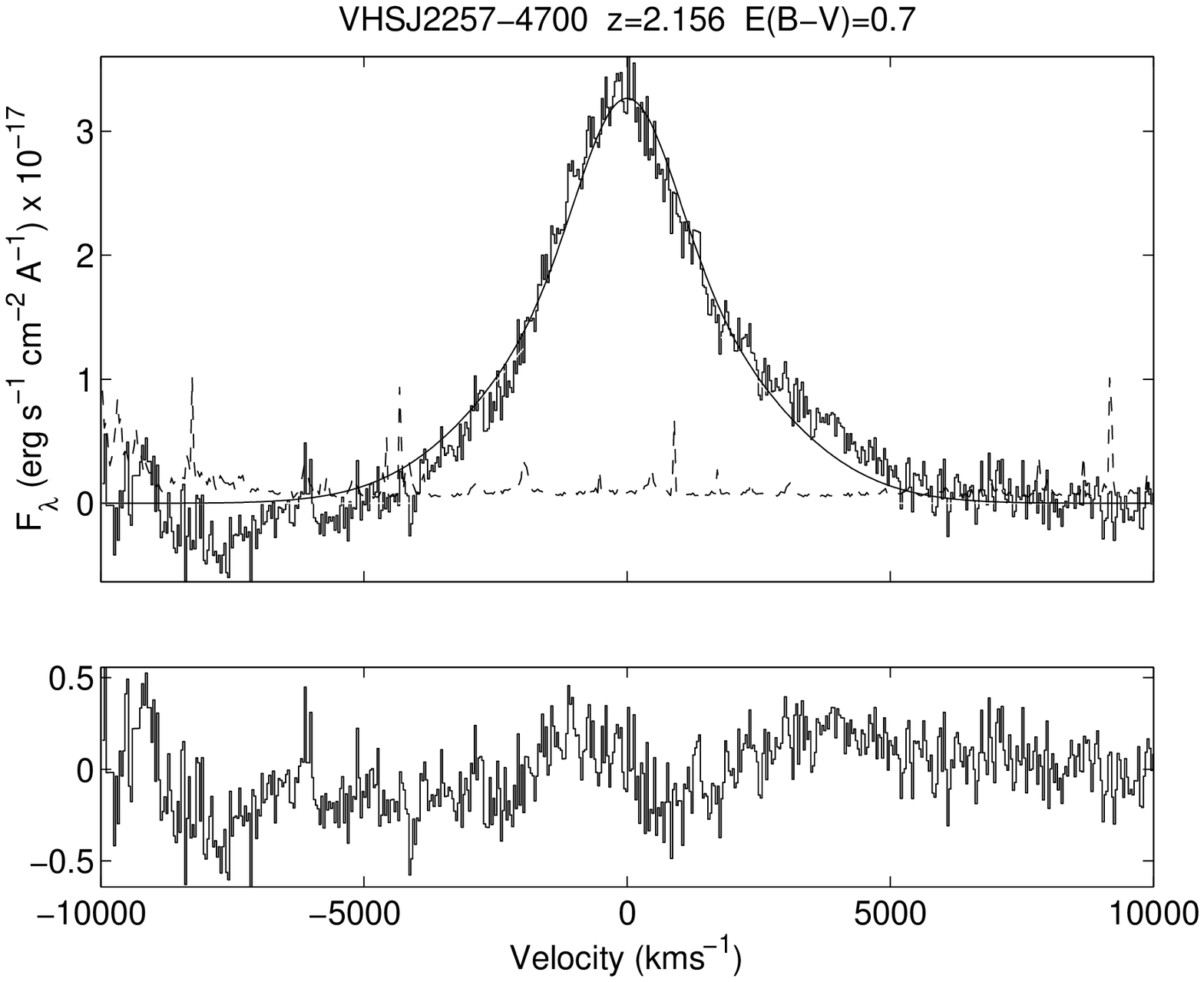} & \includegraphics[scale=0.35,angle=0]{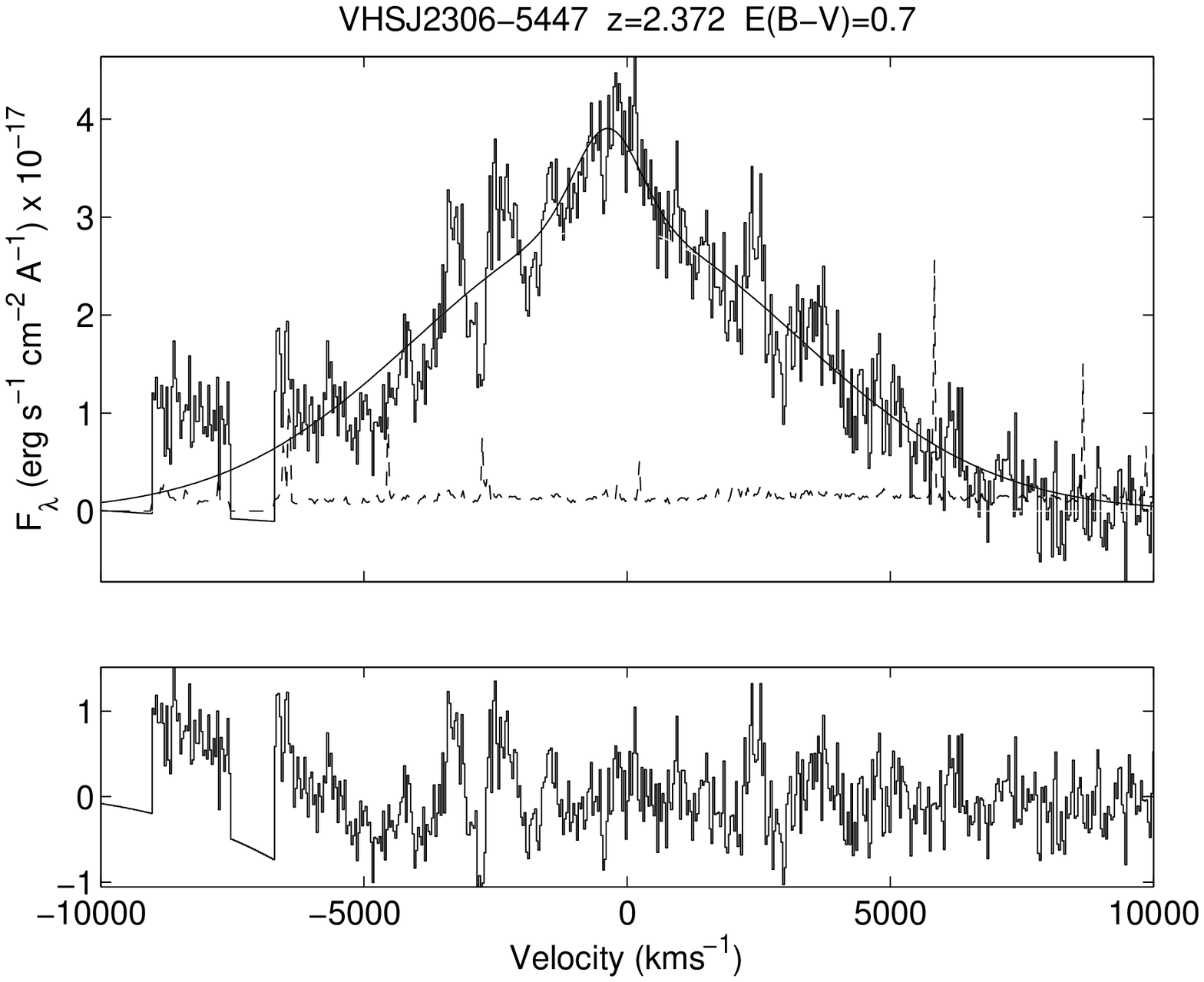} \\
\end{tabular}
\end{center}
\end{figure*}

\clearpage
\setcounter{figure}{2}
\begin{figure*}
\contcaption{}
\begin{center}
\centering
\begin{tabular}{cc}
\includegraphics[scale=0.35,angle=0]{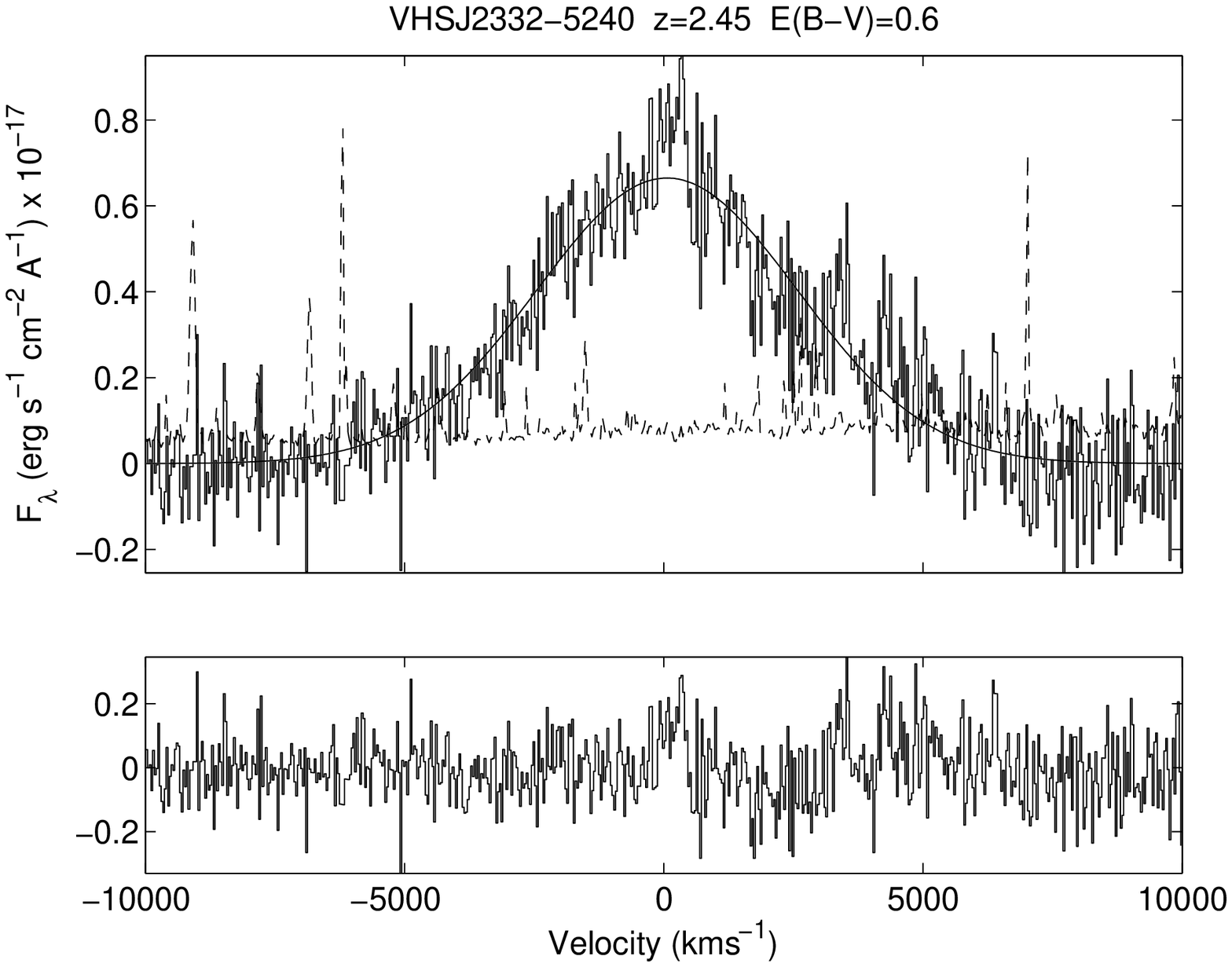}  & \includegraphics[scale=0.35,angle=0]{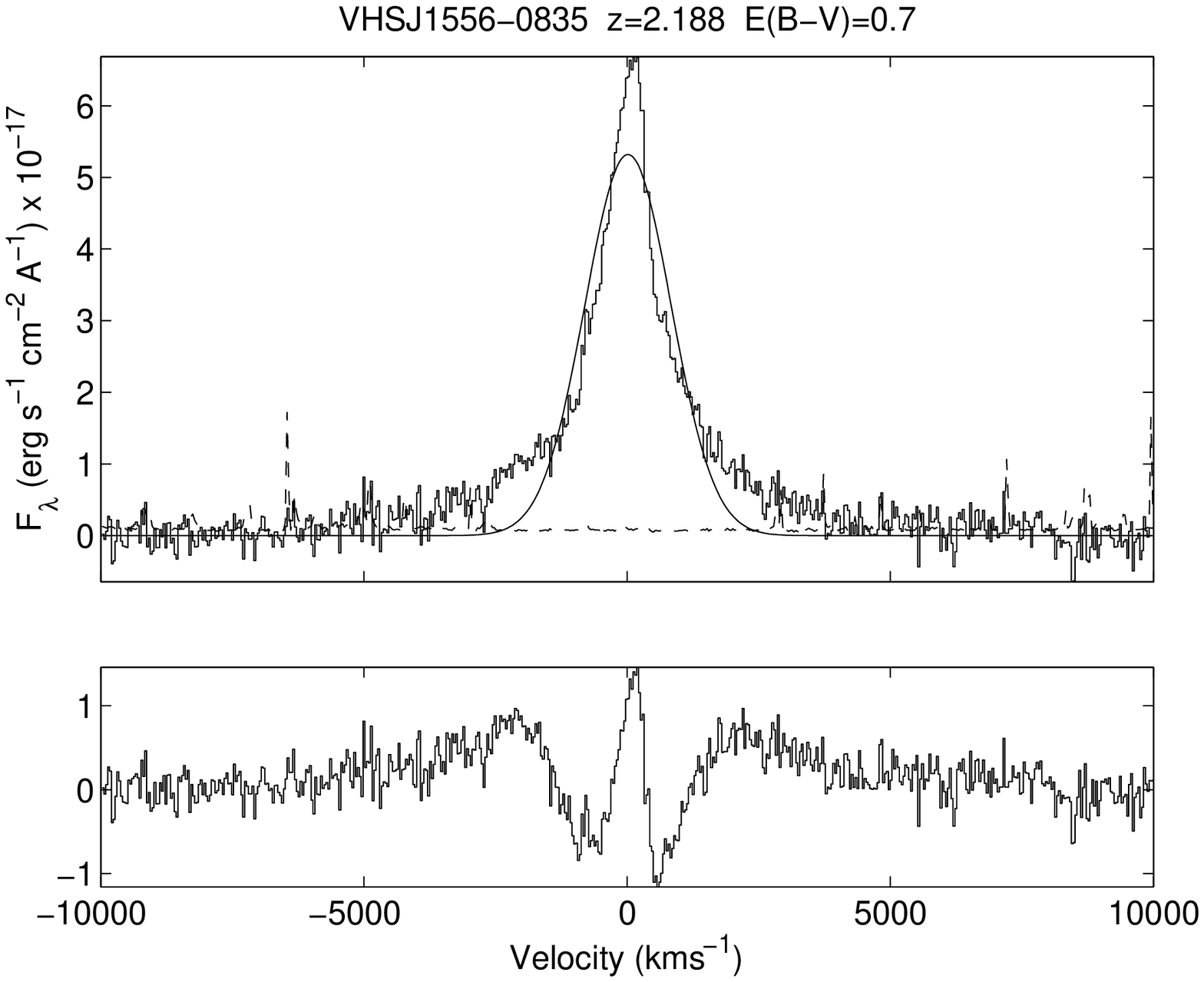} \\
\includegraphics[scale=0.35,angle=0]{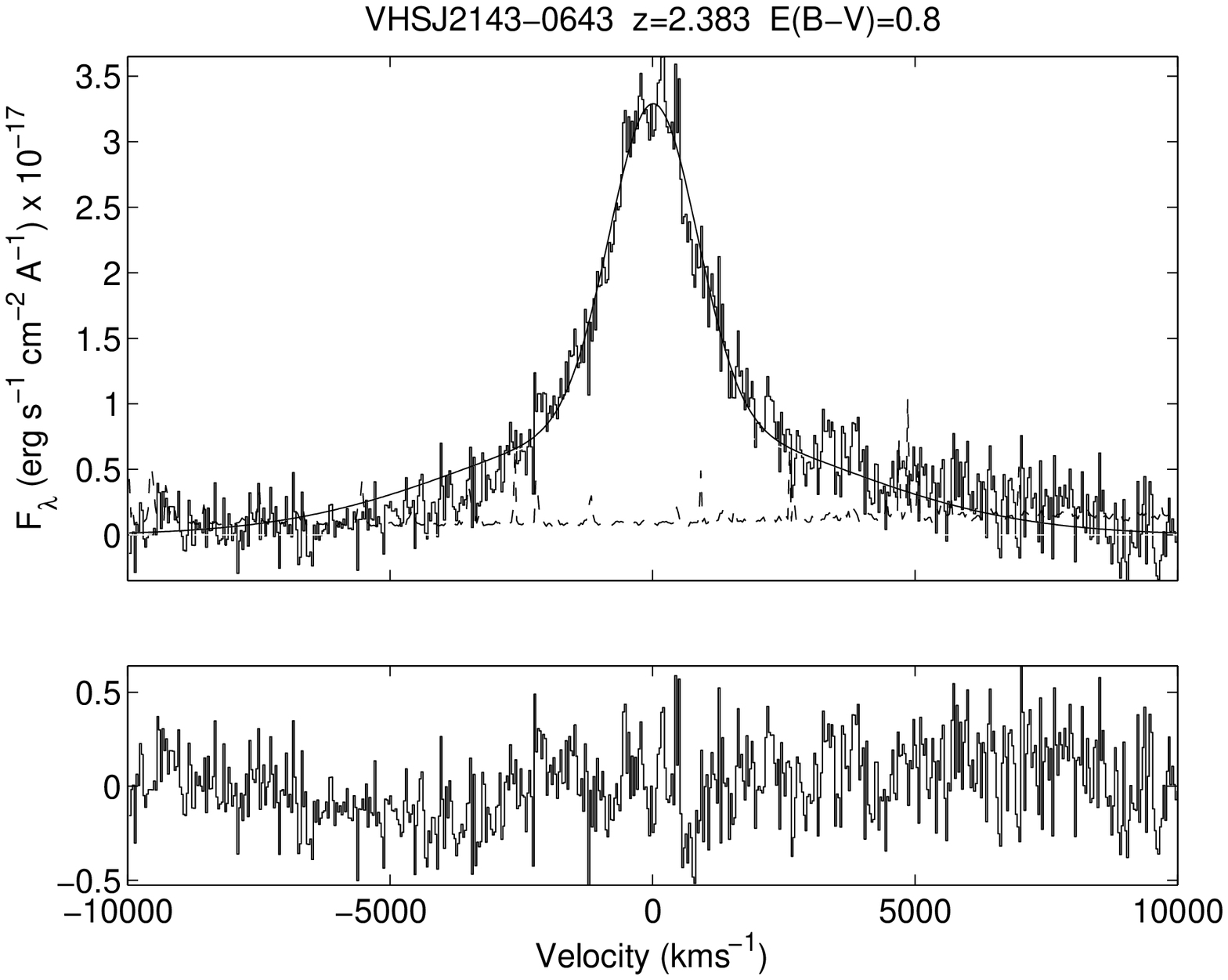} & \includegraphics[scale=0.35,angle=0]{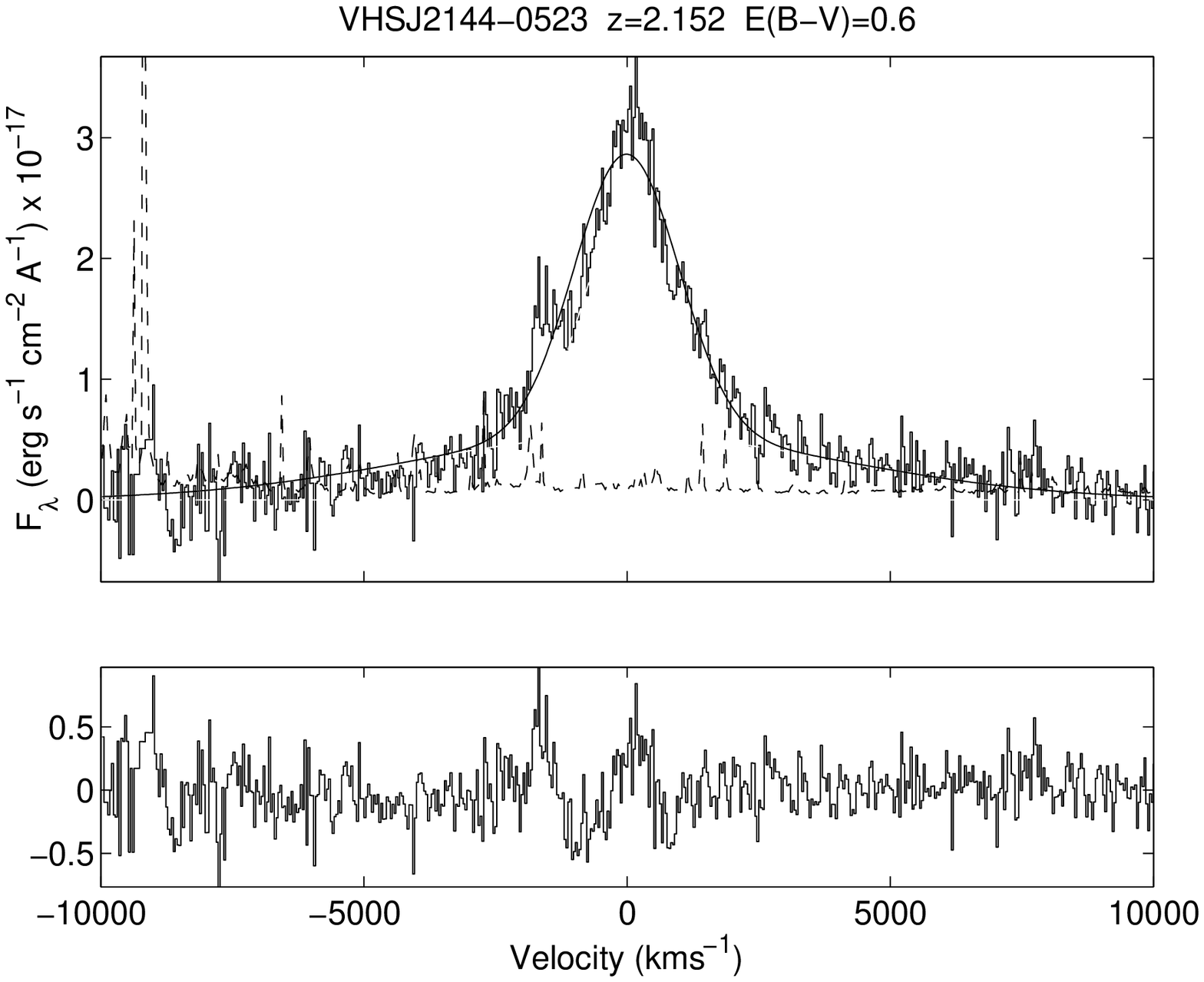} \\
\includegraphics[scale=0.35,angle=0]{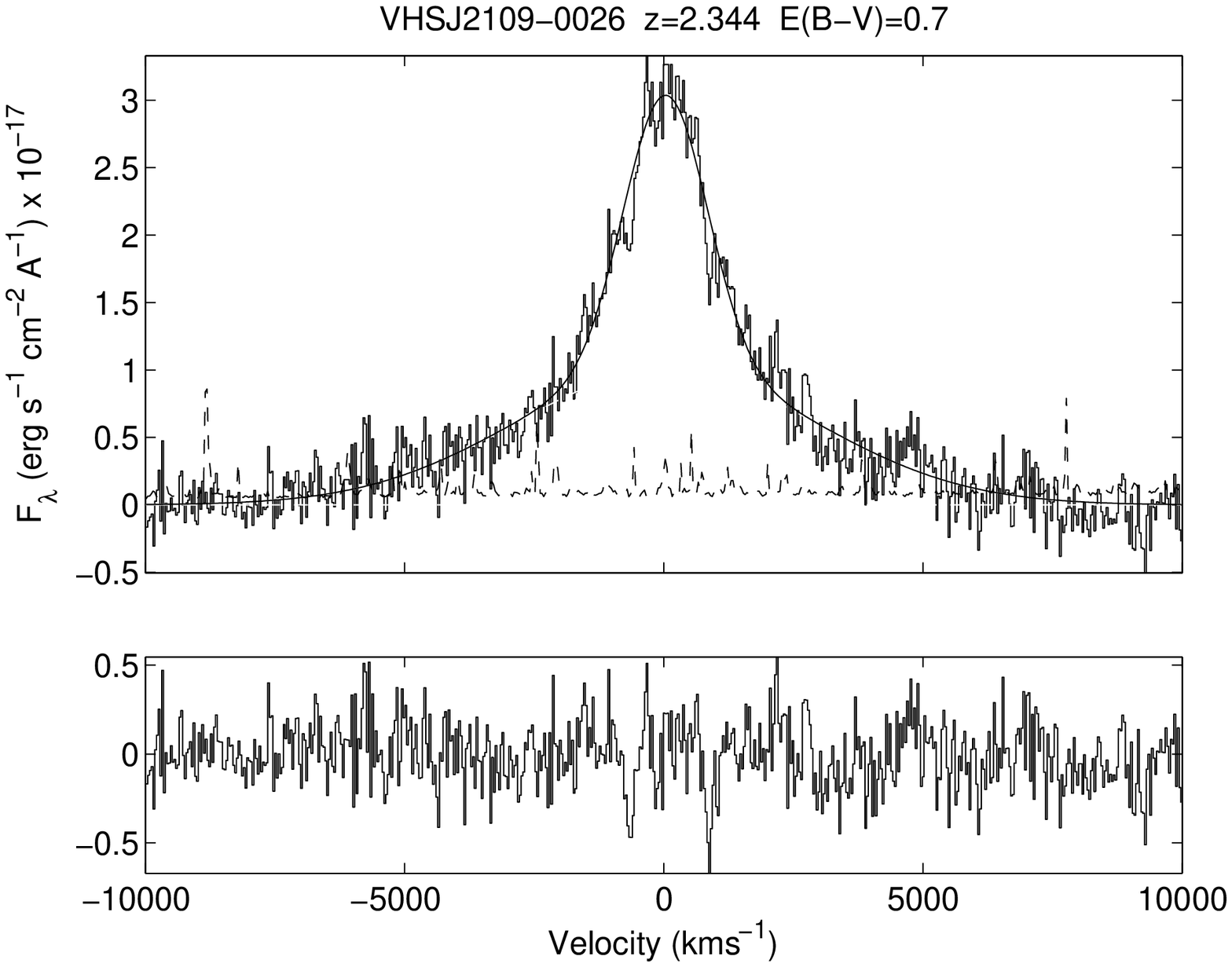} & \includegraphics[scale=0.35,angle=0]{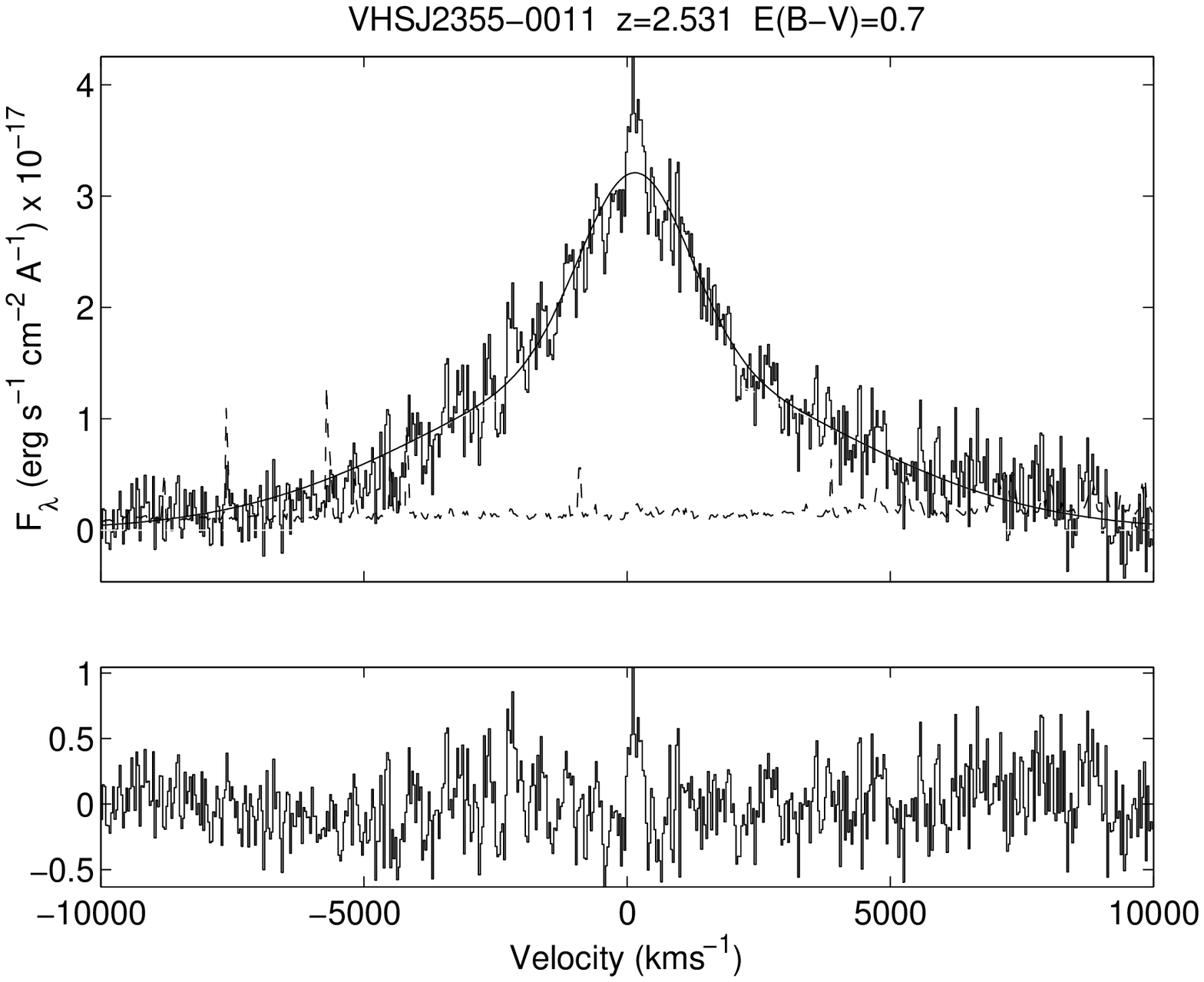} \\
\includegraphics[scale=0.35,angle=0]{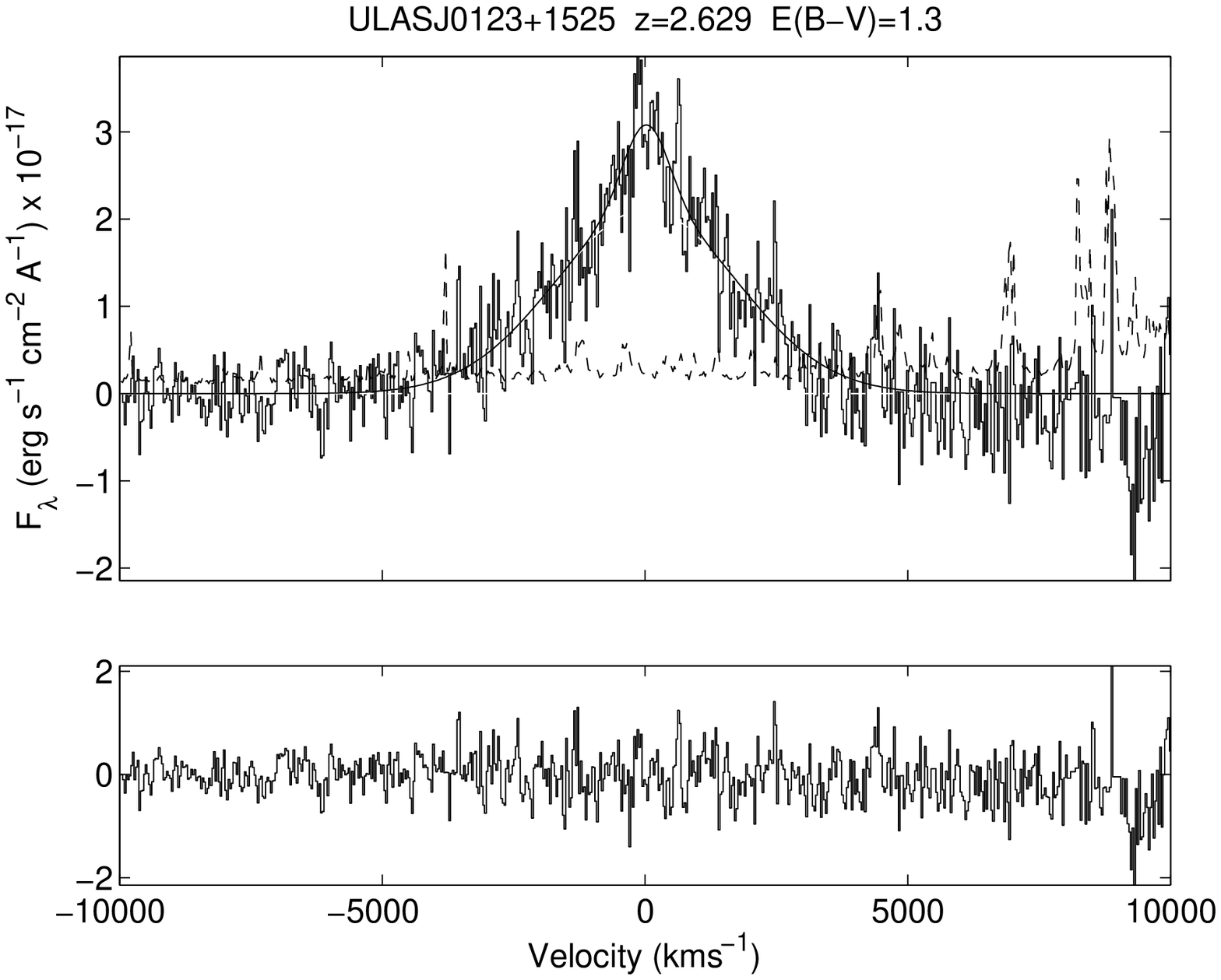} & \includegraphics[scale=0.35,angle=0]{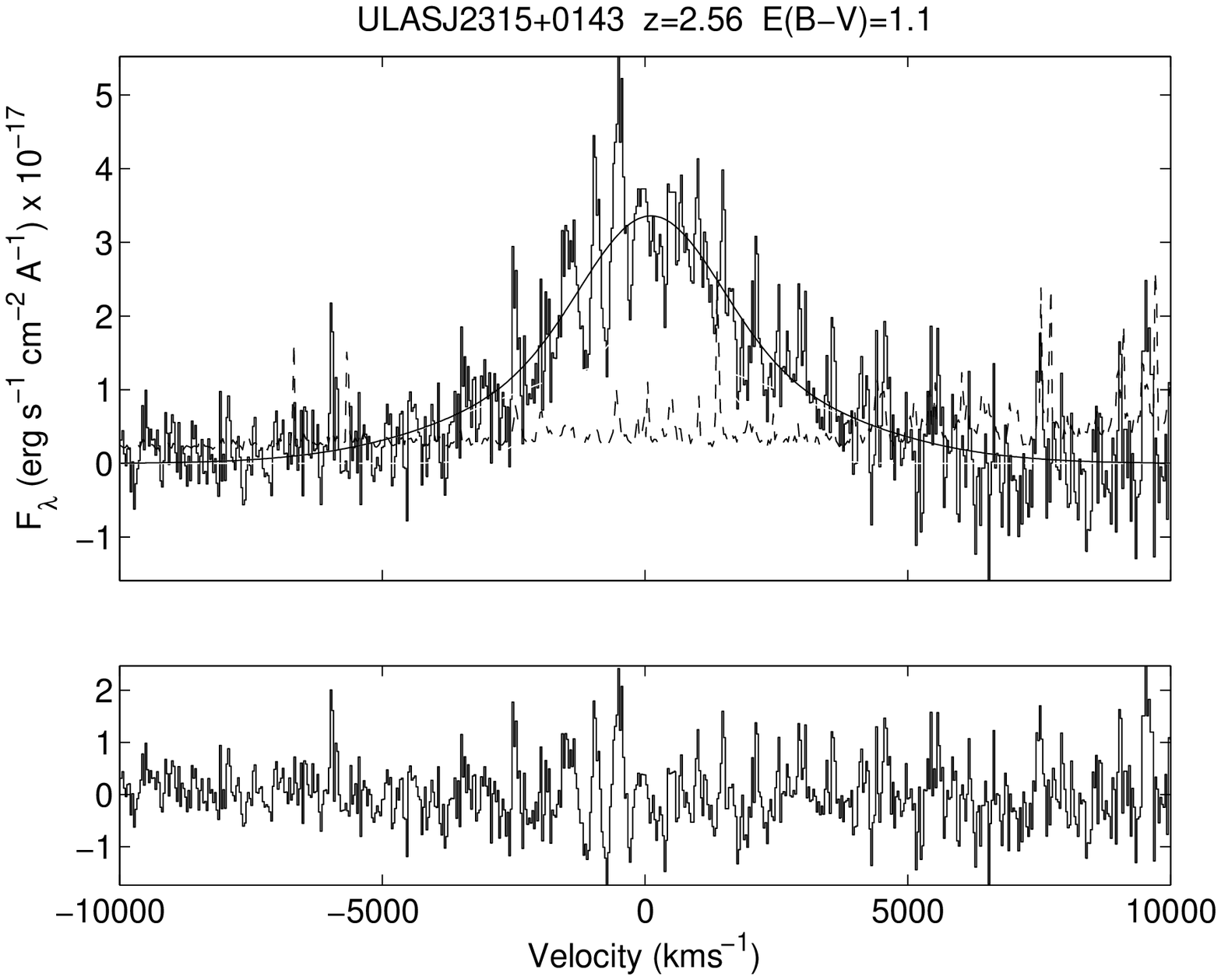} \\
\end{tabular}
\end{center}
\end{figure*}

\section{SED Model and k-corrections}

Our computations of the quasar bolometric luminosity, absolute magnitude and resulting space density relative to unobscured quasars, implicitly assume the quasar SED model described in Section \ref{sec:sed}. As this SED model is different from several others used in the literature, we provide in Table \ref{tab:kcorr} the $i$-band and $K$-band k-corrections for this model over the redshift range of our survey. 

\begin{table}
\begin{center}
\caption{K-corrections in the $i$-band and $K$-band derived using our quasar SED model from Section \ref{sec:sed}.}
\label{tab:kcorr}
\begin{tabular}{ccc}
Redshift & $k_i$ & $k_K$ \\
\hline
\hline
2.00 & -1.084 & 0.506 \\
2.10 & -1.078 & 0.287 \\
2.20 & -1.069 & 0.151 \\
2.30 & -1.061 & 0.118 \\
2.40 & -1.064 & 0.091 \\
2.50 & -1.071 & 0.092 \\
2.60 & -1.099 & 0.131 \\
2.70 & -1.153 & 0.333 \\
2.80 & -1.181 & 0.345 \\
2.90 & -1.195 & 0.321 \\
3.00 & -1.196 & 0.292 \\
\hline
\end{tabular}
\end{center}
\end{table}

\end{document}